\documentclass{jaa}
\usepackage{threeparttable}
\usepackage{amsmath}
\usepackage[utf8]{inputenc}
\usepackage{longtable}
\usepackage{graphicx}
\usepackage{natbib}
\usepackage{multirow}
\usepackage[Table,xcdraw]{xcolor}
\usepackage{rotating}
\newcommand{\apj}{Astrophysical Journal}
\newcommand{\solphys}{Solar Physics}
\newcommand{\mnras}{Monthly Notices of the Royal Astronomical Society}
\newcommand{\apss}{Astrophysics and Space Science}
\newcommand{\baas}{Bulletin of the AAS}
\newcommand{\physrep}{Physics Reports}

\newcommand{\apjl}{Astrophysical Journal Letter}

\newcommand{\aap}{Astronomy and Astrophysics}
\newcommand{\aapr}{Astronomy and Astrophysics Reviews}
\newcommand{\aj}{Astronomical Journal}
\newcommand{\procspie}{Proc. SPIE}

\usepackage{hyperref}
\hypersetup{backref=true,
    pagebackref=true,
    hyperindex=true,
    colorlinks=true,
    breaklinks=true,
    urlcolor= black,
    linkcolor= blue,
    bookmarks=true,
    bookmarksopen=false,
    filecolor=black,
    citecolor=blue,
    linkbordercolor=blue
}

\begin{document}\sloppy

\title{Hard X-ray Polarimetry $-$ An overview of the method, science drivers and recent findings}


\author{Tanmoy Chattopadhyay}
\affilOne{Kavli Institute of Astrophysics and Cosmology, 452 Lomita Mall, Stanford, CA 94305, USA.\\}


\twocolumn[{

\maketitle

\corres{tanmoyc@stanford.edu, tanmoyrng1@gmail.com}

\msinfo{---}{---}

\begin{abstract}
{The last decade has seen a leapfrog in the interest in X-ray polarimetry with a number of new polarization measurements in hard X-rays from {\em AstroSat}, POLAR, GAP, and PoGO+. The measurements provide some interesting insights into various astrophysical phenomena such as coronal geometry and disk-jet connection in black hole X-ray binaries, hard X-ray emission mechanism in pulsars and Gamma Ray Bursts (GRB). They also highlight an increase in polarization with energy which makes hard X-ray polarimetry extremely appealing. There is a number of confirmed hard X-ray polarimetry experiments which along with the existing instruments ({\em AstroSat} and {\em INTEGRAL}) makes this field further exciting. Polarization experiments may also see a significant progress in sensitivity with new developments in scintillator readouts, active pixel sensors, CZT detectors. In particular, the advent of hard X-ray focusing optics, will enable designing of compact focal plane polarimeters with a multifold enhancement in sensitivity. In this review, we will focus on the recent polarimetry findings, science potential of hard X-ray polarimetry along with possible improvements in the measurement techniques.} 

\end{abstract}


}]


\doinum{12.3456/s78910-011-012-3}
\artcitid{\#\#\#\#}
\volnum{000}
\year{0000}
\pgrange{1--}
\setcounter{page}{1}
\lp{1}

\section{Introduction}
 Since the birth of X-ray astronomy, X-ray spectroscopy, timing,
and imaging have 
met with significant advancement, while the study of the fourth parameter
of radiation or polarization remains majorly unexplored
in spite of its extremely high science throughput.
Apart from some coarse polarization detections, there have not been any
dedicated X-ray polarization experiment  
primarily due to the low sensitivity of 
polarimetry compared to spectroscopy, imaging or timing; which
results due to the extremely photon hungry nature of X-ray polarimetry.

The first successful measurement of polarization 
in X-ray astronomy dates to 1976 when an X-ray polarimeter on board 
{\em OSO-8} mission, measured $\sim$19 \% polarization at 2.6 and 5.2 keV 
for the Crab nebula \citep{weisskopf76,weisskopf78}. There were 
a number of attempts to measure X-ray polarization  with the same polarimeter as well 
as a few other space-borne and balloon-borne experiments 
\citep{angel69,wolff70,novick72,griffiths76,gowen77,silver79,hughes84} 
but these could yield only upper limits at best, due to the low 
sensitivity of these experiments. After these initial efforts, no real 
experiments to measure X-ray polarization were carried out for more than three decades. 
{Though there were some attempts to design and build the X-ray polarimeters 
\citep[e.g.][]{angel70,tsunemi92} and few concept proposals 
for space missions \citep[e.g. XPE, PLEXAS,][]{elsner97,marshall03}, 
only one instrument, Stellar X-Ray Polarimeter \citep[SXRP,][]{kaaret93} came close to a launch on the Russian Spectrum-X-Gamma (SXG) mission in 1998. However, the mission was eventually cancelled.  
Later, 
Gravity and Extreme
Magnetism Small Explorer (GEMS), a dedicated X-ray polarimetry mission \citep{jahoda10} 
was selected for launch in 2014. The mission was also eventually
cancelled due to programmatic issues
\citep[see a review on the progress and development in the X-ray polarimetry field from the initial days by][]{weisskopf18}.}
In the absence of any dedicated X-ray polarimetry experiment,
in the last decade there have been a few attempts to measure polarization for 
bright X-ray sources by standard spectroscopic instruments which have
moderate polarization measurement capability. To name a few, the most 
important results are the detection of high polarization from Cygnus X-1 and 
Crab nebula by IBIS and SPI on board {\em INTEGRAL} \citep{laurent11,jourdain12,forot08,dean08}
and Gamma Ray Bursts by {\em RHESSI} and {\em INTEGRAL} \citep{coburn03,gotz09,mcgltnn07,mcconnell09}. In most cases, due to the lower polarimetric sensitivity of the instruments, those detections are far from being conclusive \citep{rutledge04,wigger04}.  

In recent times, there are some interesting polarization measurement reports in hard X-rays from Cadmium Zinc Telluride Imager (CZTI) on board {\em AstroSat} satellite \citep{bhalerao16,rao16}, balloon borne instrument, PoGO+ \citep{chauvin16_pogo}, POLAR on board Chinese space laboratory \citep{produit18,sun16}, GAP on board IKAROS spacecraft \citep{yonetoku06} for bright hard X-ray sources \citep{vadawale17,chattopadhyay19,chauvin18a,chauvin18b,chauvin18bcrab,zhang19,yonetoku11,yonetoku12}. 
These reports along with those from last two decades 
\citep{mcconnell02,coburn03,laurent11,jourdain12,forot08,dean08,moran13,gotz09,gotz13,mcglynn07,mcglynn09} have significantly enhanced the interest in the X-ray polarimetry \citep{marin18}. With {\em IXPE}, a dedicated soft X-ray polarimetry mission \citep{weisskopf16,soffitta17} set to launch in 2021, and a couple of upcoming dedicated hard X-ray polarimetry missions \citep[e.g. {\em POLIX},][]{paul10,paul16}, 
we expect an exciting time ahead in X-ray polarimetry. 

\subsection{\bf {\textit{Hard X-ray Polarimetry: Why?}}}
\label{polhard}
Scientific importance of X-ray polarimetry has been discussed extensively in 
literature \citep{lei97,soffitta97,krawczynski11}. In Table \ref{polscience_Table}, the science drivers of X-ray polarimetry at different energy bands have been summarized for different classes of X-ray sources. 

\begin{table*}[h!]
\begin{tiny}
\begin{threeparttable}
	\caption{Summary of science potential of X-ray polarimetry at different energy bands.}\label{polscience_Table}
\begin{tabular}{p{0.5cm}p{0.7cm}p{5.5cm}p{1.5cm}p{2.7cm}p{1.4cm}p{2.8cm}}
\topline
\\
Energy & Source class & Science drivers & Predicted PF \% & Available measurements& Active missions& Upcoming/planned missions\\
\\
\midline
\\
{\bf 0.1$-$1 keV}&BH XRBs, AGNs&{BH spin, GR effects from polarized thermal disk emission \citep{dovciak08,schnittman09}}&$\sim$1-3 \%&&\\

&Blazars&{First peak of HBL $-$ Jet structure, magnetic field geometry \citep{tavecchio18,zhang14}}&a few to high \% depending on the B field geometry&&&\\

&Magnetars&{Vacuum birefringence from polarized persistent thermal emission with KT$\sim$0.5 keV \citep{heyl02}}&$>$80 \%&&&\\

&XDINS$^1$ / XRP$^2$&{M/R ratio, magnetic field geometry, angle between magnetic axis, rotation axis and viewing direction from polarized thermal emission \citep{pavlov00}}&$\sim$10-20 \%&& &\\

&Magnetic CVs&{Density profile in accretion column from polarized scattered photons off the polar accretion column \citep{Wu09}}&a few to 8 \%&&&\\
\\
\hline
\\
{\bf 1$-$10 keV}&BH XRBs, AGNs&{BH spin, inclination, study of accretion disk dynamics, GR effects from  polarized thermal disk emission \citep{dovciak08,schnittman09}}&$\sim$1-3 \%&A few \% of PF for Cygnus X-1 in high soft state \citep{long80}&&{\em IXPE} (2021) \citep{weisskopf16,soffitta17}, EXTP \citep{zhang16_extp}, SOLPEX \citep{steslicki16}\\

&Blazars&{First peak of HBL (faint in this energy range) $-$ Jet structure, magnetic field geometry \citep{zhang14,tavecchio18}}&a few to high \% depending on the B field geometry&&\\

&Magnetars&{Vacuum birefringence, signature of RICS \citep{taverna14}}&$\sim$20 \%&&&\\

&XRP&{Emission mechanism, geometry - Polar cap/Outer gap/slot gap/striped wind models \citep{weisskopf09,Harding19_book}}&a few tens of \%&$\sim$15 \% PF for Crab+nebula by PolarLight \citep{feng20}&&\\

&SNR, PWN&{Particle acceleration process in shocks, magnetic field geometry \citep{bykov09}}&$\sim$20-50 \% from different regions of SNR&$\sim$19 \% PF for Crab nebula by {\em OSO-8} \citep{weisskopf78}. $\sim$14 \% PF by PolarLight \citep{feng20}&&\\

\\
\hline
\\
{\bf 10$-$100 keV}&BH XRBs&{Corona geometry (Lamp Post/wedge/ spherical/clumpy) from polarized coronal emission \citep{schnittman10}}&$\sim$5 \%&Upper limit of a few \% for Cygnus X-1 by PoGO+ \citep{chauvin18a}&&{\em POLIX} (2021) \citep{paul10,paul16}, {\em PolariS} \citep{hayashida14}, PING-P (2025) \citep{kotov16}\\

&AGNs&{Corona and torus geometry \citep{schnittman10,goosmann11}}&$\sim$10 \%&&&\\

&Blazars&{Second peak of LBL: SSC/ EC/hadronic models \citep{mcnamara09,zhang13}}&a few \% (EC) to 50 \% (SSC)&&&\\

&XRP&{Emission mechanism, geometry - Polar cap/Outer gap/slot gap/striped wind models \citep{weisskopf09,Harding19_book}}&A few tens of \%&$\sim$21 \% PF for Crab+nebula by PoGO+ \citep{chauvin18bcrab}$, \sim$22 \% PF by {\em Hitomi} \citep{hitomi18}&&\\

&Magnetars&{Signature of RICS from polarized hard X-ray tail \citep{wadiasingh19_white}}&&&\\

&APP$^3$&{Beam shape (pencil/fan beam models) from phase resolved polarimetry near cyclotron energies. Magnetic field, accretion column geometry from accretion shock/disk scattered emission \citep{meszaros88}}&up to 30 \%&&&\\

&PWN&{Acceleration mechanism and magnetic field geometry \citep{weisskopf09,Harding19_book}}&$\sim$20-30 \%&$\sim$17.4 \% PF for Crab nebula by PoGO+ \citep{chauvin18bcrab} &&\\

&Solar flares&{Emission mechanism, electron beaming, magnetic field structure for M and X class flares \citep{bai78,leach83,zharkova10}}&$\sim$35-50 \% for flares away from Sun's center&$\sim$8-40 \% PF for 25 flares by {\em CORONAS-F} \citep{zhitnik06}&&\\

\\
\hline
\\

{\bf 100$-$1000 keV}&BH XRBs &{Contribution of jet in hard X-ray flux, jet physics, disk-jet connection \citep{zdziarski14}}&$\sim$30-60 \%&$>$50\% PF for Cygnus X-1 by {\em INTEGRAL} \citep{laurent11,jourdain12}. $>$90 \% PF for V404-Cygni by {\em INTEGRAL} \citep{laurent17}&{\em AstroSat} \citep{bhalerao16}, {\em INTEGRAL} \citep{winkler03}&{\em Daksha$^4$} (2025), {\em POLAR-2} (2024) \citep{kole19}, AMEGO \citep{amego19}, e-ASTROGAM \citep{de18}, SPHiNX \citep{pearce19}, LEAP \citep{mcconnel16_leap}, COSI-2 \citep{tomsick19}\\

&Blazars&{Second peak of LBL: SSC/EC/hadronic model \citep{mcnamara09,zhang14}}&a few (EC) to 50 \% (SSC)&&&\\

&Magnetars&{Signature of QED magneto photon splitting \citep{wadiasingh19_white}}&$\sim$20-60 \%&&&\\

&XRP&{Emission mechanism, geometry - Polar cap/Outer gap/slot gap/strpped wind models \citep{weisskopf09,Harding19_book}}&a few tens of \%&24-32 \% PF for Crab+nebula by {\em INTEGRAL}, {\em AstroSat} \citep{Jourdain19,vadawale17}&&\\

&PWN&{Acceleration mechanism and magnetic field geometry \citep{weisskopf09,Harding19_book}}&$\sim$30-60 \%&39-47 \% for Crab nebula by {\em INTEGRAL}, {\em AstroSat} \citep{dean08,vadawale17}&&\\

&GRBs&{Prompt emission mechanism (SO/SR/CD models), jet composition, dissipation mechanism \citep{toma08,gill18}}&a few \% to 80 \% PF&a few \% ($\sim$12 GRBS) to high PF for $\sim$15 GRBs by GAP, POLAR, {\em AstroSat}, {\em INTEGRAL} \citep{Kole20polar_catalog,zhang19,chattopadhyay19,yonetoku11,yonetoku12}&&\\

&Solar flares&{Emission mechanism, electron beaming, magnetic field structure for X class flares \citep{bai78,leach83,zharkova10}}&Maximum around 25 \% for flares away from Sun's center&$\sim$2 $-$ 54 \% PF for $\sim$8 flares by {\em RHESSI} \citep{garcia06,boggs06}&&\\

\hline
\end{tabular}
\begin{tablenotes}
\item[1]X-ray dim isolated neutron star, $^2$X-ray pulsar, $^3$Accretion powered pulsar, $^4$\url{https://www.star-iitb.in/research/daksha} 
\end{tablenotes}
\end{threeparttable}
\end{tiny}
\end{table*}
From Table \ref{polscience_Table}, a number of striking differences between the low energy and high energy
regimes can be highlighted $-$ 1. there is a large number of new measurements in hard X-rays, 2. predicted polarization fractions for most of the X-ray sources are relatively higher in the hard X-ray regime, 3. there is a large number of confirmed and planned polarimetry missions 
in near future along with a couple of active instruments continuing to provide useful polarization measurements.  

The field of hard X-ray polarimetry has been quite active in the last 10 $-$ 15 years which can be seen through the new developments of dedicated polarimeters like SPR-N \citep{bogomolov03,zhitnik06}, POLAR \citep{sun16}, PoGOLite and PoGO+ \citep{chauvin16_pogo}, X-Calibur \citep{guo11_2,guo13}, GRAPE \citep{bloser09,mcconnell09}, GAP \citep{yonetoku06}, PENGUIN-M \citep{dergachev09}, COSI \citep{yang18,lowell17} and a large number of new proposals to develop
hard X-ray polarimeters, e.g. COSI-2 \citep{tomsick19}, {\em POLAR-2} \citep{kole19}, {\em Daksha \footnote{\url{https://www.star-iitb.in/research/daksha}}}, AMEGO \citep{amego19} etc. There has also been a significant effort in extracting polarization information
from instruments which are not specifically meant for polarimetry measurements like {\em RHESSI} \citep{mcconnell02,coburn03,boggs06,mcconnel07}, {\em INTEGRAL} \citep{laurent11,jourdain12,dean08,forot08}, {\em AstroSat} \citep{chattopadhyay14,vadawale15} and {\em Hitomi} \citep{hitomi18}. 
Measurements from these experiments in the last two decades have provided some important scientific inputs in understanding emission mechanism and geometry in X-ray pulsars \citep{vadawale17,forot08,Jourdain19,chauvin18bcrab,hitomi18}, magnetic field structure in Pulsar Wind Nebula (PWN) \citep{vadawale17,dean08,chauvin18bcrab}, disk-jet interplay in black hole X-ray binaries (XRBs) \citep{laurent11,jourdain12,chauvin18a,chauvin18b}, hard X-ray emission mechanism in GRB prompt emission \citep{zhang19,chattopadhyay19,yonetoku11,yonetoku12,gotz09}, and emission process behind solar flares \citep{mcconnel07,garcia06,boggs06,zhitnik06}. Though in some cases, results are marred with large uncertainties and a firm conclusion is not possible, most of these findings are extremely interesting as they pose new challenges to the existing theories. One interesting and common feature in all these measurements is a systematic increase in polarization (see Fig. \ref{cygnusx1} and \ref{pfvsenergy}) within the hard X-ray band which makes this energy regime extremely promising for polarimetry experiments.

Polarization measurement in hard X-rays promises to address some of the long-standing astrophysical problems which might not be otherwise probed by spectroscopic and timing tools or even polarization measurements in soft X-rays (see 3$^{rd}$ column of Table \ref{polscience_Table}). Since hard X-rays originate in the close vicinity of the compact sources, it is likely to carry signatures of high gravitational and magnetic field close to a black hole or neutron star, e.g. accretion dynamics close to a black hole, jet launching mechanism from the black hole both in XRBs and GRBs, emission from the low altitudes of magnetar's magnetosphere, structure and dynamics of accretion column in accretion powered neutron stars. At the same time, hard X-rays originate from the highest energy particles being cooled off at faster time scales and therefore polarization in this regime can probe acceleration mechanism of particles in supernova remnants and PWNs and in the base of jets of blazars, black hole XRBs or GRBs. Fortunately, because non-thermal emission dominates over thermal emission in hard X-rays, the polarization fraction is expected to be higher compared to soft X-rays as also seen in the 4$^{th}$ column of Table \ref{polscience_Table}. This makes this regime unique as we expect more polarization measurements at these energies even from moderately sensitive polarimeters.

Because a couple of instruments (on board {\em INTEGRAL} and {\em AstroSat}) with moderate polarimetric sensitivity continues to operate and provide useful polarization information for some of the bright hard X-ray sources and GRBs, we expect the hard X-ray regime of polarimetry to continue to remain active. There is an order of magnitude increase in the activity with the new confirmed and planned experiments in near future (see 7$^{th}$ column of Table \ref{polscience_Table}).  {\em POLIX} \citep[sensitive in 10 $-$ 30 keV,][]{paul16}, {\em PolariS} \citep[sensitive in 10 $-$ 80 keV,][]{hayashida14}, {\em POLAR-2} \citep[sensitive in 50 $-$ 500 keV,][]{kole19}, {\em Daksha} (sensitive in 100 $-$ 400 keV) are expected to provide sensitive polarization measurements for a number of astrophysical sources.   

{X-ray polarization is an extremely difficult area of research. The near to be launched {\em IXPE} and {\em POLIX}, to be launched in 2021, will make measurements in soft X-rays and 
the medium energy range respectively.} Polarization measurement in hard X-rays, however, straddles the regime of thermal and non-thermal emission, particularly in accreting black holes sources and they are thought to provide inputs to the much sought after details of disk-jet synergy in powerful jet sources like GRBs, microquasars, blazars etc.  A glimpse of such details are already available by measurements with {\em AstroSat} and {\em INTEGRAL}, along with a strong trend of increasing polarisation fraction with energy. In this review, we mainly focus on this non-thermal regime and summarise the possible  meaningful polarization results for the bright hard X-ray sources from the upcoming and planned polarimetry experiments. We particularly focus more on improving the existing and upcoming results by making developments in the hardware front to increase the sensitivity of the detectors, making more careful calibration of the instruments, and putting more effort in understanding the detector systematics. The new developments in the theoretical modelling front is also discussed.  

In the next section, specific science cases of hard X-ray polarimetry (3$^{rd}$ and 4$^{th}$ column of Table \ref{polscience_Table}) have been described followed by a discussion on the new scientific findings (5$^{th}$ column of Table \ref{polscience_Table}) in section \ref{finding}   

\section{Science Drivers of Hard X-ray polarimetry}\label{polscience} 
\subsection{Black Hole XRBs} 
 
\begin{figure*}[t]
\centering
\includegraphics[height=.15\textheight]{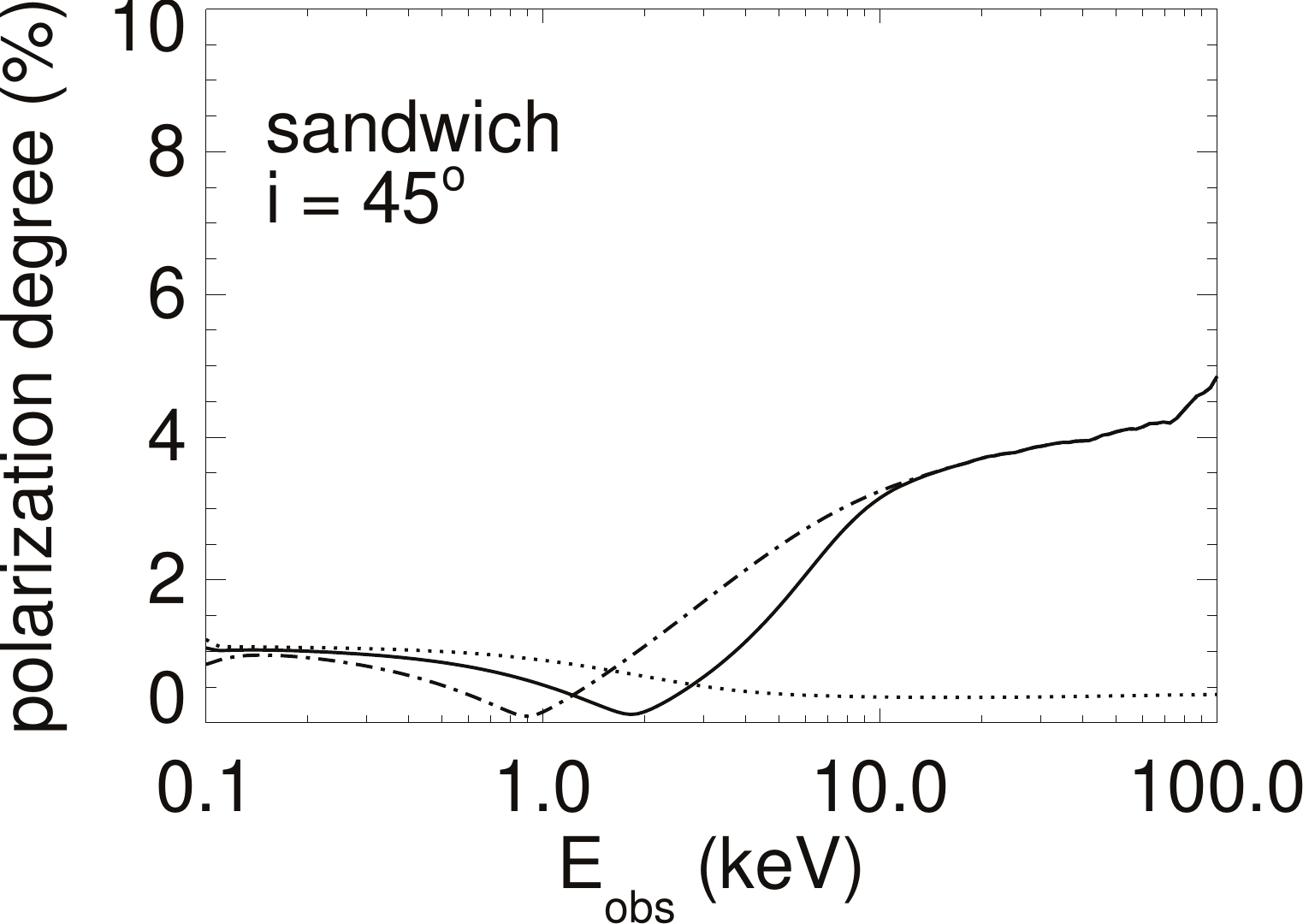}
\includegraphics[height=.15\textheight]{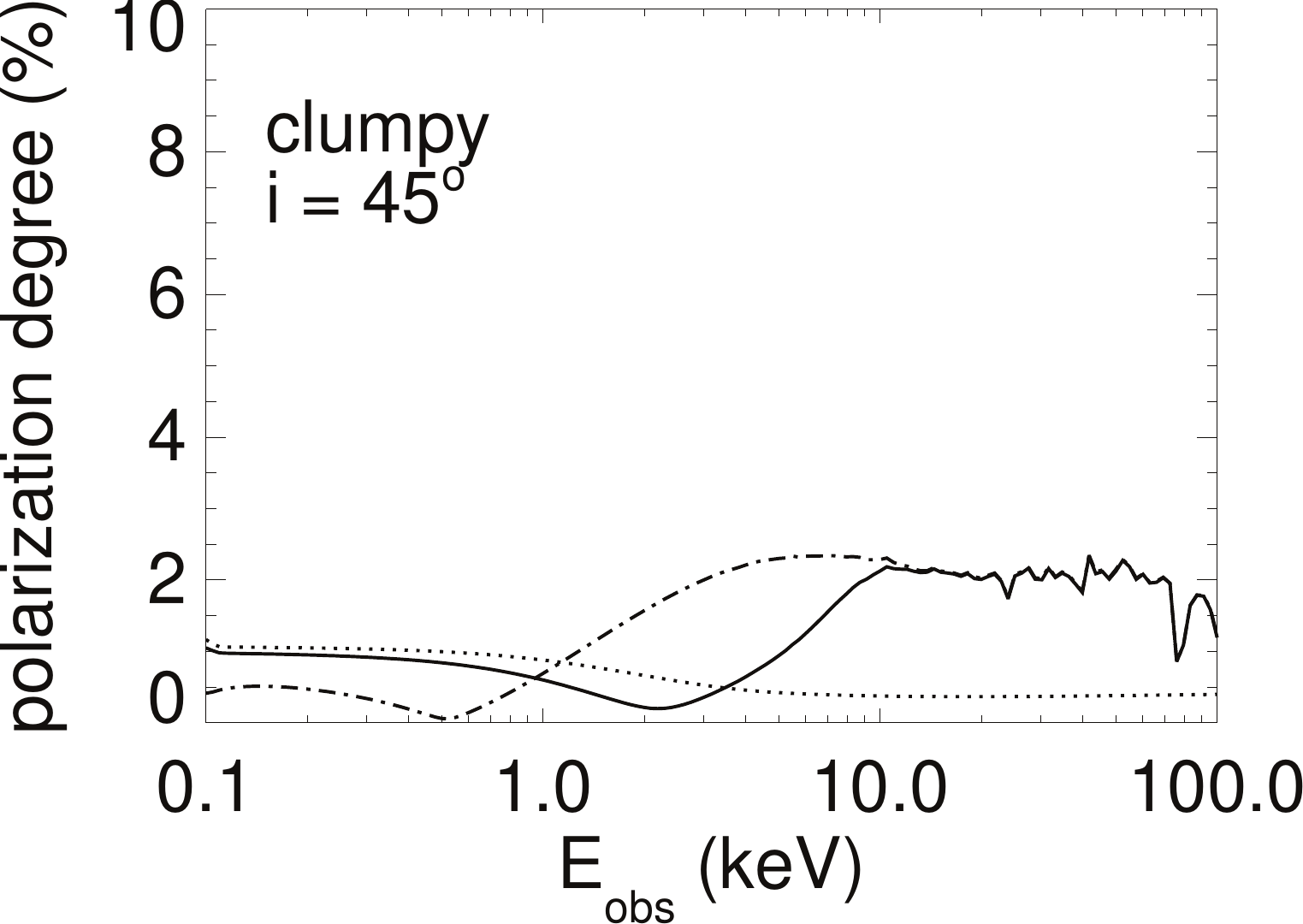}
\includegraphics[height=.15\textheight]{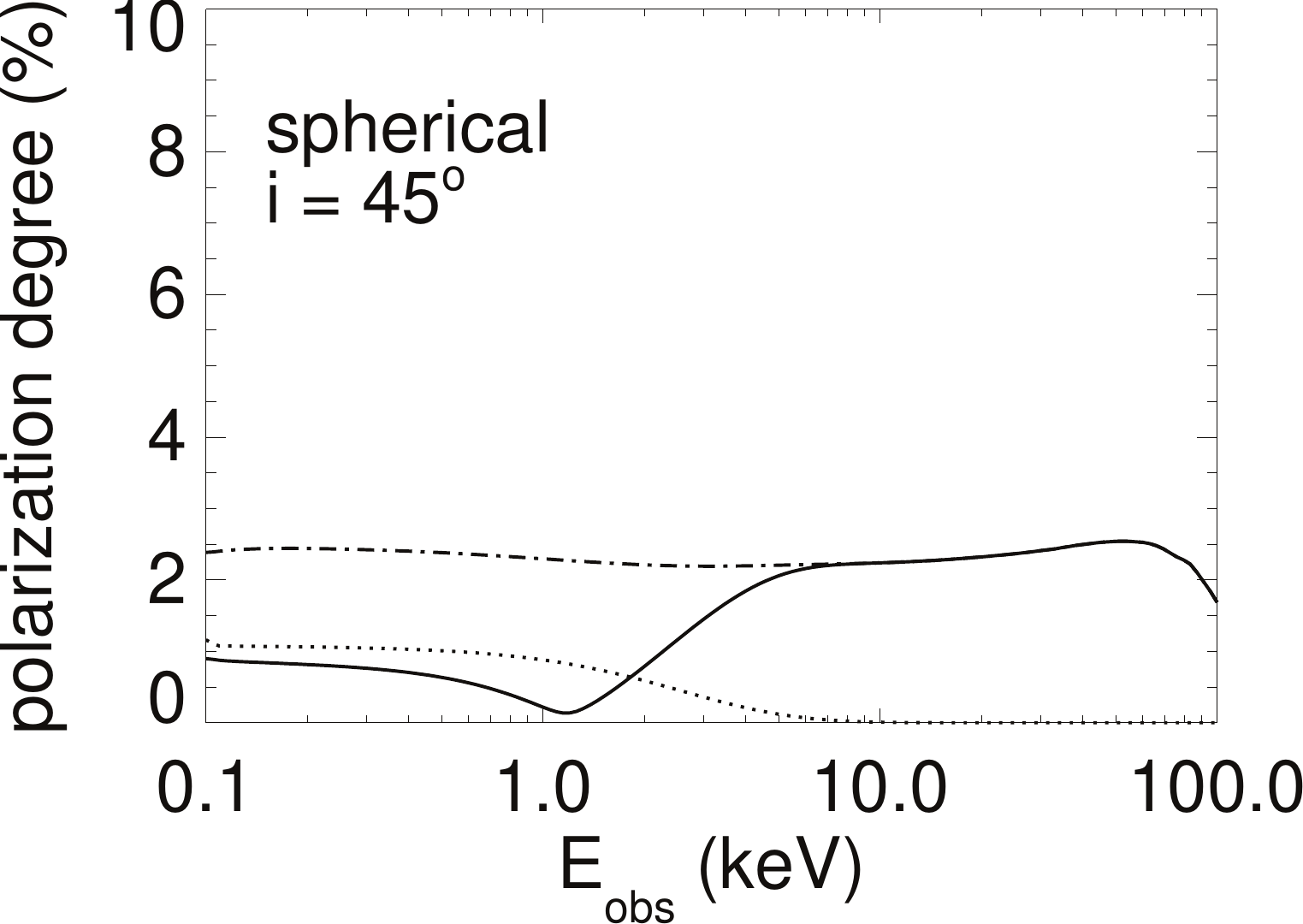}\\
\includegraphics[height=.15\textheight]{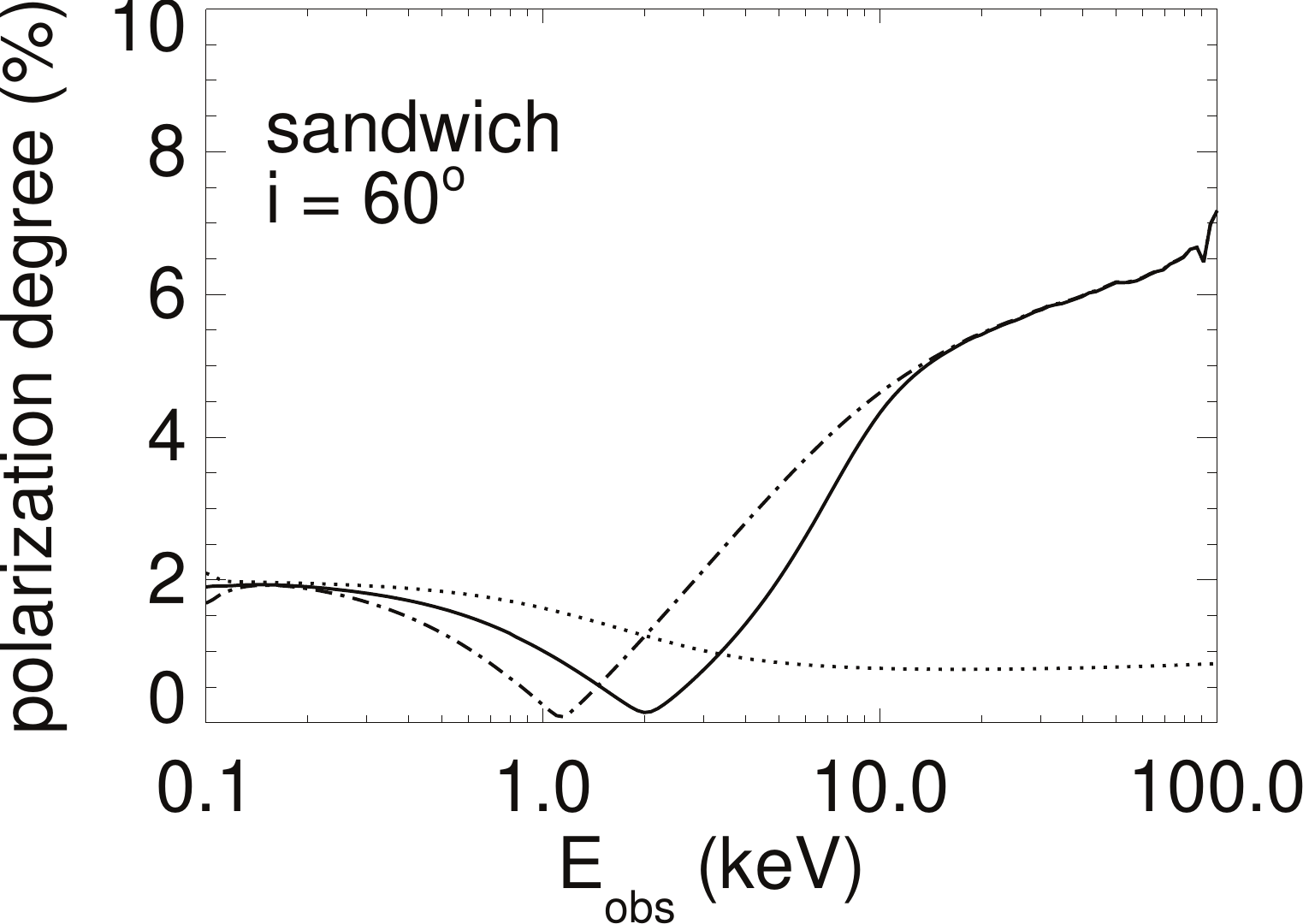}
\includegraphics[height=.15\textheight]{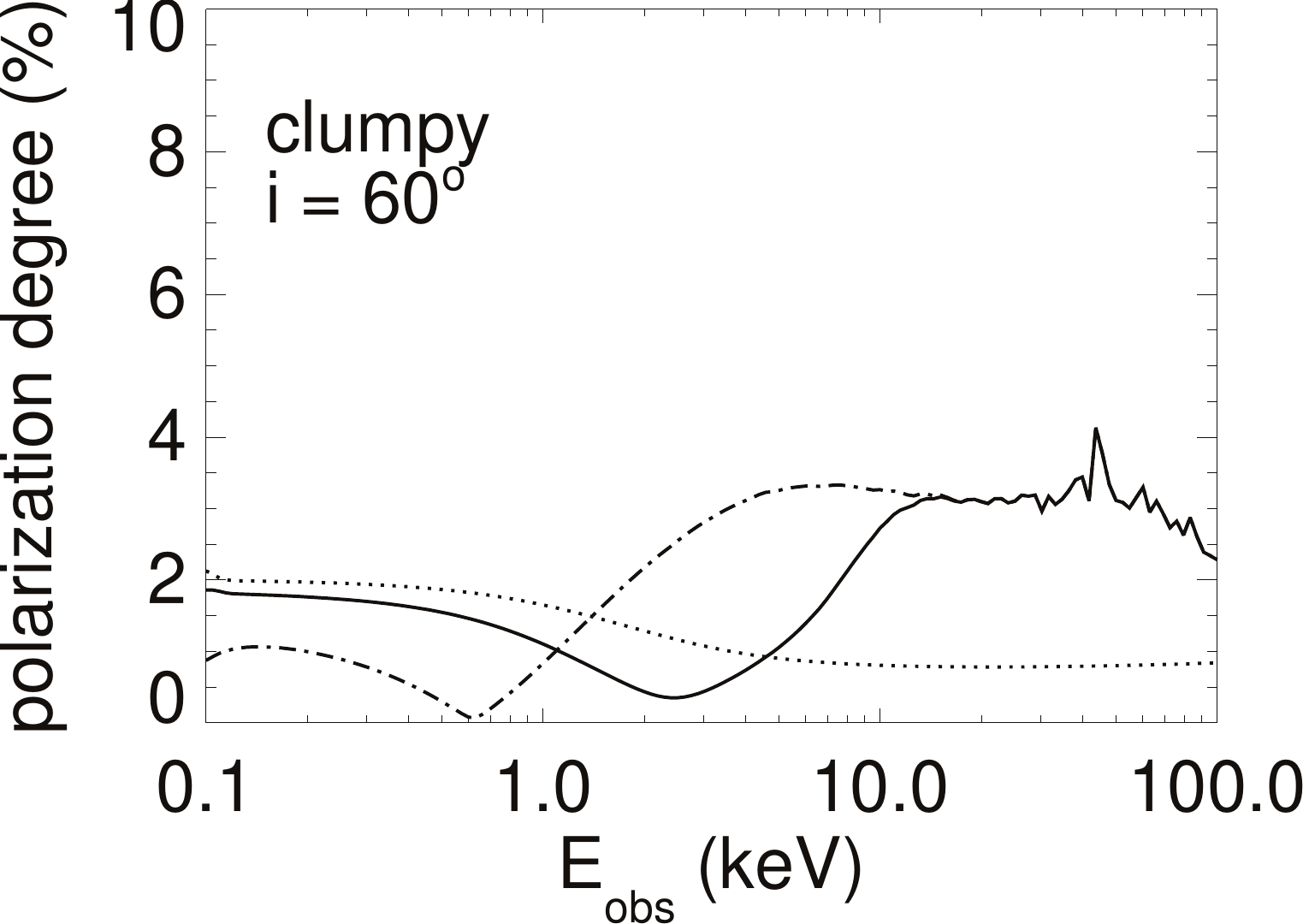}
\includegraphics[height=.15\textheight]{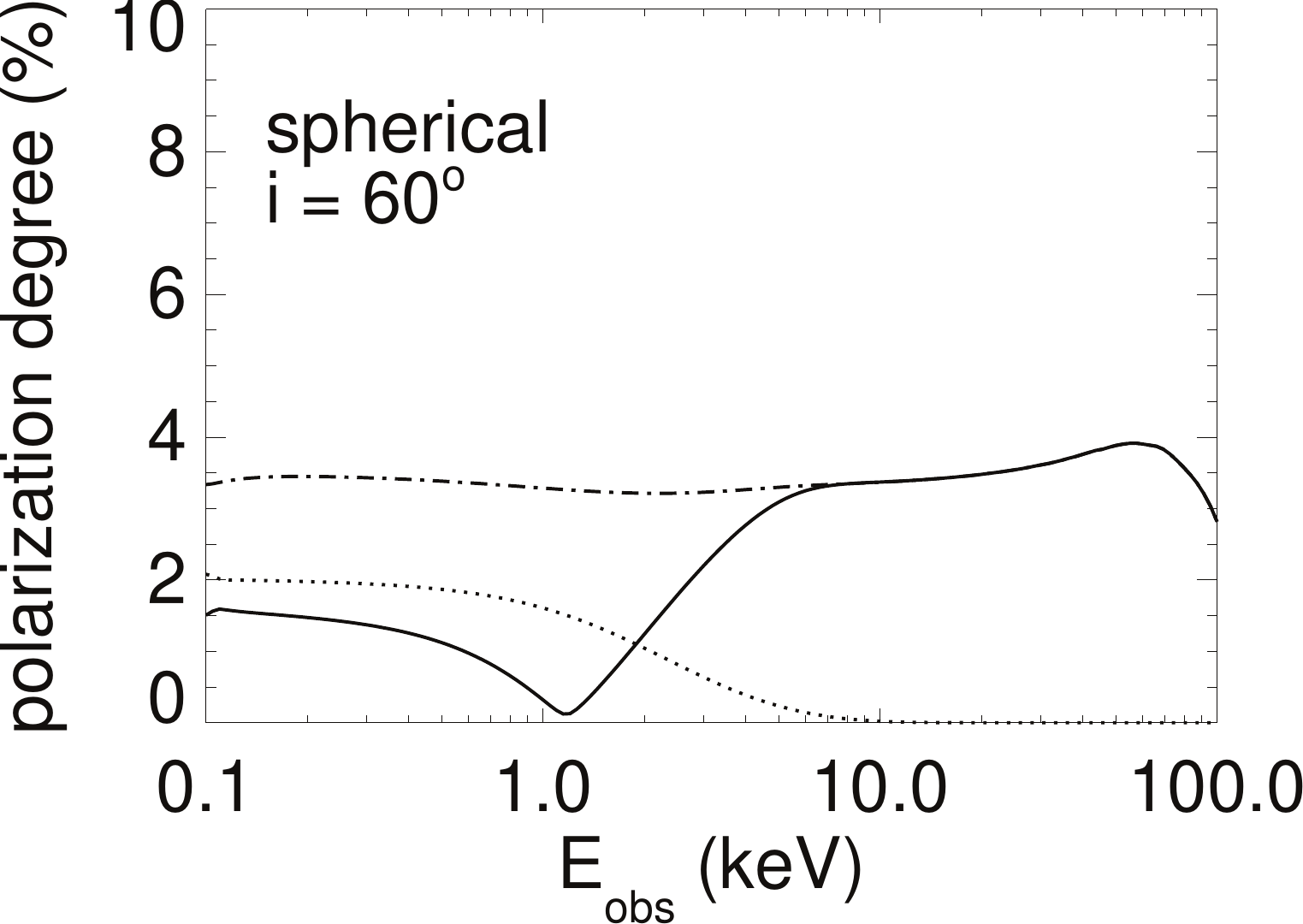}\\
\includegraphics[height=.15\textheight]{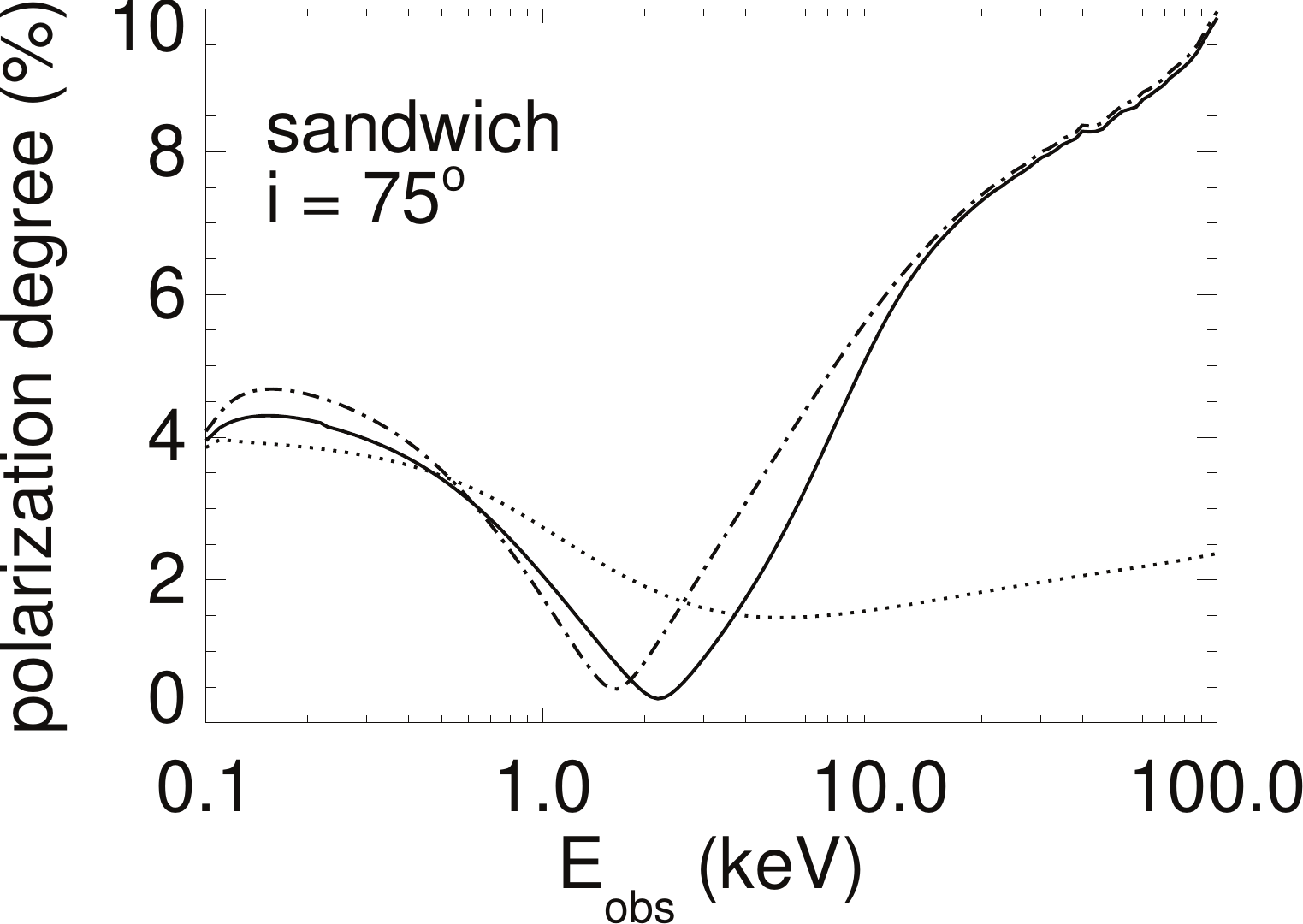}
\includegraphics[height=.15\textheight]{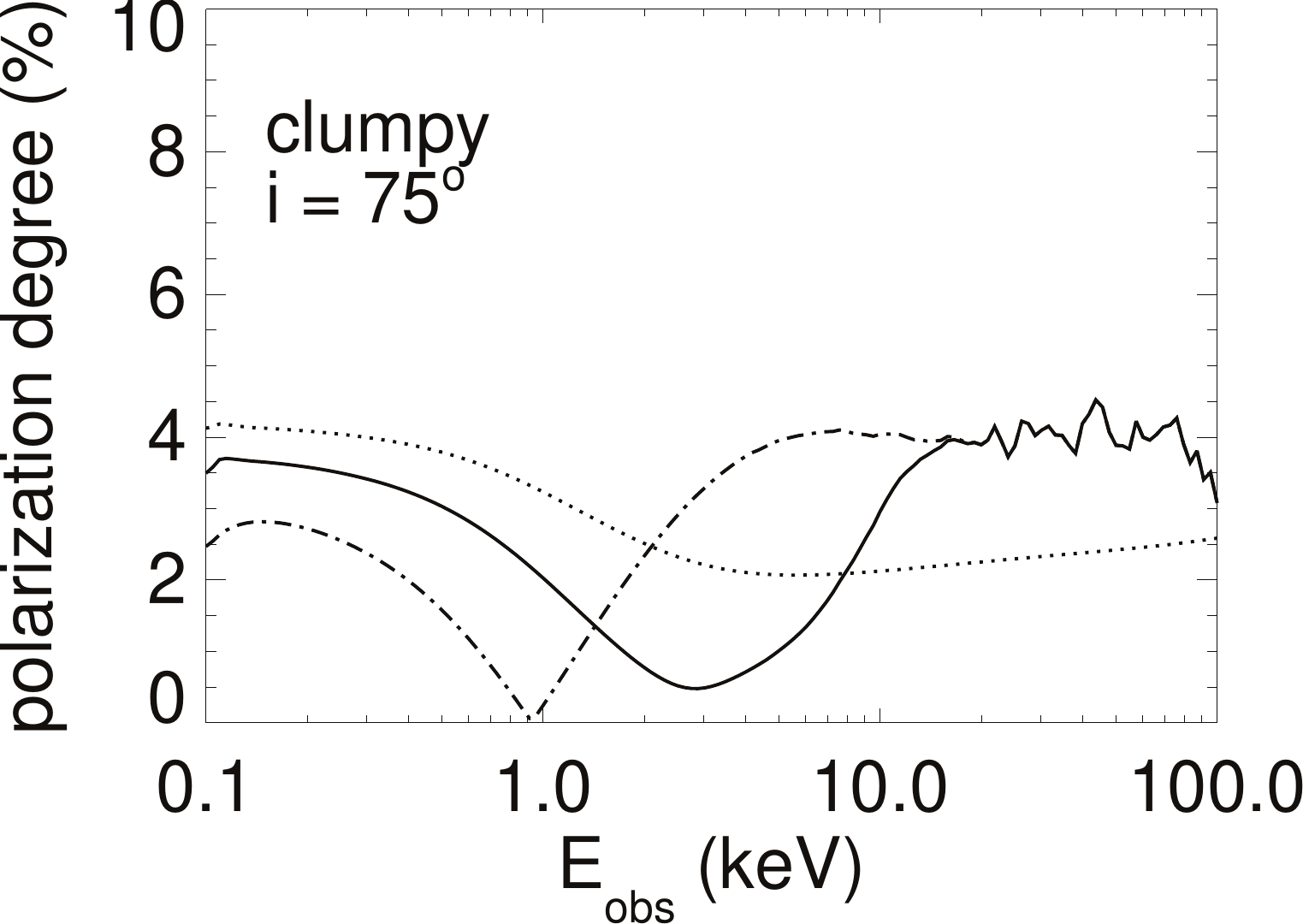}
\includegraphics[height=.15\textheight]{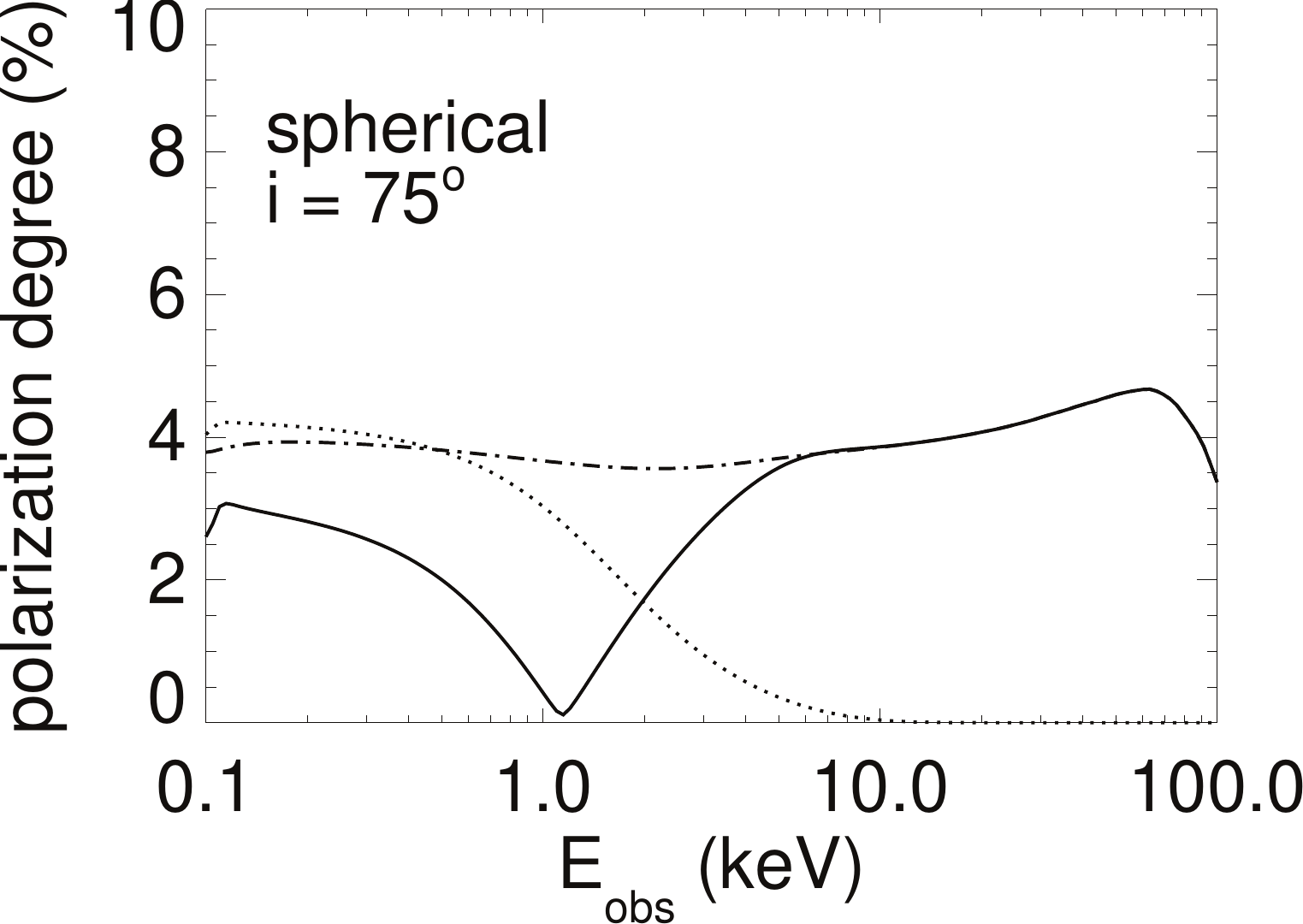}
\caption
{Polarization characteristics for various black hole coronal 
geometries as a function of observed energy and inclination. The dotted
lines represent disk emission, whereas the dot-dashed and
solid lines represent coronal and total (disk+corona) emission
respectively. Figure credit $-$ \citet{schnittman10} (\raisebox{.5pt}{\textcircled{\raisebox{-.2pt} {c}}}  AAS. Reproduced with permission)} 
\label{bh_corona}
\end{figure*}

Polarization measurements in soft X-rays from black hole XRBs can probe General Relativity (GR) effects on the thermal disk photons and is expected to constrain some of the key parameters like disk inclination and mass spin ratio of the black holes \citep{dovciak08,schnittman09}. In the hard spectral state of these systems, the thermal photons being inverse Compton scattered by the corona introduces a separate polarized spectral component at energies above 10 keV with polarization characteristics expected to be dependent on the geometry of corona. 
\citet{schnittman10} investigated the 
polarimetric signatures for various corona geometries $-$ sandwich, clumpy and spherical corona
(see Fig. \ref{bh_corona}). 
Polarization is expected to have a strong  dependence on photon energy and inclination of the system in case of a homogeneous sandwich corona, with polarization level reaching up to $\sim$10 \% at 100 keV for higher inclinations.  
On the other hand, for an inhomogeneous clumpy corona and a standard spherical corona, polarization fractions are predicted to be relatively low (3 $-$ 4 \%) with no strong dependence on energy above 10 keV. 
Sensitive energy resolved polarization measurements in the hard state of 
black hole XRBs will be a key to probe the geometries of corona.

At energies beyond 100 keV, the radiation from black hole systems in its 
low hard state is believed to be of synchrotron origin from high energy electrons in the jet
\citep{vadawale01,vadawale03,markoff01}. Recent findings of
high polarization measured for high mass black hole binary, Cygnus X-1, 
at energies spanning 
from a few hundreds of keV to a few MeVs \citep{laurent11,jourdain12} also suggest
the jet origin of the hard X-ray emission. However, multi-wavelength
SED modeling of Cygnus X-1 predicts insignificant contribution of jet in 
hard X-rays \citep{zdziarski14}.
On the other hand, there are studies reported in literature 
suggesting that radiation in hard X-rays to originate from lepto-hadronic 
corona of black holes due to synchrotron radiation, 
predicting the radiation to be highly polarized independent of its state 
\citep{romero14}. 
{More recently, \citet{Kantzas20} applied a multi-zone jet model on first ever simultaneous broadband data ranging from radio wavelengths to MeV X-rays. They interpreted the high polarization fraction seen in the
MeV band as synchrotron radiation emitted by electrons
accelerated inside the jets of Cygnus X–1 in the presence of a highly
ordered magnetic field. However, the required MeV synchrotron flux for the high polarization is only possible for a harder power-law index of accelerated electrons of 1.7 compared to that observed $\sim$2.2 by {\em INTEGRAL}-IBIS \citep{laurent11}.}  
Careful polarization measurements of black hole systems 
in both hard and soft spectral states may lead to a proper 
understanding of the origin of hard X-rays in these systems.   

\subsection{Active Galactic Nuclei}
The thermal UV photons from the disk of Active Galactic Nuclei (AGNs) are Comptonized by the corona giving rise to a power-law component in 
their spectra. The UV photons from the disk move parallel to the
disk and corona suffering a large number of interactions and upscattered to energies $>$ 10 keV \citep{schnittman10}. This causes the coronal emission to be polarized around 8 \%. 
Spectro-polarimetric studies are expected to be useful in investigating  
corona geometry in detail by constraining the number of clumps and
their over-density in case of a clumpy corona geometry.
The disk photons may also be scattered by the molecular torus. Thus, X-ray
polarimetry can also constrain the geometry of the torus \citep{goosmann11}. 
Polarization measurement of the reflected radiation from the disk or 
the torus
may also complement the reverberation studies for AGNs to study the 
geometry of the reflector by estimating the time delay between the 
direct and reflected component of radiation \citep{uttley14}.

\subsection{Neutron Stars} 
X-ray polarimetry has a wide range of applications in isolated pulsars,
accreting pulsars, magnetars and help understand the 
emission mechanism and geometry in these systems and behaviour of matter in strong 
magnetic fields \citep{weisskopf09}.

\subsubsection{Rotation powered X-ray pulsars:} 
\begin{figure}
\centering
\includegraphics[height=.25\textheight]{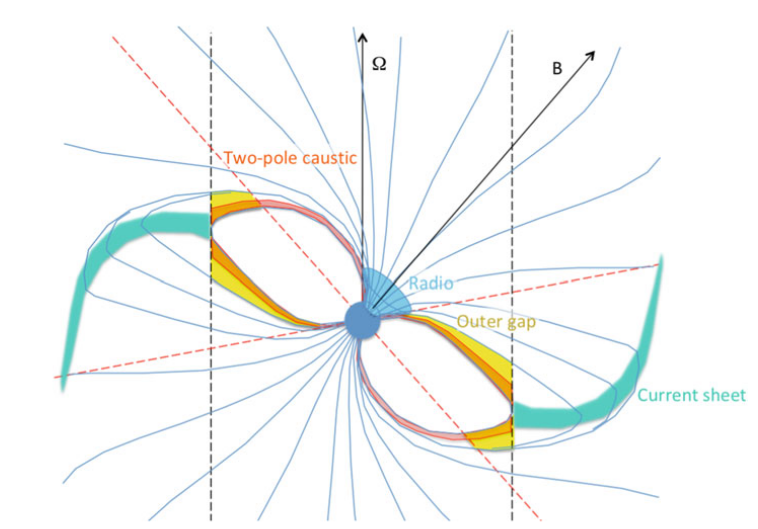}
\caption
{Geometrical illustration of various pulsar models: polar cap/radio (blue), two-pole caustic (red), outer gap (yellow), and current sheet (green) models.
The light cylinder is shown by the dashed black
lines and the dotted red lines show projections of the null-charge surface.
Different models assume different acceleration region for the electrons with distinct polarization signatures with rotation phase of the pulsars.
Figure credit $-$ \citet{Harding19_book} (reprinted by permission from Springer Nature Customer Service Centre GmbH)}
\label{pulsar_model}
\end{figure}

\begin{figure*}
\centering
\includegraphics[height=.4\textheight]{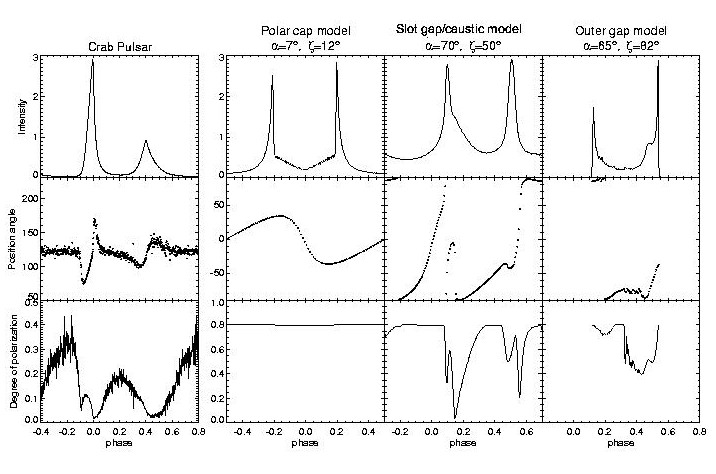}
\caption
{Polarization with the rotation phase of pulsars as predicted from various pulsar models $-$ polar gap, outer gap and slot gap models. As an example, the measured polarization with phase for Crab pulsar in optical wavelength is shown in the left most panel \citep{smith88,romani01}. 
Figure credit $-$ \citet{weisskopf09} (reprinted by permission from Springer Nature Customer Service Centre GmbH).}
\label{crab_pol}
\end{figure*}

The emission mechanism and emission site for high energy radiation from rotation powered 
pulsars are still poorly understood. There are models predicting the high energy pulsar radiation to originate directly above the polar cap
\citep[polar cap model,][]{daugherty82}, or in the outer magnetosphere
\citep[outer gap model,][]{cheng00}, or all the way from the polar cap to
the light cylinder along the last open field line
\citep[slot gap model,][]{dyks03}. A geometric description of these models is
shown in Fig. \ref{pulsar_model}. 
While these three models imply that the emission originates inside the light cylinder, current sheet model based on striped pulsar wind (shown in green) \citep{kirk02,petri05} predicts the radiation to come from outside of the light cylinder from the particles escaping through open field lines. Because of the distinct emission geometries, these models predict distinct phase dependent
polarization signatures.
Fig. \ref{crab_pol} shows variation of optical linear polarization
with pulse phase for Crab pulsar \citep{smith88,romani01}, along
with the predictions of these models. 
Phase resolved polarimetry can test these models and help 
in understanding the emission sites and emission mechanisms in isolated X-ray
pulsars.

\subsubsection{Magnetars:}
Magnetars are neutron stars with large spin periods and highest magnetic field ever observed in the universe with
polar surface values of 10$^{14-15}$ G. The huge magnetic energy reservoir powers the magnetars through
seismic activity and heating of the stellar interior \citep{duncan92}. Radiation from such an extreme magnetic and gravity fields makes the magnetars an extremely interesting object for polarimetry and other astrophysical studies. 

Magnetars spend most of the time in quiescence radiating persistent quasi-thermal emission with KT $\sim$0.5 keV. They show occasional short outbursts (a few to hundreds in number with duration of $\sim$0.1 s) $-$ famously known as `Magnetar outbursts' where the radiation extends all the way to sub-MeV region with a few to thousand times increase in the flux (refer to review papers on magnetars by \citet{Rea10,Turolla15,Mereghetti15,Kaspi17,esposito18}). {\em INTEGRAL}, {\em CGRO}, {\em RXTE}-HEXTE, {\em NuSTAR} have detected hard X-ray tail for a number of magnetars in quiescence (there are 7 such magnetars) with a possible cut off energy lying in the sub-MeV region \citep{enoto17,hartog08}. 

One possible explanation behind the hard X-ray component on top of the thermal emission assumes Resonant Inverse Compton scattering \citep[RICS, see][]{wadiasingh17,Beloborodov12,baring06} of the soft thermal photons off the relativistic e$^-$/e$^+$ population at the lower altitudes where the density of the soft X-ray photons and magnetic field are relatively high. RICS predicts a flat spectrum in hard X-rays with a high linear polarization making the hard X-ray region extremely promising for polarimetry studies, particularly for the seven magnetars detected with bright hard X-ray tails in their quiescence. Detection of high polarization in hard X-rays will be a clear signature of RICS meachanism in magnetars. A phase resolved spectro-polarimetry study is supposed to constrain the angle between the rotation and magnetic axis because at different viewing angles with respect to the magnetic axis the electron population and beaming geometry are expected to vary \citep{wadiasingh19_white}.

One interesting aspect of magnetar's hard X-ray polarimetry is the first ever possible confirmation of QED effects in strong magnetic field. Magnetic photon splitting of Gamma rays in strong magnetic field ($\gamma+B \longrightarrow \gamma+\gamma$) is one explanation of the sub-MeV spectral cut off seen in the magnetars. Polarization strongly depends on the photon splitting modes (splitting of $\perp$(X) mode or both $\perp$(X) and $\parallel$(O) modes). Therefore, spectro-polarimetry study in hard X-rays near the sub-MeV cut off region of magnetars offers a unique opportunity to test the QED effects in strong magnetic field \citep{wadiasingh19_white}.    

\subsubsection{Accretion powered pulsars:}
In accretion-powered pulsars, theoretical models predict high polarization
owing to the high magnetic field ($10^{12-13}$ G) in those systems, particularly for emission perpendicular
to the magnetic field. Phase resolved polarization study can therefore be used to 
determine the beam shape of the pulsar. 
For example, for the pencil beam model, the oscillations in polarization fraction 
are expected to be out of phase with pulse phase, whereas for the fan 
beam model, the opposite signature is predicted \citep{meszaros88}. 
This effect is more prominent at energies near the cyclotron resonances. 
Many accretion-powered pulsars have been found to exhibit cyclotron features 
in the energy range of 15 $-$ 50 keV. A polarimeter sensitive in this cyclotron line populated energy regime is expected to distinguish between the pencil and
fan radiation patterns in these systems.

Millisecond X-ray pulsars are accretion-fed systems where the pulsar is spun
up to high rotation speed with period of a few milliseconds. 
Polarization at higher energies in these systems derives from Compton
scattering of photons in accretion shock \citep{viironen04} or possibly
from the accretion disk \citep{sazonov01}. Polarization measurements 
from these sources may test these various models and put tighter constraints 
on various geometrical parameters like orbital and dipole axis inclinations.    

\subsection{Gamma Ray Bursts} 
Despite the detection and study of a large number of Gamma Ray Bursts (GRBs) in the last decade \citep{gehrels12} by Swift \citep{gehrels04,barthelmy05} and Fermi \citep{meegan09}, the
prompt emission mechanism has not yet been well understood \citep{kumar15} owing to the unpredictable nature of the GRBs. 
GRB prompt emission is widely believed to be of synchrotron origin from high energy electrons in the jet \citep{meszaros93}. Apart from synchrotron, other possible mechanisms are inverse Compton
scattering \citep{ghisellini99,ghisellini00,lazzati04}, thermal blackbody radiation, presumably of photospheric origin \citep{ryde04,peer11,basak15,iyyani15}.
Measurement of hard X-ray/Gamma ray polarization has the potential to shed light on the GRB prompt emission \citep{covino16,mcconnell16}.

\begin{figure}
\centering
\includegraphics[height=.25\textheight]{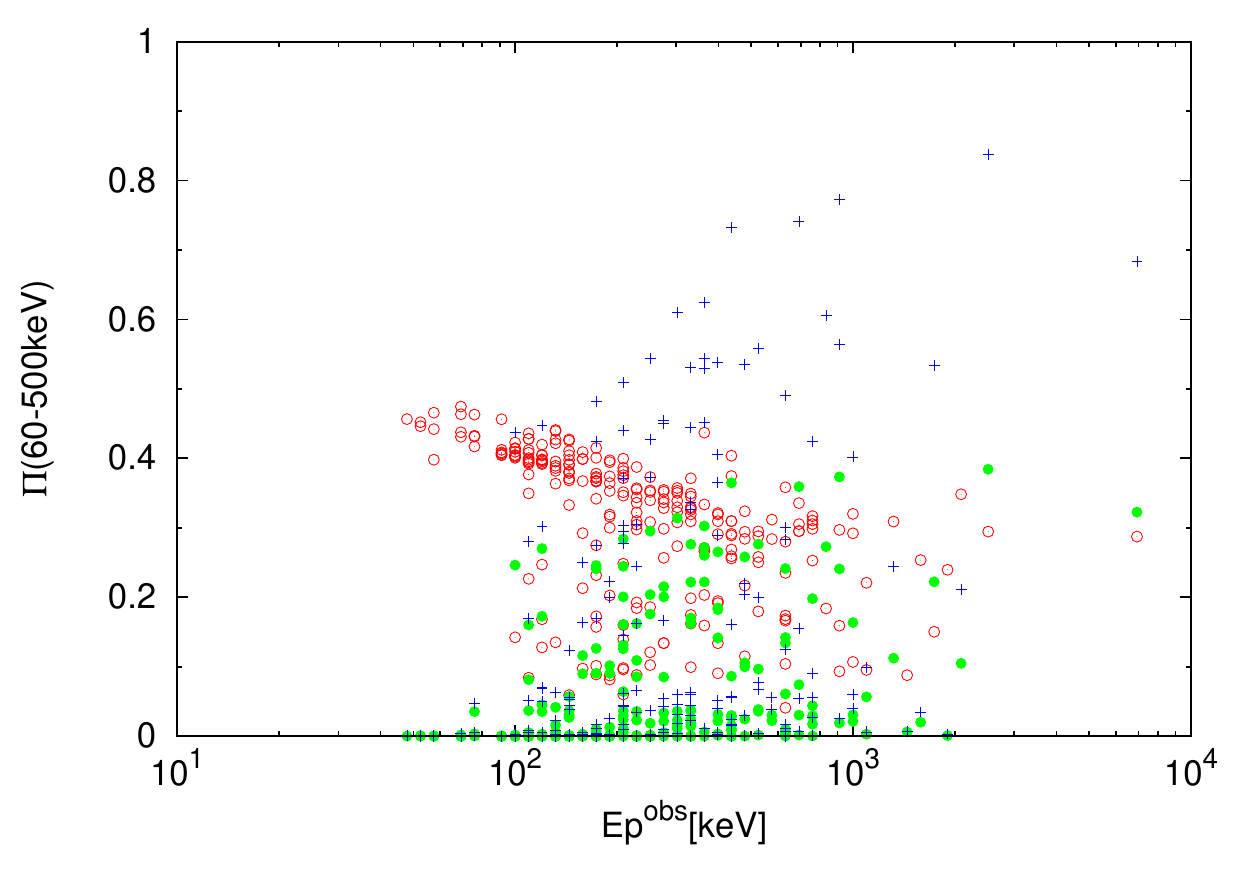}
\caption
{Predicted distribution of polarization fraction with E$_{peak}$ for a large sample of GRBs in 50–500 keV energy range for various prompt emission polarization models $-$ Synchrotron radiation in ordered magnetic field (red), Synchrotron radiation in random magnetic field (green) and Comptonization of the soft X-ray photons  (blue). Figure credit $-$ \citet{toma08} (\raisebox{.5pt}{\textcircled{\raisebox{-.2pt} {c}}}  AAS. Reproduced with permission)}
\label{stat_pol_toma}
\end{figure}

There are various prompt emission polarization models, namely, synchrotron 
emission from relativistic electrons energized in internal shocks
within the jet either in globally ordered magnetic field derived from the
central engine \citep[Synchrotron Ordered or `SO model',][]{lyutikov03,nakar03,granot03} or in random magnetic 
field generated in the
shock plane within the jet \citep[Synchrotron Random or `SR model',][]{medvedev99} or emission due to Comptonization
of the soft photons \citep[Compton Drag or `CD model',][]{shaviv95} by the relativistic jet. Polarization is expected to be high in SO model for a wide range of viewing angles ($>$20 \%).
On the other hand,
polarization in the SR and CD model will be dependent on the 
geometry of the viewing angle, as for certain viewing angles, net 
polarization remains. High polarization ($\sim$40 \%) can be achieved from SR model if the jet is narrow and viewed along the edge \citep{waxman03,medvedev07,sari99,ghisellini_lazzati99}. Similar to SR model, CD can also result in high polarization (60 $-$ 100 \%) if the jet is
narrow and is observed along the edge \citep{lazzati04}.
Possibility of such a geometric
configuration favorable for high polarization is relatively small and therefore the GRBs without such favorable viewing geometry are expected to be unpolarized according to the geometric models (SR and CD).
\citet{toma08} showed that statistical distribution of prompt emission polarization for a large sample of GRBs
may efficiently distinguish the various theoretical models (see Fig. \ref{stat_pol_toma}). 

Study of spectral and temporal evolution of the polarization parameters for individual GRBs is equally important in the complete understanding of the GRB jets and emission mechanisms. \citet{mcconnell19,zhang11,gill18} discuss these aspects for matter and magnetic field or poynting flow dominated jets and the dissipation mechanisms in the jets.

\subsection{Blazars} 
Hard X-ray polarimetry for blazars may probe the origin
of the second characteristic emission peak in their spectral energy
distribution \citep{zhang17}. For low energy peaked blazars, the low energy 
peak occurs at the optical regime whereas the high energy peak occurs in MeVs. 
The optical emission is believed to be due to synchrotron radiation from the relativistic electrons. 
On the other hand, there exists two different models in explaining the origin of the high energy peak.  
According to the Synchrotron Self Compton model (SSC model, \citet{celotti94}), the optical synchrotron photons are inverse Compton scattered off the relativistic electrons to high energies. 
SSC model predicts around 30 \% polarization for uniform magnetic field with 
polarization direction identical to that for optical emission. External Compton 
model (EC), on the other hand, assumes a different origin for the seed photons where either the accretion disk photons or photons from
broad line region or dusty molecular torus are upscattered to high energies. Polarization fraction in this scenario is expected to be significantly lower compared to the SSC model ($<$5 \%) \citep{mcnamara09}. 

Besides the leptonic models, there exists a completely different 
approach based on lepto-hadronic models which can produce equally 
good fits to the SEDs of blazars \citep{zhang13}. In the hadronic models, although the source of the low energy peak 
is same as that for the 
leptonic models, the high energy peak is believed to originate from the high energy proton
induced radiation mechanisms. Because of the dominance of synchrotron
radiation in hadronic models, a relatively higher level of polarization is expected compared
to that for leptonic models \citep{zhang13}.

\subsection{Solar Flares}
Solar flares are the powerful events due to magnetic reconnection
in Sun's corona, accelerating the electrons towards the chromosphere. 
Radiation in soft X-rays is due to thermal heating at the reconnection
site and are therefore expected to be unpolarized. However, 
because of anisotropies in electron distribution, the thermal radiation
may have a low level of polarization \citep{emslie80}. 
\begin{figure*}
\centering
\includegraphics[height=.3\textheight]{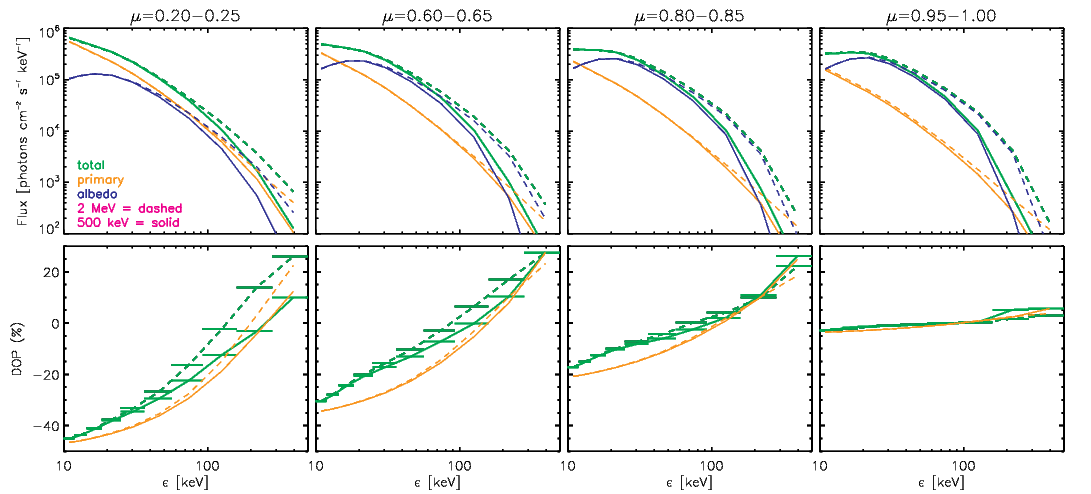}
\caption[Spatially integrated flux and polarization for solar flares]
{Variation of flux (top row) and spatially integrated polarization 
(bottom row) with observed energy for an extremely beamed electron 
distribution in solar flares. $\mu$ (= $\cos\theta$) refers to the
direction of emission where $\theta = 0^\circ$ is the local solar
vertical. Green, orange and blue denote the total source, primary
and albedo components respectively, whereas the solid and dashed lines 
refer to electron cut off energy of 500 keV and 2 MeV respectively. 
Figure credit $-$ \citet{jeffrey11} (reproduced with permission\raisebox{.5pt}{\textcircled{\raisebox{-.2pt} {c}}} ESO).}
\label{sol_pol}
\end{figure*}

Hard X-rays, on the other hand, is believed to be from non-thermal Bremsstrahlung emission
by the high energy electrons and thus expected to be highly polarized
with degree of polarization depending on multiple factors like beaming of the electrons, magnetic field structure, back-scattering of the photons from the 
photosphere and location of the flare on the disk \citep{bai78,leach83,zharkova10,jeffrey11}. For example, Fig. \ref{sol_pol} shows polarization from various components of the hard X-ray emission of solar flares for different emission sites and electron energy distributions.  
Because of sufficient photon flux in X-rays, solar flares are the 
potential targets for X-ray polarimetry, particularly in hard X-rays. 

It can be seen that hard X-ray polarimetry has a lot of science potential for all classes of X-ray sources. However, there have not been any dedicated attempt to utilize this potential for a long time. Only recently in the last couple of decades, there are some efforts to measure polarization for some of the bright hard X-ray sources. In the next section, we will discuss those measurements and implications on understanding the nature of emission mechanism in X-ray sources.

\section{Recent scientific findings of Hard X-ray polarimetry} \label{finding}

\subsection{Cygnus X-1}
Cygnus X-1 is a famous bright galactic high mass black hole X-ray binary system, mostly seen in its hard spectral state. After a few initial attempts back in 1970s to explore the polarization properties of the source  \citep[e.g. Bragg polarimeter on board {\em OSO-8} measured low polarization around 2.4 \% and 5.3 \% at 2.6 and 5.2 keV respectively in its high soft state, see][]{long80}, lately, there are some valiant attempts to measure polarization of Cygnus X-1 in hard X-rays, both in the coronal regime (a few 10s of keV to $\sim$100 keV) and the suspected jet regime (above 100 keV) \citep{vadawale01,markoff01,vadawale03}.

PoGO+ \citep{chauvin16_pogo,friis18}, a dedicated balloon borne hard X-ray polarimeter found the source to be unpolarized in its hard state (12 $-$ 18 July 2016) in 20-180 keV \citep{chauvin18a,chauvin18b}. They placed an upper limit of 5.6 \% at position angle of 154$\pm$31$^\circ$. They also found upper limits of 11.6 \%, 2.2 \%, and 2.9 \% assuming position angle perpendicular to the accretion disk ($\pm$155$^\circ$), parallel to the disk ($\pm$330$^\circ$), and position angle aligned to that of {\em INTEGRAL} results ($\pm$222$^\circ$, see below) respectively. These results favour the hard X-ray flux to be originating from an extended  corona which has a small fraction of reflection component in the hard X-ray band \citep{chauvin18a, chauvin18b}. These results also give an upper limit of polarization for the jet component around 5 $-$ 10 \% when the emission from the extended corona was assumed to be accompanied by a synchrotron jet at energies $<$200 keV.     

\begin{figure*}
\centering
\includegraphics[height=.2\textheight]{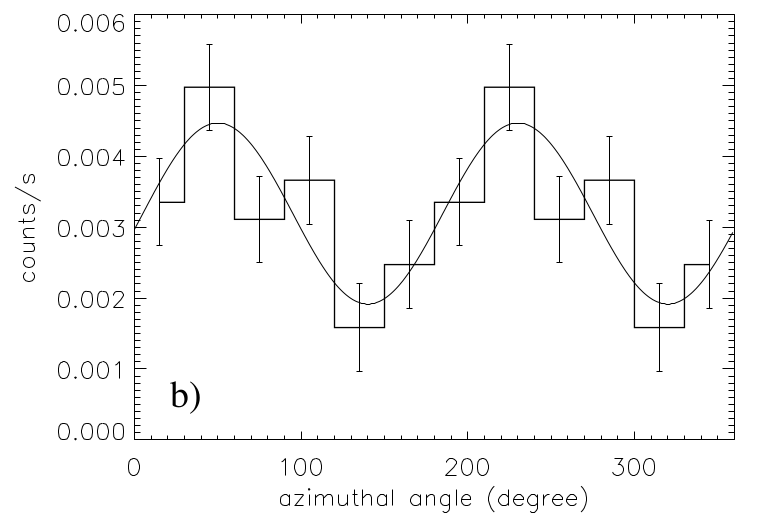}
\includegraphics[height=.2\textheight]{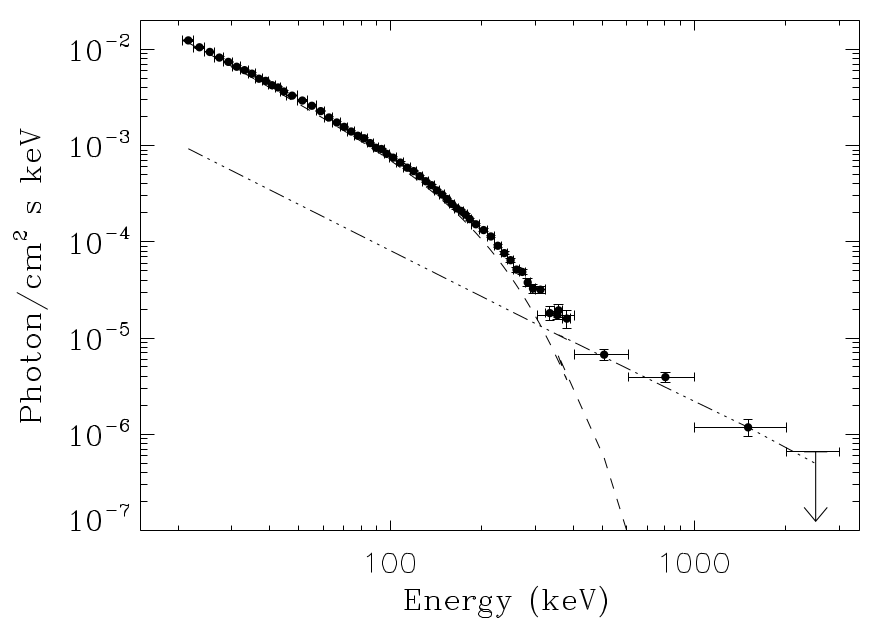}
\caption
{Polarization measurement of Cygnus X-1 using the IBIS instrument on board the {\em INTEGRAL} satellite. Polarization signal is clearly seen in higher energies  (450 $-$ 2000 keV), as can be seen in the modulation curve (left panel) indicating a separate  jet component, as is evident from the energy  spectrum (right panel). 
Figure credit $-$ from \citet{laurent11}. Reprinted with permission from AAAS.}
\label{cygx1_ibis}
\end{figure*}

On the other hand, at higher energies, IBIS \citep{ubertini03}, on board {\em INTEGRAL} estimated  high polarization for this source (67$\pm$30 \%) at 220$\pm$15$^\circ$ (north-east in anti-clockwise) in 400 $-$ 2000 keV, while emission in 250 $-$ 400 keV was found to be weakly polarized \citep{laurent11}. 
SPI \citep{vedrenne03} on board {\em INTEGRAL}, later confirmed the measurements independently. High polarization (76$\pm$15 \% at 222$\pm$3$^\circ$) was estimated in 230 $-$ 850 keV. An upper limit of 20 \% was reported in 130 $-$ 230 keV \citep{jourdain12}.  
The results were interpreted as synchrotron origin of high energy emission from the relativistic electrons in the jet. This was further corroborated by simultaneously observed two distinct spectral components of the source, a thermal comptonization component at energies below 200 keV and a power-law component beyond 200 keV, mostly due to synchrotron radiation from the jet (see Fig. \ref{cygx1_ibis}). 

\begin{figure*}
\centering
\includegraphics[height=.4\textheight]{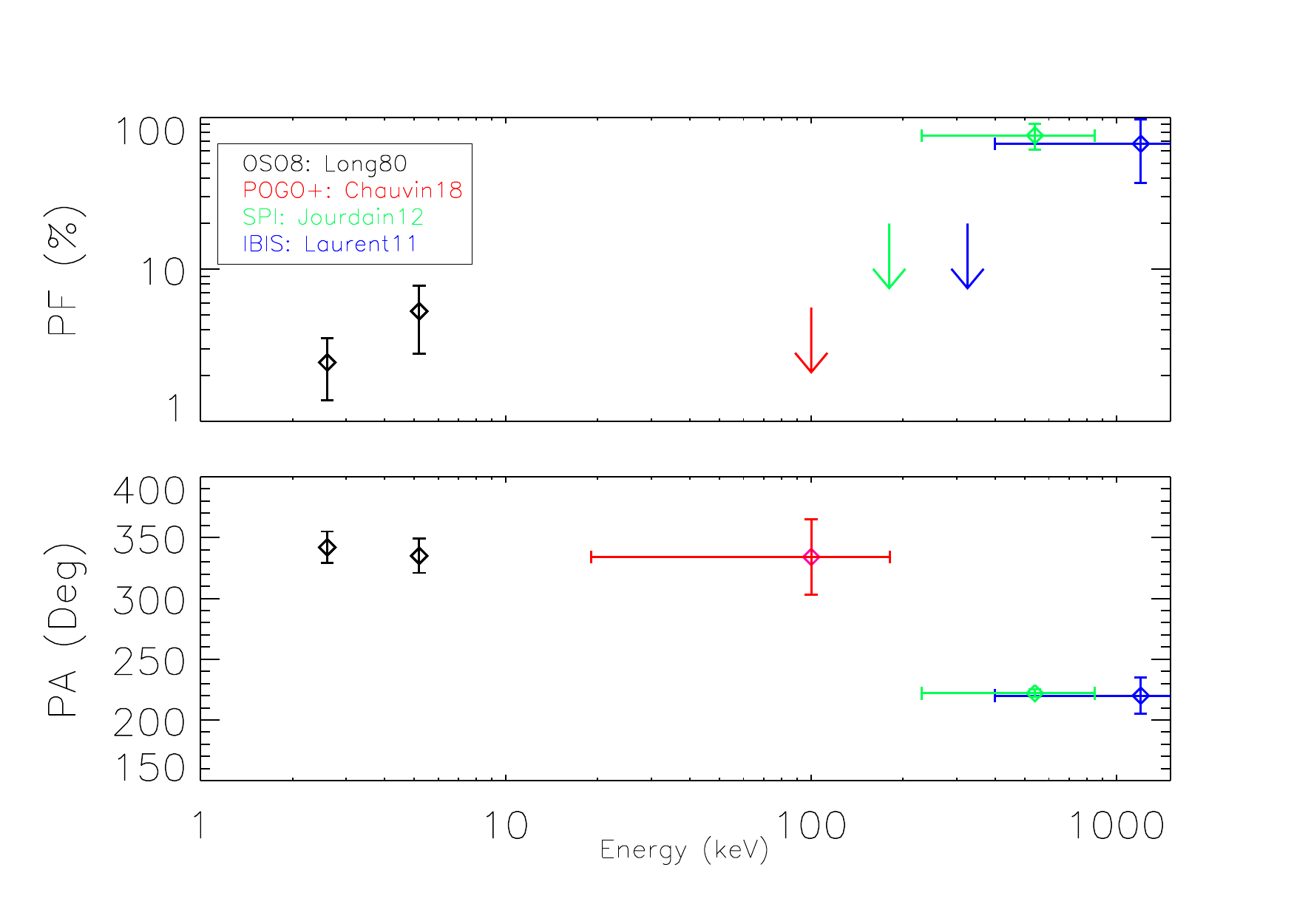}
\caption
{Summary of polarization measurements (fraction in top panel and angle in bottom panel) of Cygnus X-1 by various instruments like IBIS and SPI on board {\em Integral}, {\em OSO-8}, and PoGO+. Except for the {\em OSO-8} measurements, all other measurements correspond to the hard spectral state of the source. When plotted against observed energy of measurement, we see a clear increase in the polarization fraction and a swing in polarization angle at higher energies.}
\label{cygnusx1}
\end{figure*}

In recent years, Cadmium Zinc Telluride Imager \citep[CZTI,][]{bhalerao16} on board {\em AstroSat} has attempted a detailed energy resolved polarization in the hard state of Cygnus X-1. They found the source to be polarized at 38$\pm$7 \% with a position angle of 221$\pm$2$^\circ$ (North-East, anti-clockwise) with a hint of increase in polarization fraction (from 17 \% to 58 \%) in 100 $-$ 500 keV (private communication, Chattopadhyay et al. 2021, under preparation). This further validates increasing jet contribution at energies above 100 keV compared to the coronal counterpart. Fig. \ref{cygnusx1} shows the measured PF and PA for the source as a function of energy from all these measurements. 

Jet origin of the high energy photons can be further validated by exploring the soft or intermediate state polarization of the source as the jet is supposed to be quenched in the soft state. 
Both {\em INTEGRAL} and {\em AstroSat} have tried to explore the soft and intermediate state for polarization. However due lower flux in these states, the measurements are associated with large uncertainties. More data from these observatories and upcoming missions in the future will be useful in establishing the hard X-ray emission models of black hole XRBs.   

\subsection{V404-Cygni}
After a 25 years of quiescence, V404 Cygni which is a low mass X-ray binary with a black hole of $\sim$12 solar mass entered into outburst mode exceeding its Eddington luminosity. The high flux offered a great opportunity to explore polarization properties of the source in a broad frequency band. IBIS on board {\em INTEGRAL} observed the source in the month of June, 2015 for 10 times with exposures greater than 100 ks in each observation. The source was found to be unpolarized for all the observations except for that on June 20 $-$ 22, 2015. The emission was found to be highly polarized at 95$\pm$35 \% level with position angle of 160$\pm$15$^\circ$ in 450 $-$ 2000 keV \citep{laurent17}. It is to be noted that optical and NIR studies of polarization from this source during the outburst reveals a variable polarization with position angle around 170$^\circ$ and favours the jet origin of the optical and NIR flares \citep{shahbaz16_v404}.   

\subsection{Crab pulsar and nebula} Crab is the brightest persistent X-ray source in the sky. From early 1970s, there have been multiple attempts to measure polarization of Crab using a several rocket experiments. 
The first confirmation of Crab polarization was
made by \citet{weisskopf78} using {\em OSO-8} Bragg polarimeter. For the off-pulse nebula (non-pulsating) region of the light curve, they reported a polarization fraction (PF)
and polarization angle (PA) of 19.2$\pm$1.0 \% and 156.4$\pm$1.4$^\circ$  at 2.6 keV and
19.5$\pm$2.8 \% and 152.6$\pm$4.0$^\circ$ at 5.2 keV respectively. 

Following this historic measurement, polarization measurement of Crab was attempted a multiple times in other energy bands.
However, the detection of polarization of Crab in hard X-rays came almost after 30 years of the {\em OSO-8} detection. From observations of Crab spanning in 2003 $-$ 2006 by SPI on board {\em INTEGRAL}, \citet{dean08} reported a PF of 47$\pm$10 \% at 123$\pm$11$^\circ$ in the energy of 100 $-$ 1000 keV for the off-pulse nebula. For the phase averaged Crab signal, the PF and PA were found to be 28$\pm$6 \% and 117$\pm$9$^\circ$ respectively in 130 $-$ 440 keV \citep{chauvin13}. In the subsequent years, SPI reported improvements in the detection of PF and PA: 24$\pm$4 \% and 120$\pm$6$^\circ$ after accumulation of data spanning over 15 years between 2003 and 2018 \citep{Jourdain19}.    
{\em INTEGRAL}-IBIS, sensitive in a similar energy range as SPI also measured the phase averaged polarization of Crab $-$ $47.0\substack{+19 \\ -13}$ \% at  100.0$\pm$11.0$^\circ$ in 200 $-$ 800 keV \citep{forot08}. They found the off-pulse region to be highly polarized ($>$72 \%) at 120.6$\pm$8.5$^\circ$. 

In recent years, PoGO+, SGD \citep{tajima10} on board {\em Hitomi}, and CZTI on board {\em AstroSat} reported statistically significant polarization detections in hard X-rays.
CZTI, sensitive at energies $>$100 keV like {\em INTEGRAL}, reported statistically the most significant polarization detection of Crab \citep{vadawale17}. They reported a similar level of polarization as {\em INTEGRAL} $-$ 32.7$\pm$5.8 \% at 143.5$\pm$2.8$^\circ$ for phase averaged Crab in 100 $-$ 380 keV. For the off-pulse region, a PF and PA of 39$\pm$10 \% and 140.9$\pm$3.7$^\circ$ were reported respectively.
PoGO+, sensitive from a few tens of keV, detected a PF and PA of 20.9$\pm$5.0 \% and 124.0$\pm$0.1$^\circ$ respectively for phase averaged Crab
and $17.4\substack{+8.6 \\ -9.3}$ \% and 137.0$\pm$15.0$^\circ$ for the
off-pulse region over the energy range of 20 $-$ 160 keV \citep{chauvin18}.
SGD, sensitive in the same energy range as PoGO+, reported a similar PF and PA of 22.1$\pm$10.6 \% and $110.7\substack{+13.2 \\ -13}^\circ$ respectively \citep{hitomi18}. 

\begin{figure*}
\centering
\includegraphics[height=.25\textheight]{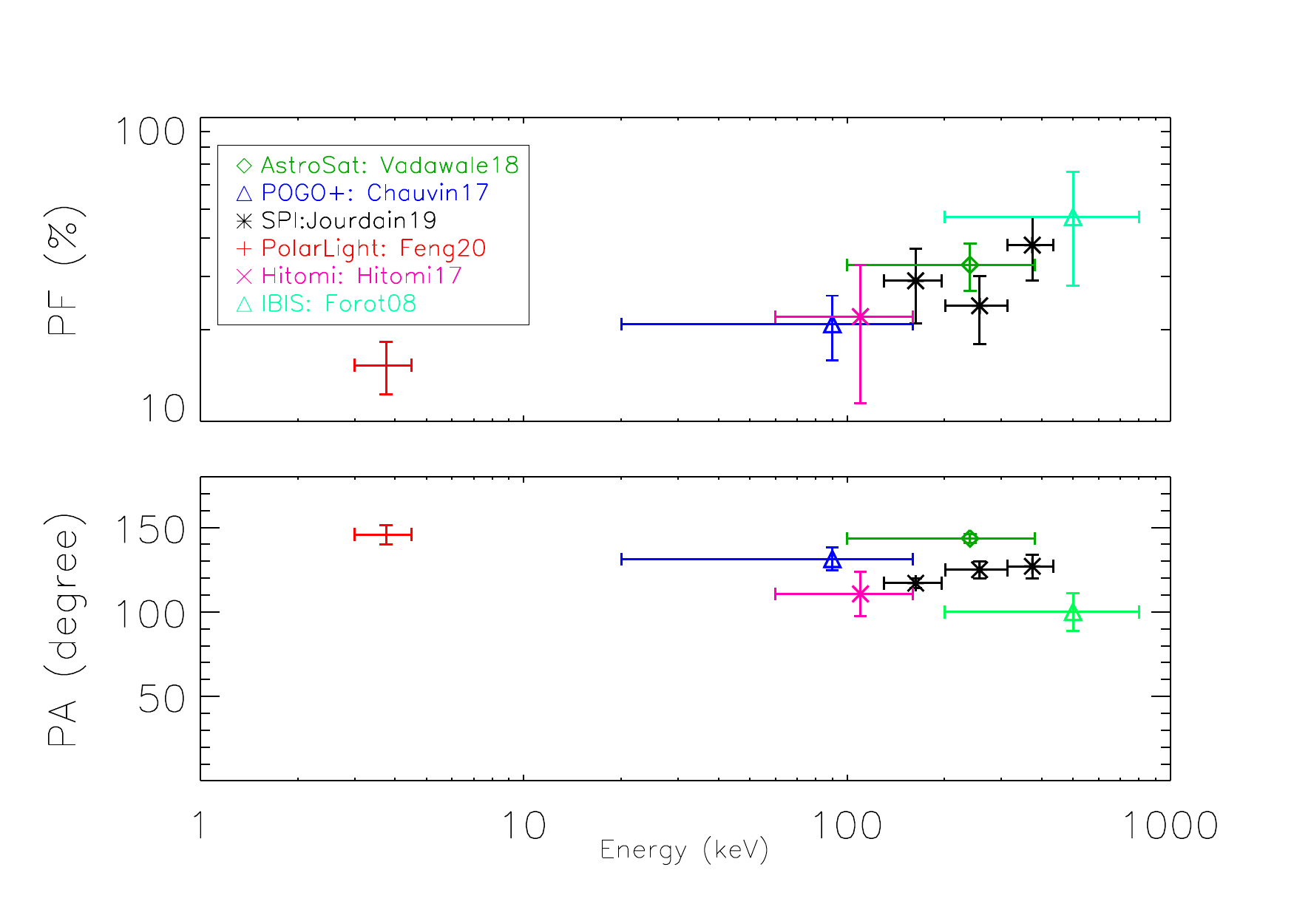}
\includegraphics[height=.25\textheight]{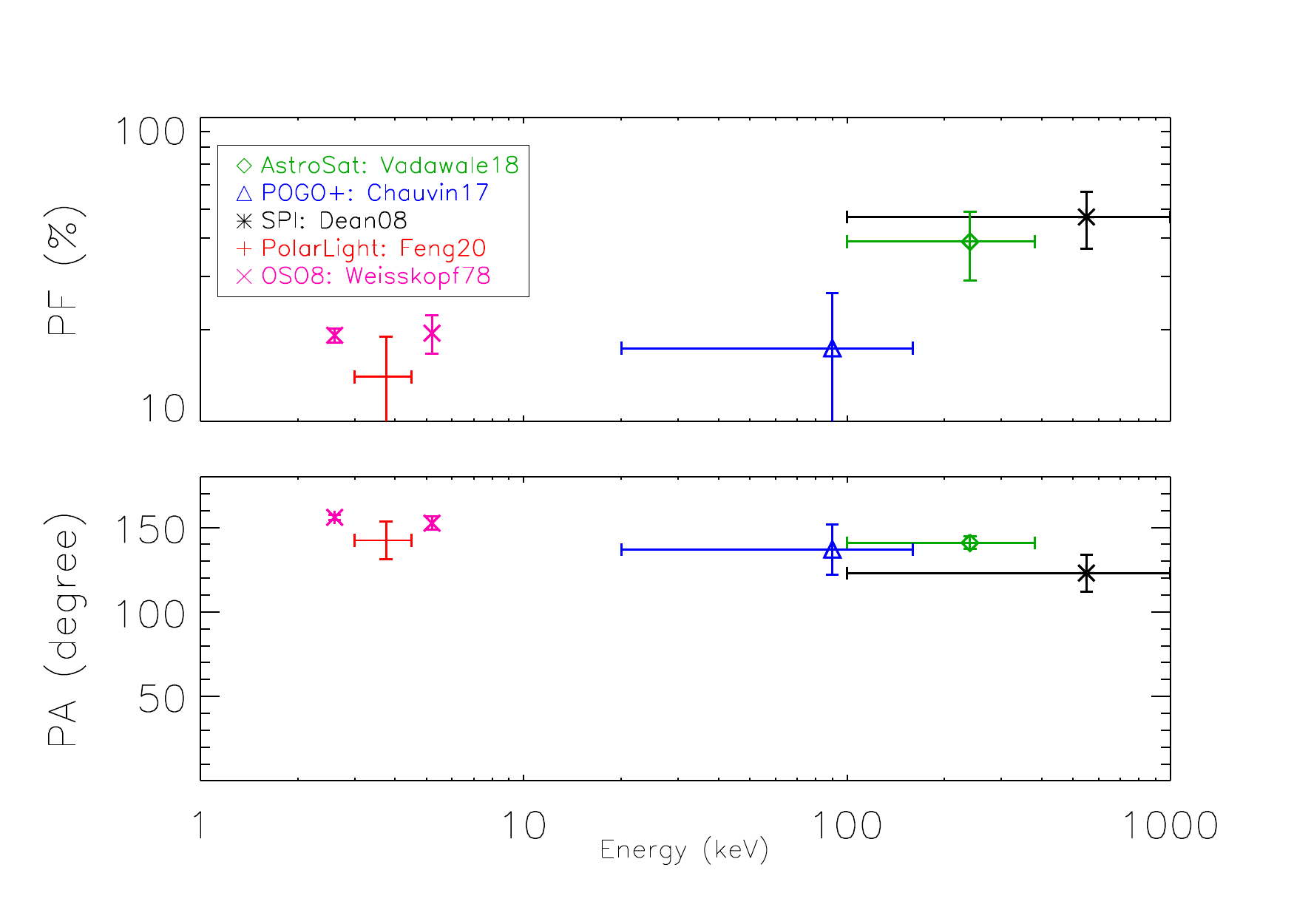}
\caption{
Summary of polarization measurements (fraction in top panel and angle in bottom panel) of total Crab (left) and off-pulse nebula (right) by various instruments like CZTI on board {\em AstroSat}, IBIS and SPI on board {\em Integral}, SGD on board {\em Hitomi}, PoGO+ and PolarLight. When plotted against observed energy of measurement, we see a hint of increase in the polarization fraction in both phase-averaged Crab and nebula.}
\label{pfvsenergy}
\end{figure*}

All these observations (summarized in Fig. \ref{pfvsenergy}) are in agreement with the optical measurements of Crab polarization    \citep[9.8$\pm$0.1 \% and 109.5$\pm$0.1$^\circ$ for phase averaged Crab and 9.7$\pm$0.1 \% and 139.8$\pm$0.2$^\circ$ for the nebula; for more details, see][]{slowikowska09}.  
The high polarization across the broad energy range indicates synchrotron emission from radio to MeV region from a magnetically ordered compact emission site. The measurements support the theory that electrons trapped in a toroidal magnetic field produce synchrotron radiation with polarization angle parallel to the spin axis of the pulsar \citep{nakamura07}. This is consistent with the X-ray images taken from {\em NuSTAR} showing toroidal ring region dominating the emission in X-rays \citep{madsen15}. 

The increase in polarization fraction for the off-pulse/nebula (and nubula+pulsar) could be because of the fact that the higher energy radiation comes from regions closer to the pulsar whereas the lower energy measurements encompass the entire nebula and pulsar. Emission from varying morphological features associated with the Crab pulsar and nebula smears the polarization at lower energies. 

\begin{figure}
\centering
\includegraphics[height=.25\textheight]{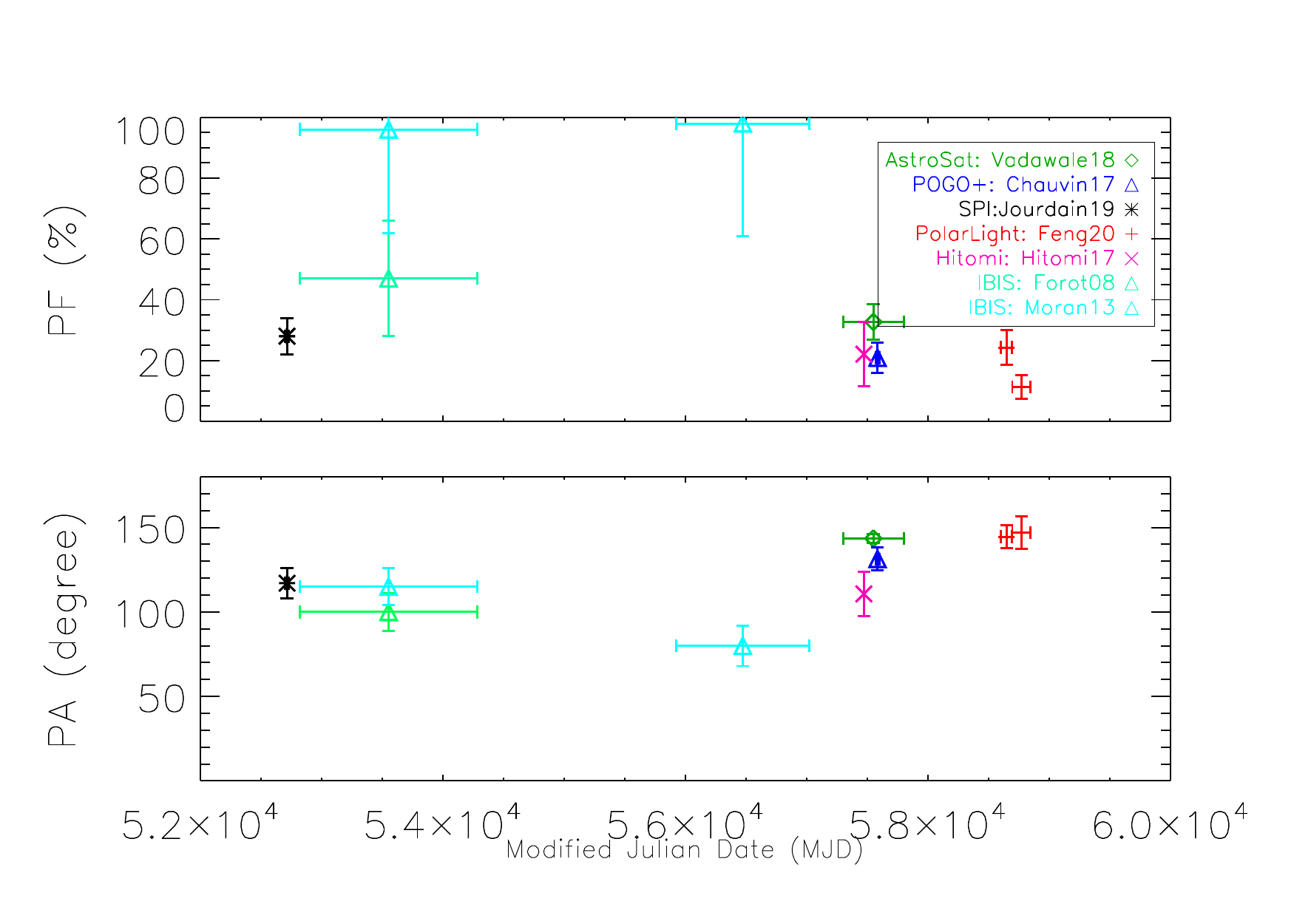}
\caption{Evolution of polarization fraction (top) and angle (bottom) for phase averaged Crab (or total Crab) with time of measurement by various instruments.}
\label{pfstability}
\end{figure}

Another interesting aspect is the stability of the source in terms of polarization given the fact that Crab is extremely stable photometrically and spectroscopically (see Fig. \ref{pfstability}). 
There are some recent reports claiming detection of change in the polarization parameters of Crab. \citet{moran13} detected a change in the polarization angle of Crab using {\em INTEGRAL}-IBIS from 115$\pm$11$^\circ$ during observations between 2003 and 2007 to 80$\pm$12$^\circ$ during 2012 $-$ 2014 observation while the polarization fraction remains consistently high. The change is attributed to the strong GeV flaring activities detected in Crab during that period which suggests a possible change in magnetic field orientation during the flares. More recently, \citet{feng20} using 
PolarLight, a gas pixel detector based soft X-ray polarimeter (sensitive in 2 $-$ 8 keV) on board a cubesat platform, reported a drop in polarization fraction from 24.3$\pm$5.7 \% at 144.5$\pm$6.7$^\circ$ to $11.3\substack{+3.7 \\ -3.8}$ \% at 146.9$\pm$9.6$^\circ$ of the pulsed emission of Crab in 3 $-$ 4.5 keV after a glitch in the pulsar detected on 23 July 2019. The nebular polarization was found to be unaffected. {The change in polarization seen in Crab could be because of possible change in the co-rotating magnetosphere configuration after the glitch.} 

\begin{figure*}
\centering
\includegraphics[height=.3\textheight]{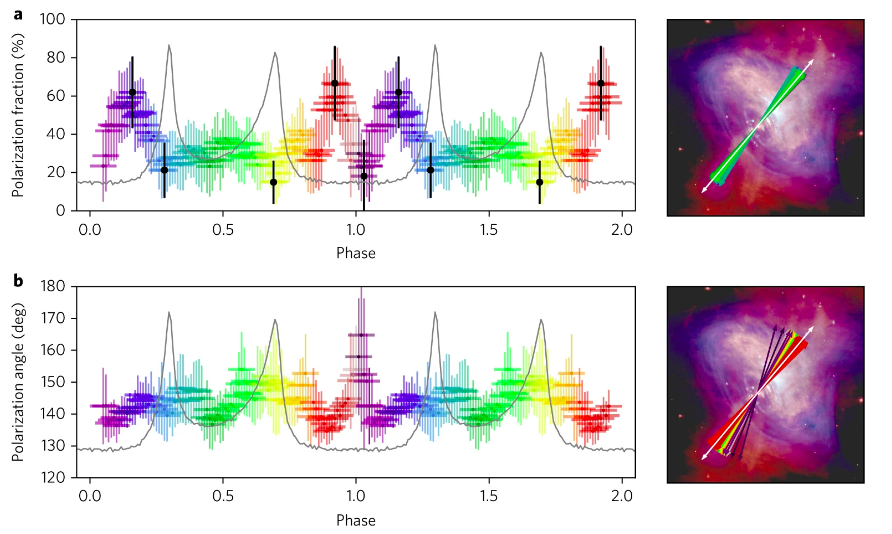}
\caption
{Dynamic phase resolved polarization of Crab pulsar+nebula from CZTI on board {\em AstroSat} in 100 $-$ 380 keV (a. Polarization fraction, b. Polarization angle) in total $\sim$800 ks of Crab observation. Polarization fraction is seen to change across the pulse peaks and in the off-pulse region. Figure credit $-$ from \citet{vadawale17}. Reprinted by permission from Springer Nature Customer Service Centre GmbH.}
\label{phasecrab}
\end{figure*}

As mentioned the previous section, phase resolved polarization study has the potential to provide useful insights of the acceleration site of the electrons in pulsars. There is only one hard X-ray phase resolved polarization study reported by \citet{vadawale17} using {\em AstroSat}-CZTI in 100 $-$ 380 keV (see Fig. \ref{phasecrab}). 
There is a change in polarization angle and fraction reported across the pulse peaks which is also seen in the optical phase resolved polarization study of Crab by \citet{slowikowska09} supporting the Two Pole Caustic (TPC) and striped wind models of acceleration. However, difference in polarization properties in two peaks in hard X-rays impose further constraints in the models. The off-pulse region also shows significant variation in polarization which is difficult to explain within the existing models. More observational data as well as further exploration in the theoretical models in future will be able to constrain the acceleration sites.   

\subsection{GRB polarization from POLAR, AstroSat, GAP, COSI and INTEGRAL}
The past decade has seen a several attempts to measure the X-ray/gamma-ray polarization of GRBs. Instruments such as {\em RHESSI} \citep{coburn03}, IBIS \citep{gotz13,gotz14} and 
SPI \citep{mcglynn07,kalemci07,mcglynn09} on board {\em INTEGRAL}, as well as
BATSE \citep{willis05} on board {\em CGRO} have reported several cases of strong polarization \citep[see a review by][] {mcconnell16}. {However, because these instruments are not optimized for polarimetry, the results are limited by statistical and systematic uncertainties and have been questioned sometimes \citep{rutledge04,wigger04}.}

One notable measurement by IBIS on board {\em INTEGRAL} for GRB 041219A shows a change in polarization from the first peak (no polarization) to high polarization around 43$\pm$25 \% in 200 $-$ 800 keV \citep{gotz09}. Analysis on finer time scale shows a change in PF and PA within the first peak resulting in dilution in the polarization signal. 

Later, GAP \citep{yonetoku06}, a dedicated large field of view (hereafter FOV) Compton polarimeter flown 
in 2011, provided polarization measurements for three bright GRBs $-$ GRB 100826A, GRB 110301A,
GRB 110721A in 70 $-$ 300 keV \citep{yonetoku11,yonetoku12}. They reported high polarization fraction for GRB 110301A ($\sim$70 \%) and GRB 110721A ($\sim$84 \%) in the full burst period while for GRB 100826A (time integrated PF of 27 \%), a $\sim$90$^\circ$ change in polarization angle was reported between the first and second interval at 99.9 \% confidence level.   

More recently, 
POLAR \citep{sun16,orsi11}, a dedicated GRB polarimeter launched in 2016, provided precise polarization measurements for 14 GRBs \citep{zhang19,Kole20polar_catalog}. They reported lower level of polarization fraction for the GRBs in the full burst intervals, between 5.9 and 21 \%, hinting at unpolarized nature of GRBs in general. For GRB 170114A, a change in polarization angle was reported by 105$^\circ$ between the first and second interval of the burst \citep{zhang19}. The time evolution of polarization properties across the bursts is thought to be a generic feature resulting in the unpolarized nature of the POLAR detected GRBs.

CZTI on board {\em AstroSat} reported polarization measurements for 11 bright GRBs from the first year of its operation in 100 $-$ 300 keV \citep{chattopadhyay19}. For six of the GRBs, high polarization fractions ($>$50 \%) were measured whereas for the remaining GRBs, upper limits were reported. The selection of energy and GRB intervals were optimized in order to get the best detection of polarization. Some of the bright GRBs were analyzed in more details and were found to have time and energy dependent polarization properties. GRB 160821A showed a $\sim$90$^\circ$ flip in polarization angle from the initial few seconds to the peak region of the burst \citep{vidushi19}. GRB 171010A was also found to have a change in polarization angle across the burst \citep{chand18b}. For GRB 171010A \citep{chand18b} and GRB 160802A \citep{chand18a}, changes in polarization properties across the peak energy of the bursts were noticed. 

COSI \citep{yang18}, a balloon borne hard X-ray spectro-polarimeter, flown in 2017, detected GRB 160530A. An upper limit of $\sim$40 \% was reported for this GRB \citep{lowell17} in 200 keV to 2 MeV. 

{Last few years have been extremely fruitful in the field of GRB polarimetry science. Measurements from POLAR and {\em AstroSat} have almost tripled the GRB polarization sample in last five years. Roughly, one half of the sample possesses significant amount polarization either in the full burst interval of the GRBs or in some cases, only in some selected intervals of the prompt emission, suggesting synchrotron radiation in an ordered magnetic field generated close to the center of explosion and carried by ejecta in a poynting flow dominated jet \citep{lyutikov03,nakar03,granot03}.
In the other half of the sample, we see low or no polarization when integrated over the full burst interval (all the reported measurements are summarized in Fig. \ref{pol_epeak_measurement}). 
\begin{figure*}[htb]
\centering
\includegraphics[height=.4\textheight]{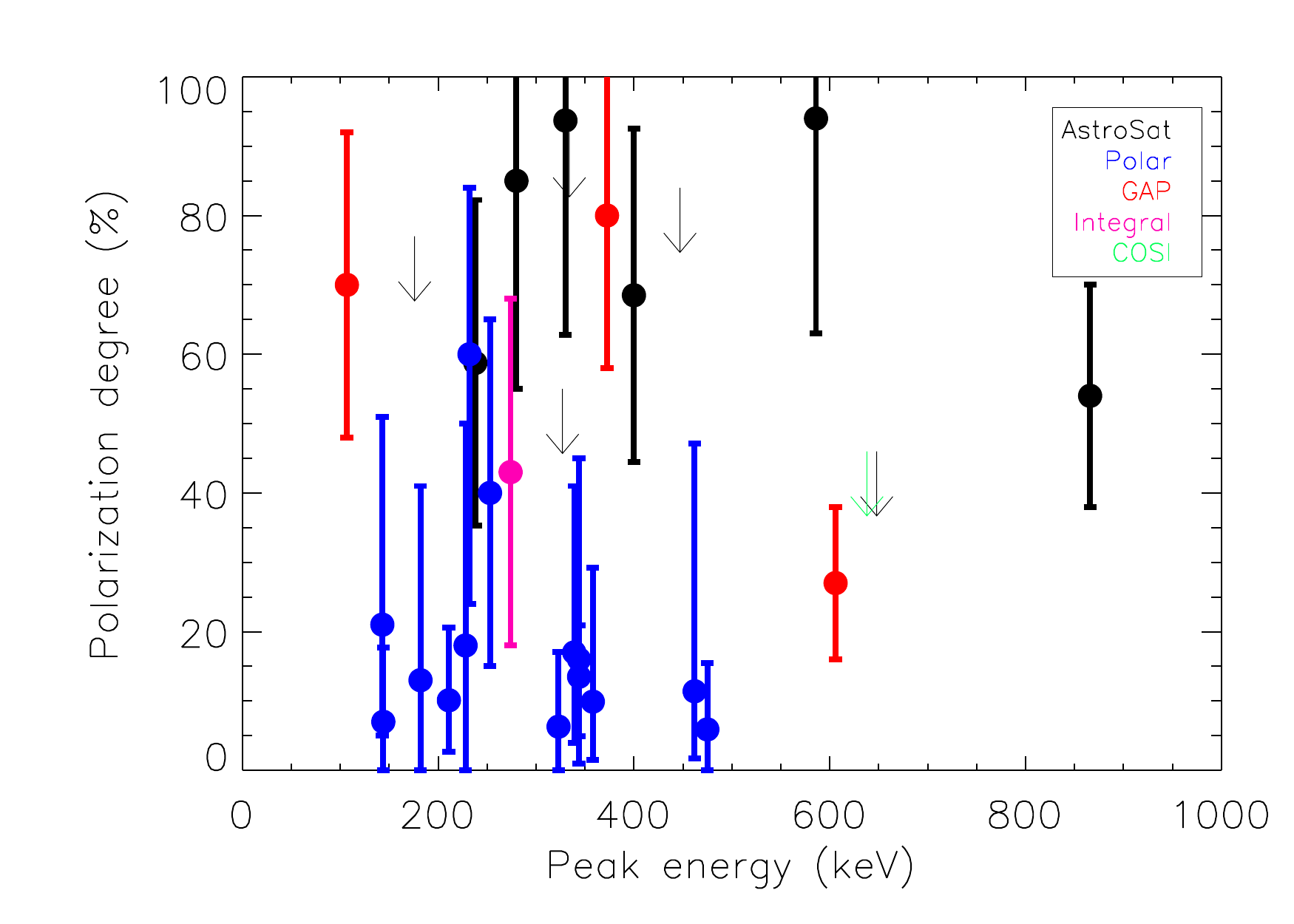}
\caption
{Measured distribution of polarization fraction with E$_{peak}$ for GRBs detected by instruments - CZTI on board {\em AstroSat} \citep[black,][]{chattopadhyay19}, POLAR \citep[blue,][]{Kole20polar_catalog}, GAP \citep[red,][]{yonetoku11,yonetoku12}, IBIS on board {\em INTEGRAL} \citep[pink,][]{gotz09} and COSI \citep[green,][]{lowell17}.}
\label{pol_epeak_measurement}
\end{figure*}
There is a strong indication that evolution of polarization parameters across the bursts has been responsible for the unpolarized nature of the GRBs when integrated across the burst \citep{vidushi19,chand18b,yonetoku11,gotz09,zhang19}. There is also a hint of change in polarization with energy, particularly across the peak energy \citep{chand18a,chand18b}.
A firm conclusion on the prompt emission models from polarization sample study still requires more precise measurements for a larger sample of GRBs. 
While POLAR has stopped its operation on March 31, 2017, {\em AstroSat} is still active and is supposed to provide a large sample of GRB polarization along with those detected by {\em INTEGRAL}.} 

\subsection{Polarization of solar flares from RHESSI and CORONAS-F}
In the recent past, there are some reports on the measurement of polarization of solar flares in hard X-rays using {\em RHESSI} and SPR-N on board {\em CORONAS-F} \citep{mcconnel07}. While {\em RHESSI} is primarily a spectroscopic instrument, \citet{boggs06} and \citet{garcia06} attempted to extract polarization information at energies above 100 keV utilizing the scattering of photons between the Germanium detectors. They measured a polarization fraction ranging from 2 $-$ 54 \% for M and X class flares that occurred from Sun's center to limb. SPR-N, a dedicated Thomson polarimeter sensitive in 20 $-$ 100 keV \citep{bogomolov03}, on board {\em CORONAS-F}, on the other hand reported polarization measurement for 25 M and X class flares with an upper limit of 8 $-$ 40 \% \citep{zhitnik06}. For one of the flares occurred on 29$^{th}$ October 2003, they detected a significant
polarization degree that increases from about 50 \% at 20  $-$ 40 keV energy band to more than 70 \% in the 60 $-$ 100 keV band. Due to the large uncertainties in the measurements, any firm conclusion was not possible. 

We can see that though the Sun is very close to us, the hard X-ray polarimetry measurements are very sparse. This is because the Sun is very faint in hard X-rays except during large flares and even then the emerging spectra are very steep giving very few photons in hard X-rays. In fact many GRBs are brighter than bright solar flares. But, a good measurement of hard X-ray polarization from the Sun will be very useful to develop the techniques of figuring out the source emission mechanism based on hard X-ray polarimetry because many of the source parameters are well constrained (for example, the distance to the Sun is well measured). CZTI of {\em AstroSat} was operational during the current solar minimum and hopefully it will measure hard X-ray polarisation of a few solar flares in the next few years when the solar activity is likely to pick up.

Currently we only have a glimpse of the science that could be achieved from polarization measurements and it appears that focusing on the higher energy part of the emission is likely to be more rewarding. This energy range, however, is an extremely difficult area even for spectroscopic measurements. The hard X-ray focusing techniques, used in {\em Hitomi} and {\em NuSTAR}, is likely to be the mainstay of observations in energy ranges up to about 80 keV. 
{To optimally use the scarce observational platforms that would be available in this energy range, it is important to develop sensitive polarimetry configurations with the best possible detection technologies available today. 
At energies above 80 keV, unlikely to be accessible to focusing techniques in the near future, conventional open detectors like CZT-Imager are the most effective and we touch upon possible improvements, like that contemplated in the forthcoming Indian mission {\em Daksha}. We will develop these arguments through section \ref{pol_measurement} to \ref{pol_instrument}, starting with the basics of X-ray polarimetry measurement methods in the next section. }

\section{Hard X-ray Polarimetry: How?}
\label{pol_measurement}
\subsection{The basics}
Since polarization is not a directly measurable quantity, its
measurement requires conversion to some observable 
quantity. The common
feature of any X-ray interaction with matter, is the dependence of 
cross-section on polarization, giving rise to a 
variable intensity in number of 
photons or electrons with respect to the  polarization direction.  
Based on this, there are three basic techniques to extract 
polarization information, namely, Compton / Rayleigh scattering, photo-electron 
imaging and Bragg reflection \citep{kaaret14}, where the variability
in intensity is fitted with a suitable modulation function, with
amplitude of modulation (measure of the degree of polarization) being 
obtained from non-linear regression. In all these processes
the detected polarization signal on the detector plane can be described as
\begin{equation}
S=\bar{S}[1+a_0\cos2\left(\phi-\phi_0\right)],
\label{c1q1}
\end{equation}
where, $\phi$ is the azimuthal angle with respect to a reference axis on 
the detector plane and $\bar{S}$ is the mean number of events / counts 
in $\phi$ bins. It is evident from Eq. \ref{c1q1}, that 
the distribution of the events, as discussed earlier, is modulated 
with $\phi$ having an 
amplitude $a_0$ and position angle of $\phi_0$, where $a_0$ is 
proportional to the degree of linear polarization.

However, in presence of noise (which we assume to be of Poisson distribution), 
there is a certain probability, $P(a,\phi)$, to measure an amplitude of $a$ and 
phase $\phi$, even though the actual amplitude and position
angle in the source signal are $a_0$ and $\phi_0$ respectively, given by,
\begin{equation}
P(a,\phi)=\dfrac{Na}{4\pi} \exp\left[-\dfrac{N}{4}\left(a^2+a_0^2-2aa_0\cos(\phi-\phi_0)\right)\right],
\label{c1q2}
\end{equation}
where, $N (= n\bar{S}$, $n$ is the number of $\phi$ bins) is the total 
number of detected events. Since, modulation 
is always positive definite, even if the source is unpolarized ($a_0$ = 0), 
there is still a finite probability to measure an amplitude $a$ (i.e. 
$P(a) \ne 0$). 

Sensitivity or the minimum polarization that the 
instrument will be able to detect, can be established by 
estimating the value 
of modulation amplitude for unpolarized source signal ($a_0$ = 0),
which is exceeded by chance with 1 $\%$ probability, i.e.
\begin{equation}
\dfrac{N}{2} \int_{a_{1\%}}^{\infty}{a\,\exp\left[-\dfrac{Na^2}{4}\right]\,da}=0.01\;.
\label{c1modampl}
\end{equation}
From Eq. \ref{c1modampl}, we get modulation amplitude for unpolarized 
source,
\begin{equation}
a_{1\%}=\dfrac{4.29}{\sqrt{N}}.
\label{c1q3}
\end{equation}
Eq. \ref{c1q3} leads to the Minimum Detectable Polarization (MDP) or sensitivity of the instrument in terms of source and background 
rates ($R_{src}$ and $R_{bkg}$ respectively) and modulation amplitude 
for 100 $\%$ polarized signal ($\mu_{100}$), 
\begin{equation}
\centering
MDP_{99\%} = \frac{4.29}{R_{src}\,\mu_{100}}\,\sqrt{\frac{R_{src}\,+\,R_{bkg}}{T}},
\label{clq4}
\end{equation}
where, $T$ is the time exposure for polarization measurement. 

\begin{figure}[t]
\includegraphics[height=.25\textheight]{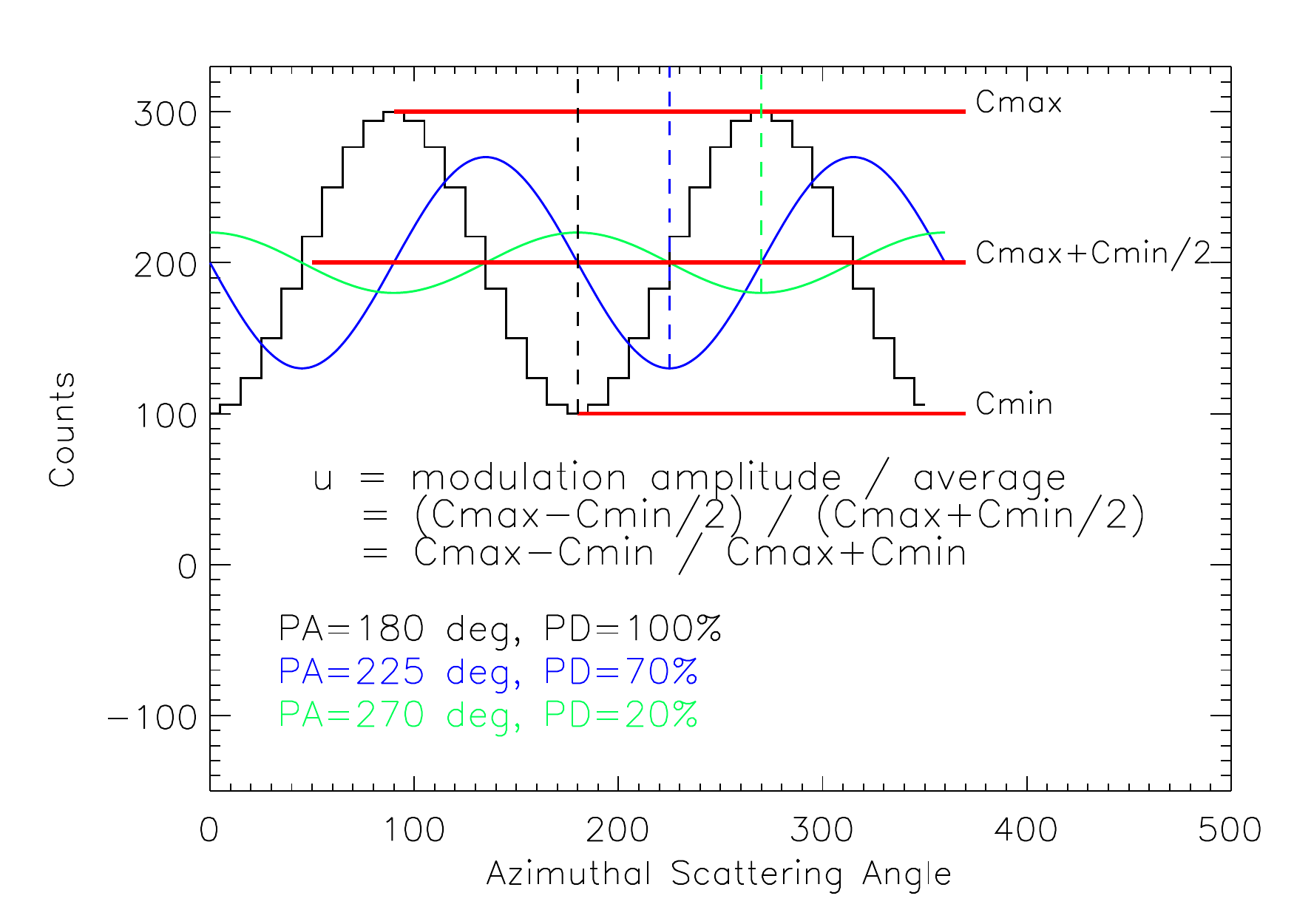}
\caption{An example of how modulation signature looks for highly polarized, medium polarized and low polarized radiation (black, blue, green respectively).}
\label{modcurve_gen}
\end{figure}

Once the modulation curve is obtained for any unknown
polarized radiation, the conventional way to 
measure polarization fraction, $P$, is to first obtain the 
modulation amplitude from the
modulation curve (with $C_{max}$ and $C_{min}$ being the maximum and minimum 
number of counts in the modulation curve as shown in Fig. \ref{modcurve_gen}),
\begin{equation}
\mu=\dfrac{C_{max}-C_{min}}{C_{max}+C_{min}},
\end{equation}
and then normalize it with respect to the modulation factor for 100 $\%$
polarized beam, $\mu_{100}$, such that, 
\begin{equation}
P=\dfrac{\mu}{\mu_{100}}.
\end{equation}    

\begin{figure}
\includegraphics[height=.25\textheight]{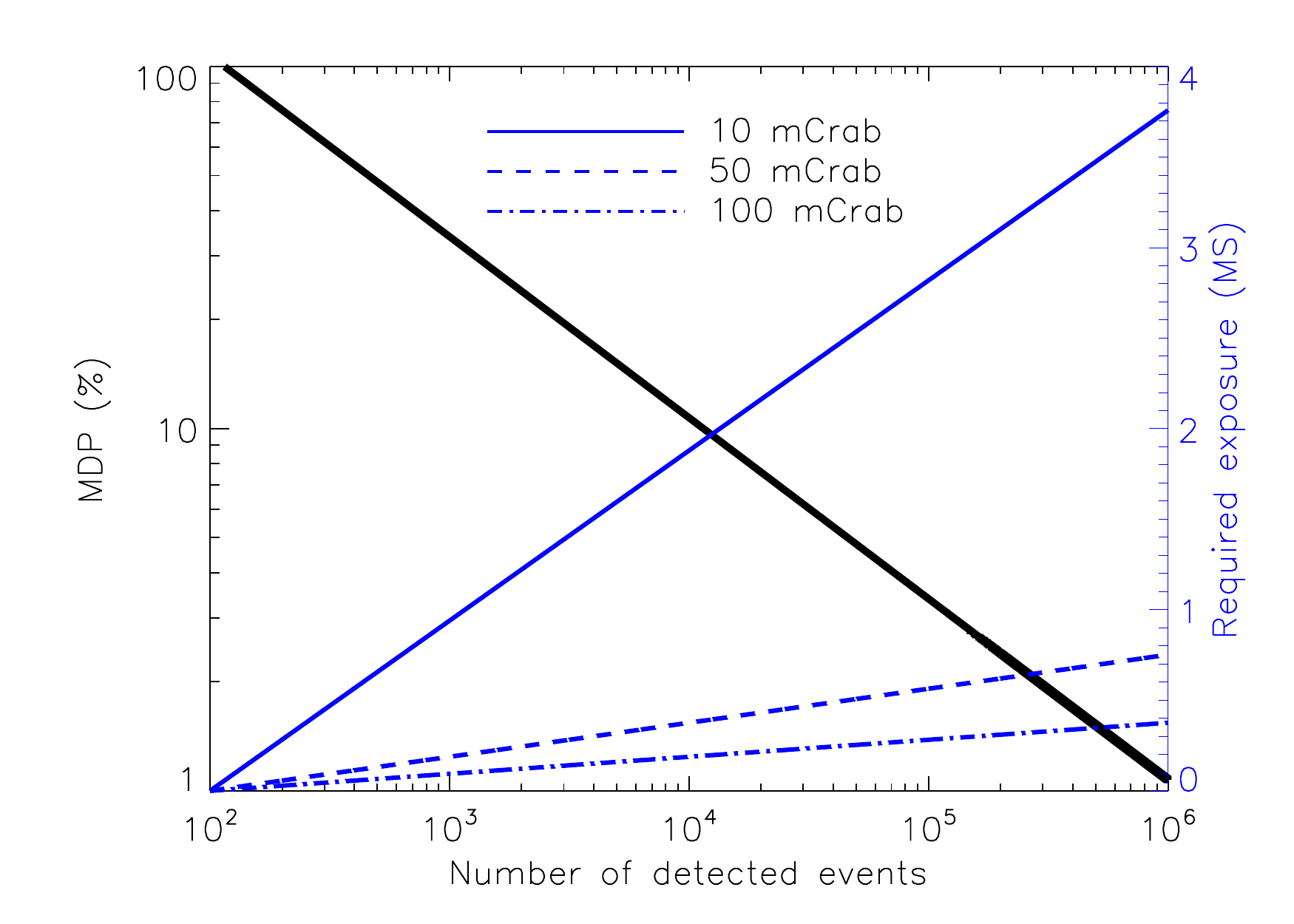}
\caption{An example of Minimum Detectable Polarization (MDP) and required number of photons to be detected in absence of background. The MDP values are estimated assuming a polarimeter with an effective area of 80 $cm^2$ and $\mu_{100}$ of 0.4 averaged over 20 $-$ 1000 keV. The blue lines in the right axis correspond to the required time exposures for various source intensities.}
\label{mdp_time}
\end{figure}

$\mu_{100}$ is typically estimated from simulations or experiments. For any polarization measurement technique, $\mu_{100}$ and efficiency 
should be as high as possible to have sensitivity well above the
expected degree of polarization from the astrophysical sources (see Eq. \ref{clq4}). Number of detected photons required to estimate various MDP levels in absence of any background are shown in Fig. \ref{mdp_time}. 
{With polarimetric efficiency, $\epsilon_{pol}=0.4$ and $\mu_{100}=0.4$ averaged over 20 $-$ 1000 keV,
an MDP of 10 \% requires detection of $\sim$10$^4$ photons or an exposure of 2.5 million seconds for a 200 cm$^2$ collecting area polarimeter for a 10 mCrab source.} The numbers shown in Fig. \ref{mdp_time} are an order of magnitude higher than that required for imaging and spectroscopy studies which makes X-ray polarimetry extremely difficult.       
\subsection{The available techniques}
There are mainly three conventional techniques available to convert the polarization of radiation into some measurable quantity $-$ scattering, photo-electric, and Bragg reflection. Here we will briefly discuss the basics of the measurement techniques followed by a comparison in hard X-rays. For more details on the techniques, readers are suggested to see \citet{kaaret14}.

\subsubsection{Scattering polarimetry}
\begin{figure}
\centering
\includegraphics[height=.2\textheight]{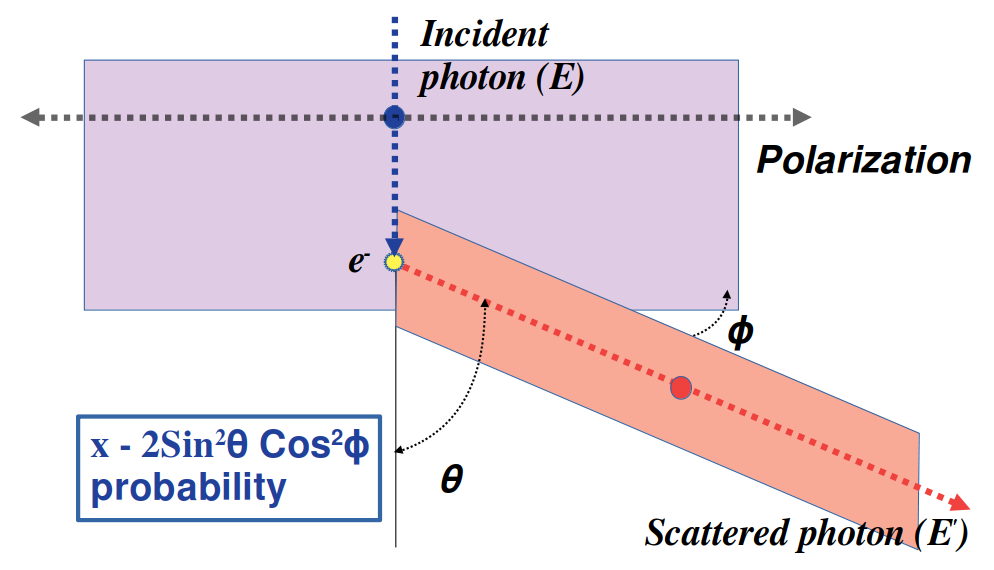}
\caption{Schematic of Compton scattering geometry. $\theta$ and $\phi$ represent polar and azimuthal angle of scattering respectively. 
The primary photon is preferentially scattered perpendicular to the polarization vector giving rise to an asymmetry in the distribution of azimuthal angle with $x - 2\sin^2{\theta}\cos^2{\phi}$ probability. Value of
`x' depends on the incident and scattered photon energies, given by E and $E'$ respectively (x $=$ 2 for $E = E'$).}
\label{scat_geom}
\end{figure}

\begin{figure}
\centering
\includegraphics[height=.5\textheight]{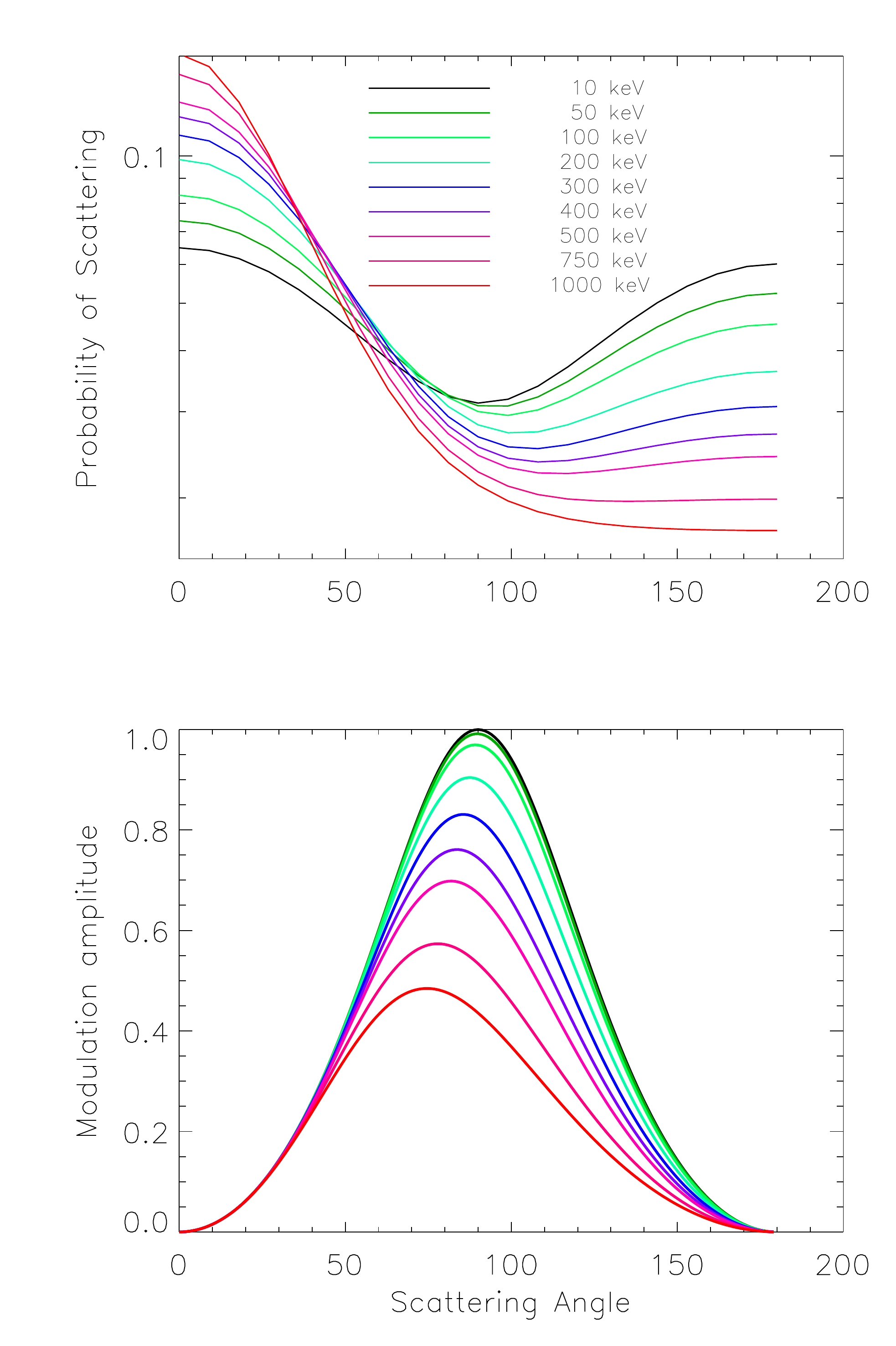}
\caption{Probability of Compton scattering and modulation amplitude as a function of  scattering angle and incident photon energies (shown in different colors) are shown in the top and bottom panel respectively. We see the probability of scattering is minimum for $\theta = 90^\circ$ where we expect highest modulation amplitude which results in an overall low/moderate level of modulation factor in Compton polarimeters.}
\label{comp_cross}
\end{figure}

Scattering polarimetry is based on Compton or Rayleigh scattering, where 
the photon is scattered off an electron and imparts 
either a small energy to the electron (Compton scattering) or travels 
with the same energy (Rayleigh scattering). The differential cross-section 
for Compton scattering is given by 
Klein-Nishina formula \citep{heitler54},
\begin{equation}
\label{c1q6}
{\dfrac{d\sigma}{d\Omega}}={\dfrac{r_{e}^{2}}{2}}\left({\dfrac{E^{'}}{E}}\right)^{2}\left({\dfrac{E^{'}}{E}}+\dfrac{E}{E^{'}}-2\sin^{2}\theta\cos^{2}\phi\right),
\end{equation}
where $E$ and $E'$ are the incident and scattered photon energies 
respectively given by,
\begin{equation}
\label{c1q7}
{\dfrac{E^{'}}{E}}={\dfrac{1}{1+\dfrac{E}{mc^{2}}\left(1-\cos\theta\right)}}.
\end{equation}
$r_{e}$ is the classical electron radius, $m$ is the mass of electron, 
$\theta$ is the polar scattering angle, and $\phi$ is the azimuthal angle between the polarization direction and scattering plane as shown in Fig. \ref{scat_geom}. Cross-section for 
Rayleigh scattering is obtained from Eq. \ref{c1q6} with $E^{'}$ made
equal to $E$. In both Compton and Rayleigh scattering, the 
distribution of the scattered
photons with azimuthal angle, $\phi$, is modulated by $\cos^2\phi$ function. 
It is evident that
the amplitude of modulation is maximum for $\theta = 90^\circ$, however, the probability of scattering is 
found to be minimum at 90$^\circ$ compared to that for forward and 
back-scattering (see Fig. \ref{comp_cross}). This makes scattering polarimeters 
to have moderate or low modulation factors as compared to Bragg and 
photoelectric polarimeters (discussed later). 

\begin{figure}
\centering
\includegraphics[height=.25\textheight]{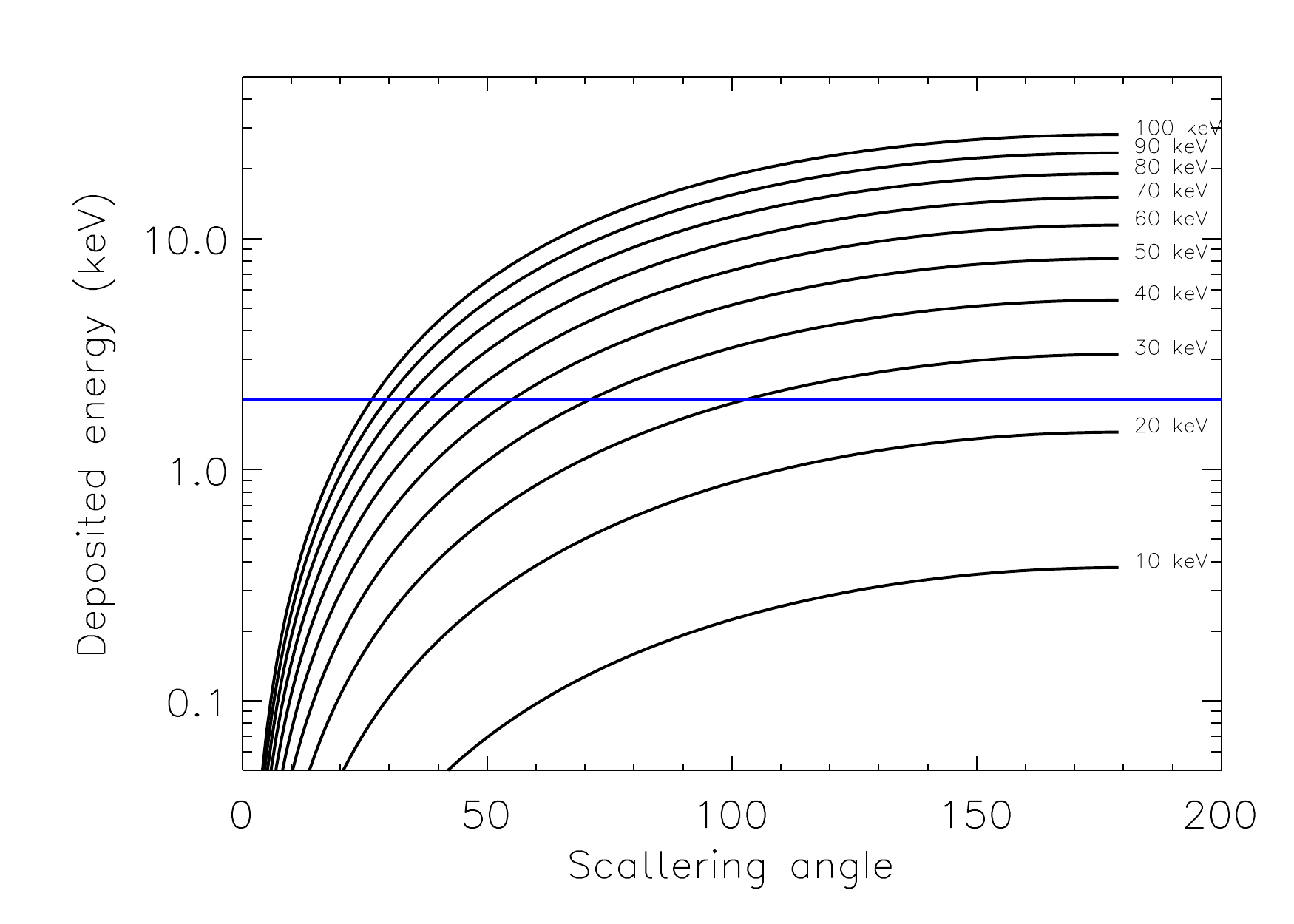}
\caption{Electron recoil energy deposited in the scatterer in Compton scattering as a function of scattering angle and different photon energies. A lower energy threshold of 2 keV in the scatterer only allows active coincidence between scatterer and absorbers for energies greater than $\sim$25 keV. The lower energy threshold in the plastic determines the threshold energy of the Compton polarimeter and its polarimetric sensitivity.}
\label{enedep_comp}
\end{figure}

Scattering polarimeters, being based on recording of the photons scattered at 
various azimuthal angles, they consist of scatterers surrounded by an array of absorbers in order to absorb the 
scattered photons. An important feature of Compton polarimetry 
is the extremely low background in comparison to the Rayleigh mode,
which is achieved due to the requirement of simultaneous detection 
of both the primary Compton scattering event in the scatterer as 
well as the detection of the scattered photon by the surrounding 
absorber. Since, the energy transferred to the electron in a scattering 
event is typically a small fraction of the incident photon energy, 
scattering polarimeters working in Compton mode are unable to work at 
lower energies. This is demonstrated in Fig. \ref{enedep_comp} where the deposited energy in a Compton scattering event is shown as a function of scattering angle and incident photon energies.     
A lower energy threshold of 2 keV in the scatterer (shown by blue horizontal line in Fig. \ref{enedep_comp}) only allows photons of energies 25 keV or above to deposit sufficient energy in the scatterer to achieve temporal coincidence between the scatterer and absorber.   
On the other hand, since Rayleigh polarimeters do 
not require temporal coincidence, they are sensitive to lower energies as well, 
where the lower energy cut off depends on the turn over between the
photoelectric and Rayleigh scattering probabilities.   

\subsubsection{Photoelectric polarimetry}
\begin{figure}
\centering
\includegraphics[height=.2\textheight]{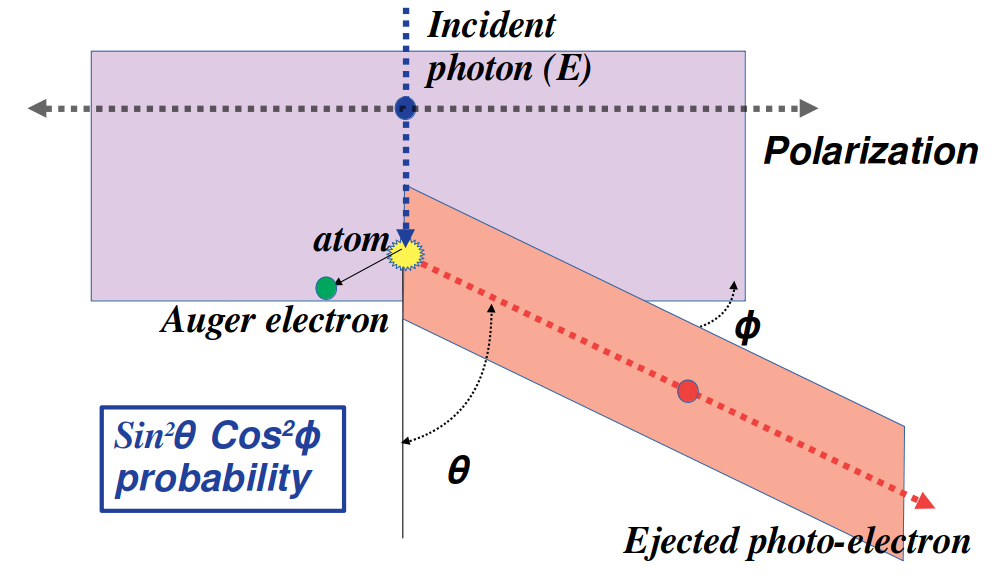}
\caption{Schematic of Photo-electric absorption geometry. $\theta$ is the polar angle of the ejected k-shell electron and $\phi$ is the angle between polarization vector and electron ejection plane. The photo-electron is preferentially ejected along the polarization direction of the photon giving rise to an asymmetry in the ejected electron distribution which is proportion to $\sin^2{\theta}\cos^2{\phi}$. Each photo-absorption event is associated with an Auger electron emission shown by the green sphere here.}
\label{photo_schematic}
\end{figure}

In photo-absorption of the X-ray photons, the k-shell photo-electrons 
are preferentially ejected in the direction of polarization, constituting the basic asymmetry in the azimuthal angle 
distribution. Cross-section of photoelectric absorption is given by,
\begin{equation}
{\dfrac{d\sigma}{d\Omega}}=\dfrac{r_{e}^{2}Z^5}{137^4}{\left(\dfrac{mc^2}{E}\right)}^{7/2}\dfrac{4\sqrt{2}\sin^2\theta\cos^2\phi}{{(1-\beta\cos\theta)}^4},
\label{q2}
\end{equation}
where $\theta$ is the polar angle between the direction of incoming photon and 
ejected k-shell electron and $\phi$ is the azimuthal angle of the ejected 
electron with respect to the polarization vector (see Fig. \ref{photo_schematic}). Modulation in the ejected 
angle distribution is maximum for $\theta = 90^\circ$ (see Eq. \ref{q2}). 
Since at energies of a few keV, the photo-electrons are preferentially 
emitted at 90$^\circ$ polar angle, modulation amplitude is expected 
to be higher for photoelectric polarimeters compared to the 
scattering polarimeters. 
Furthermore, since at lower energies, most of the photons interact via 
photoelectric absorption, polarimetric efficiency
for the photoelectric polarimeters is high in soft X-rays making it 
intrinsically more sensitive 
instrument compared to scattering or Bragg polarimeters at lower energies. However, at 
higher energies, Compton polarimeters are more sensitive due to the increase in
scattering probability of photons. Therefore, these two techniques 
are sensitive in different energy ranges and are 
complimentary to each other.

\begin{figure}
\centering
\includegraphics[height=.26\textheight]{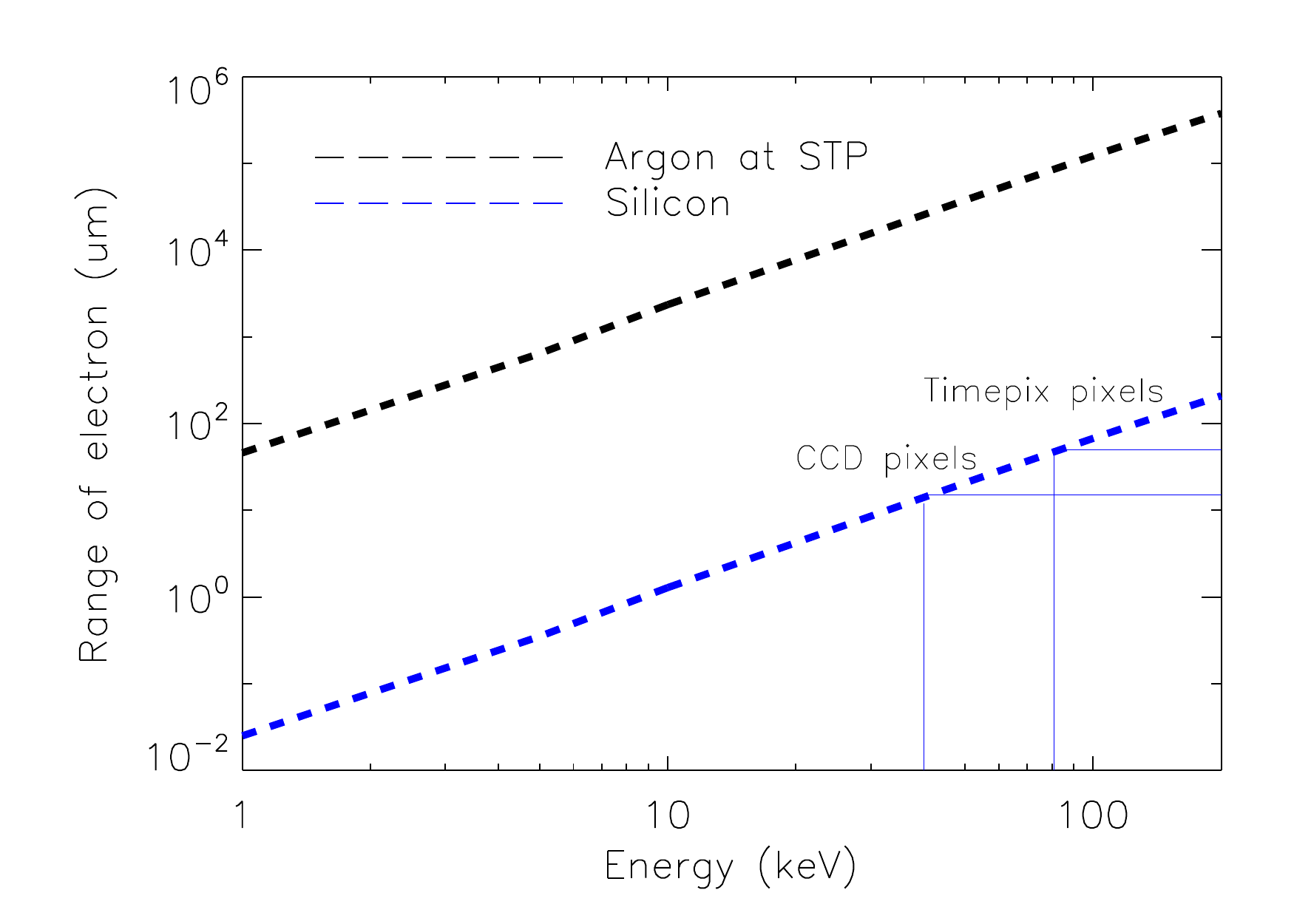}
\caption{Range of electron computed as a function of energy in solid (Silicon) and gases (argon at STP). Horizontal lines represent typical pixel size of 15 $\mu$m in CCD and 55 $\mu$m in Timepix detectors to illustrate that photo-electron track imaging in CCDs and Timepix detectors is possible at very high energies where track length is significantly larger than the pixel sizes.}
\label{track_pixel}
\end{figure}

\begin{figure}
\includegraphics[height=.12\textheight]{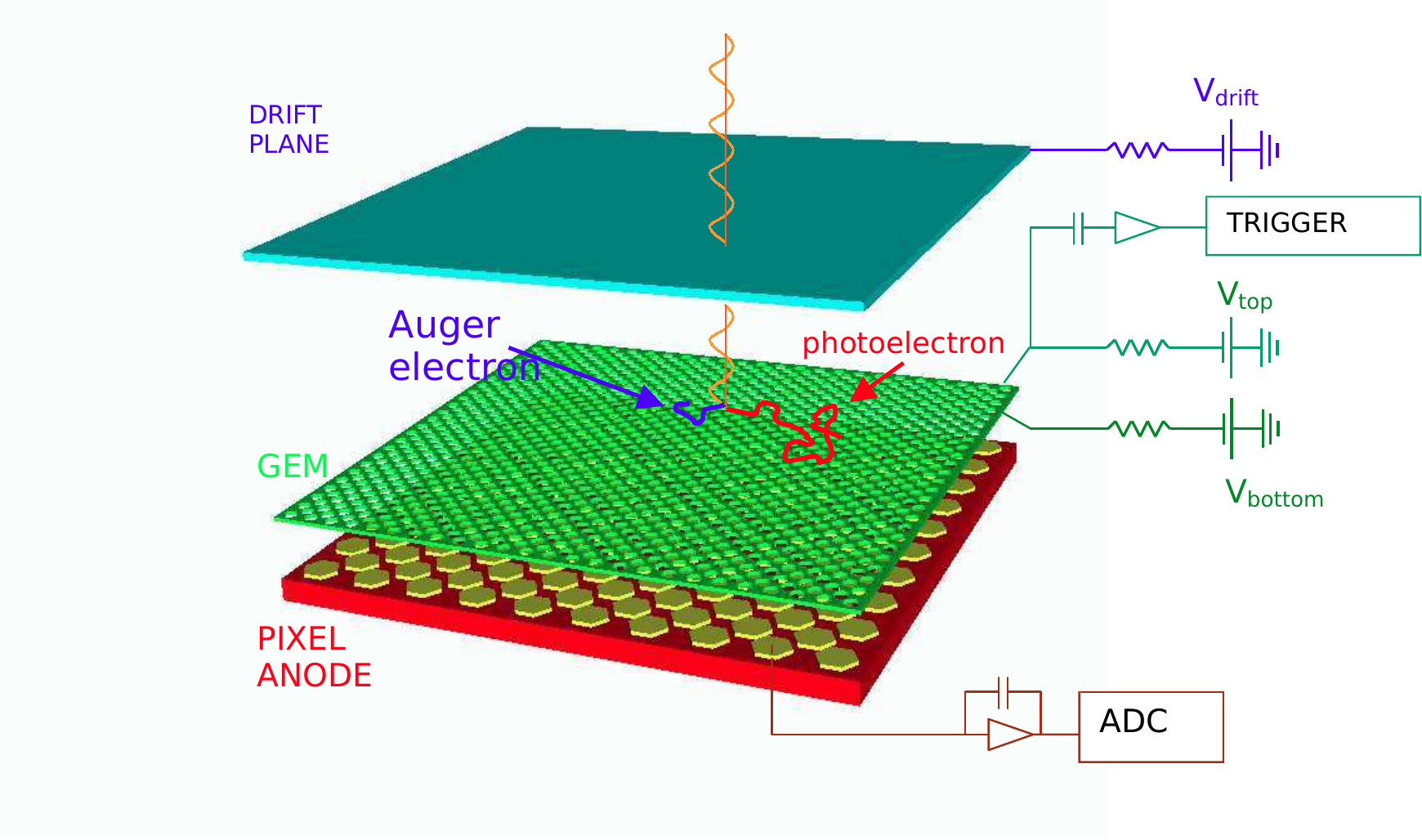}
\includegraphics[height=.12\textheight]{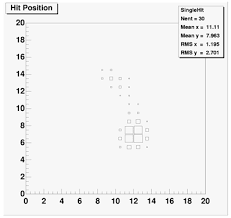}
\caption{Schematic of a GEM based Gas Pixel Detector (GPD). Figure credit $-$ \citet{costa01} (Reprinted by permission from Springer Nature Customer Service Centre GmbH).}
\label{gpd}
\end{figure}

\begin{figure}
\includegraphics[height=.13\textheight]{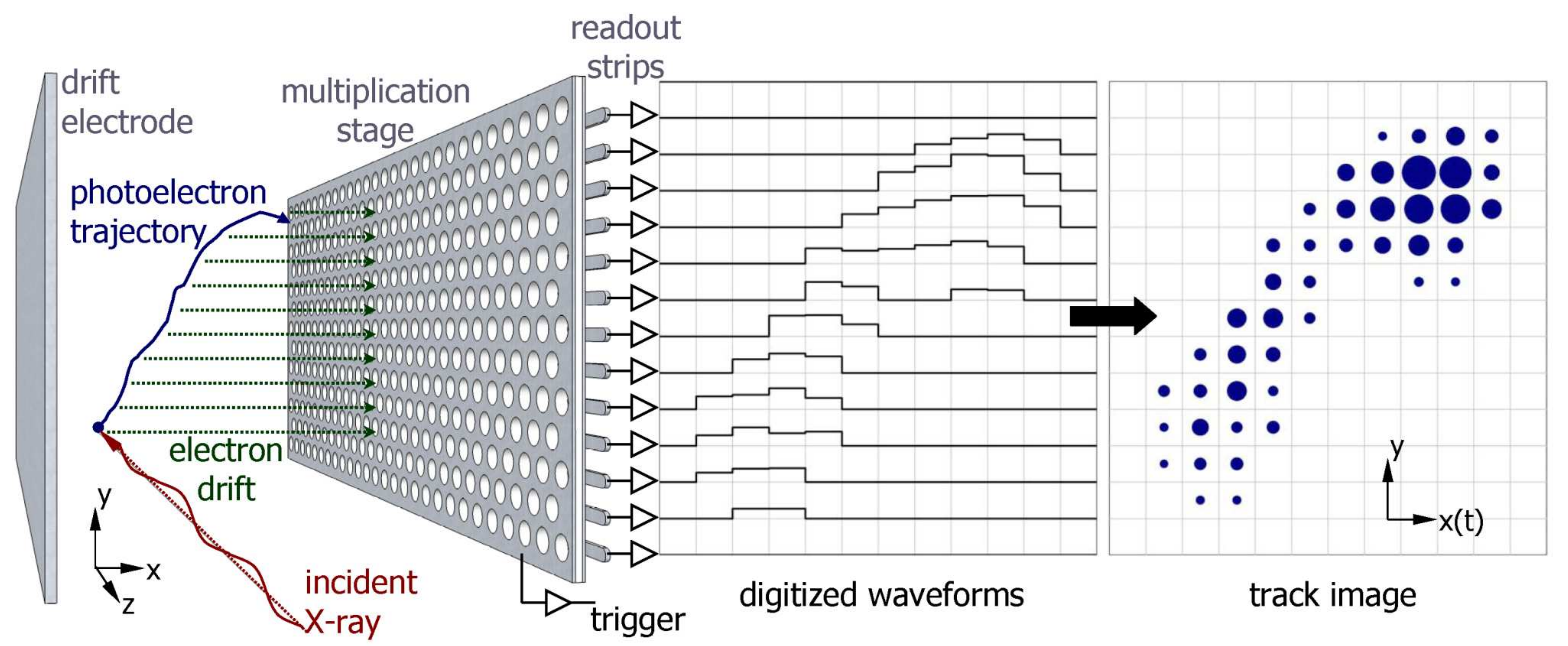}
\caption{Schematic of a GEM based Time Projection Chamber (TPC) polarimeter. Figure credit $-$ reprinted from \citet{black07} with permission from Elsevier.}
\label{tpc}
\end{figure}

\citet{tsunemi92} and \citet{michel08} discuss the method to image 
photo-electron track in pixelated semiconductor detectors. 
In semiconductor materials, the
photo-electron tracks are very small ($\sim$1 $\mu$m for 10 keV electron). 
Imaging these short photo-electron tracks requires pixel sizes much less than
the track length. With current solid state detectors having pixel sizes of
a few tens of $\mu$m, it is extremely difficult to image the photo-electron
tracks, making these detectors insensitive to polarization measurements. Fig. \ref{track_pixel} compares the photo-electron track lengths in solids and gases with typical pixel sizes available in today's state-of-the-art solid state detectors like Charge Coupled Devices \citep[CCDs,][]{lumb91_ccd,gruner02_ccd,Lesser15_ccd} where the typical pixel size is around 20 $\mu$m with substrate thickness of $\sim$100 $\mu$m or Timepix sensors \citep{llopart07} with typical pixel size of $\sim$55 $\mu$m and relatively thicker substrate of 300 $\mu$m. Photo-electron imaging in CCDs or Timepix sensors is therefore possible only at very high energies. 
Since in gases, photo-electron tracks are typically of 
the order of a few millimeter, Gas Electron Multiplier (GEM) based gas detectors 
\citep{sauli97} are expected to be more sensitive to imaging photo-electron
tracks, where the image is either formed 
by two dimensional readout anode pixels in Gas Pixel Detectors 
(GPD, \citet{costa01,bellazzini06,bellazzini07}, see Fig. \ref{gpd}) or with one dimensional 
readout strips in Time Projection Chambers (TPC, \citet{black07}, see Fig. \ref{tpc}), 
where the other dimension is obtained from the drift time of the
electrons. 

\subsubsection{Bragg reflection polarimetry}
\begin{figure}
\centering
\includegraphics[height=.24\textheight]{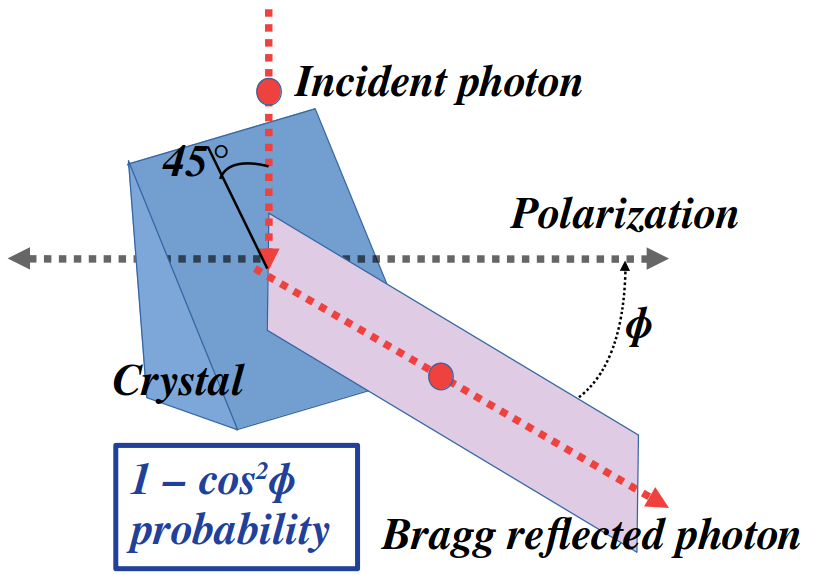}
\caption{Schematic of Bragg reflection geometry. A crystal preferentially reflects the incident photons perpendicular to the polarization direction giving rise to an asymmetry in the azimuthal angle ($\phi$) distribution which is proportional to 1-$\cos^2\phi$ for a Bragg reflection angle of 45$^\circ$.}
\label{bragg_schematic}
\end{figure}

\begin{figure}
\centering
\includegraphics[height=.25\textheight]{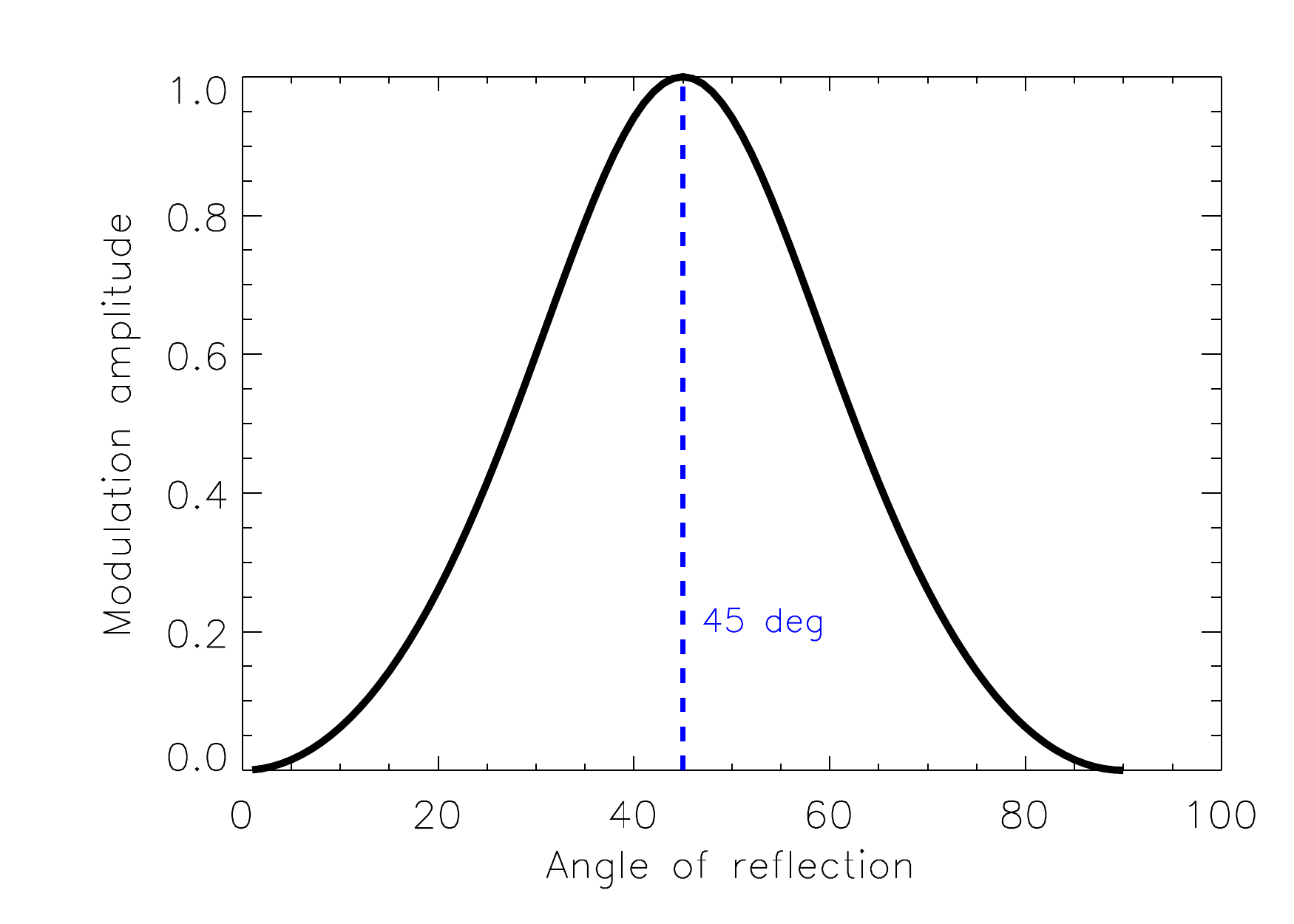}
\caption{Modulation factor in Bragg reflection as a function of angle of reflection. Modulation factor is close to unity for Bragg reflection angle of 45$^\circ$.}
\label{mod_bragg}
\end{figure}

Bragg crystal polarimeter \citep{toraskar75,novick75,silver90} 
utilizes the polarization dependence of Bragg reflection,
where the photons are preferentially reflected perpendicular to the 
polarization direction. The integrated reflectivity, $\Delta \theta(E)$, of a crystal is given by
\begin{equation}
  \Delta \theta(E) =  \frac{N^2F^2{r_0}^2\lambda^3}{2\mu(E)} \bigg[\frac{1}{\sin(2\theta_B)} - \frac{\sin(2\theta_B)}{2} \bigg(1 + \cos(2\phi)\bigg)\bigg],
  \label{eqbragg}
\end{equation}
where, N is the number of scattering cells per unit volume, F is the structure factor, $r_0$ is the classical electron radius, $\mu$ is the absorption coefficient, $\theta_B$ is the Bragg reflection angle, $\lambda$ is the wavelength of radiation, and $\phi$ is the azimuthal angle between the electric vector and the reflection plane of radiation (see Fig. \ref{bragg_schematic}). 
Since modulation in the azimuthal reflection is 
found to be maximum at reflection angle of 45$^\circ$,
a crystal kept at angle 45$^\circ$ to the incident X-ray radiation  
surrounded by a proportional counter in order to absorb the reflected 
X-rays, constitute a good polarization analyzer. Both the crystal 
and the detector are rotated around the incident flux direction to 
obtain count rates as a function of azimuthal angle. Such a system 
provides modulation factor close to unity irrespective of the incident beam energy (see Fig. \ref{mod_bragg}). 
However perfect atomic 
crystals reflect 
X-rays in a narrow energy band extending over a small 
fraction of an eV, resulting in a poor polarimetric 
efficiency, making it insensitive to X-ray polarimetry measurements.
Ideally imperfect crystals that are mosaic of small crystal domains
with random orientations provide higher effective widths (few eVs) and
therefore more suitable for Bragg polarimetry. 


Bent crystals can also be used
to focus the X-rays onto a small detector in order to minimize the background \citep{schnopper69}. The Bragg polarimeter on board {\em OSO-8}
used a parabolic
mosaic graphite reflector \citep{novick75} which obtained the most precise 
polarization measurement of Crab so far in soft X-rays \citep{weisskopf76,weisskopf78}. A multi-layer optic using the same Bragg principle can also be used as a polarimeter. Using a combination of a number of gratings and multi-layer planes, it is possible to develop broadband sensitive soft X-ray polarimeters in the sub-keV region \citep[$<$1 keV, see][]{marshall10,marshall18}.   

\subsection{\bf {\textit{Comparison of the techniques in hard X-rays}}}
The polarimetry techniques discussed above have their relative 
advantages and disadvantages. Bragg reflection, despite of achieving 
high modulation factor, work only at discrete 
energies or in a narrow sub-keV band which results in a low polarimetric sensitivity. 
Compton
scattering polarimeters have a moderate modulation factor and 
polarimetric efficiency but are unable to work at lower energies 
where X-ray flux from the sources is expected to be relatively higher. However, advantage of 
Compton polarimeters is that it can work in a broad energy range in hard X-rays. 
On the other hand, gas-based photoelectric polarimeters (GPD, TPC) possess 
high modulation factor and efficiency in soft X-rays but the efficiency drops fast at higher energies. With changes in the design and gas density or pressure, these
gas detectors, TPC in particular, are expected to work at higher energies. With the development in the detector technology, 
it might also be possible to image photo-electron tracks in hard X-rays in Silicon or CdTe based semiconductor detectors with very small spatial segmentation. 

\begin{figure*}
\centering
\includegraphics[height=.3\textheight]{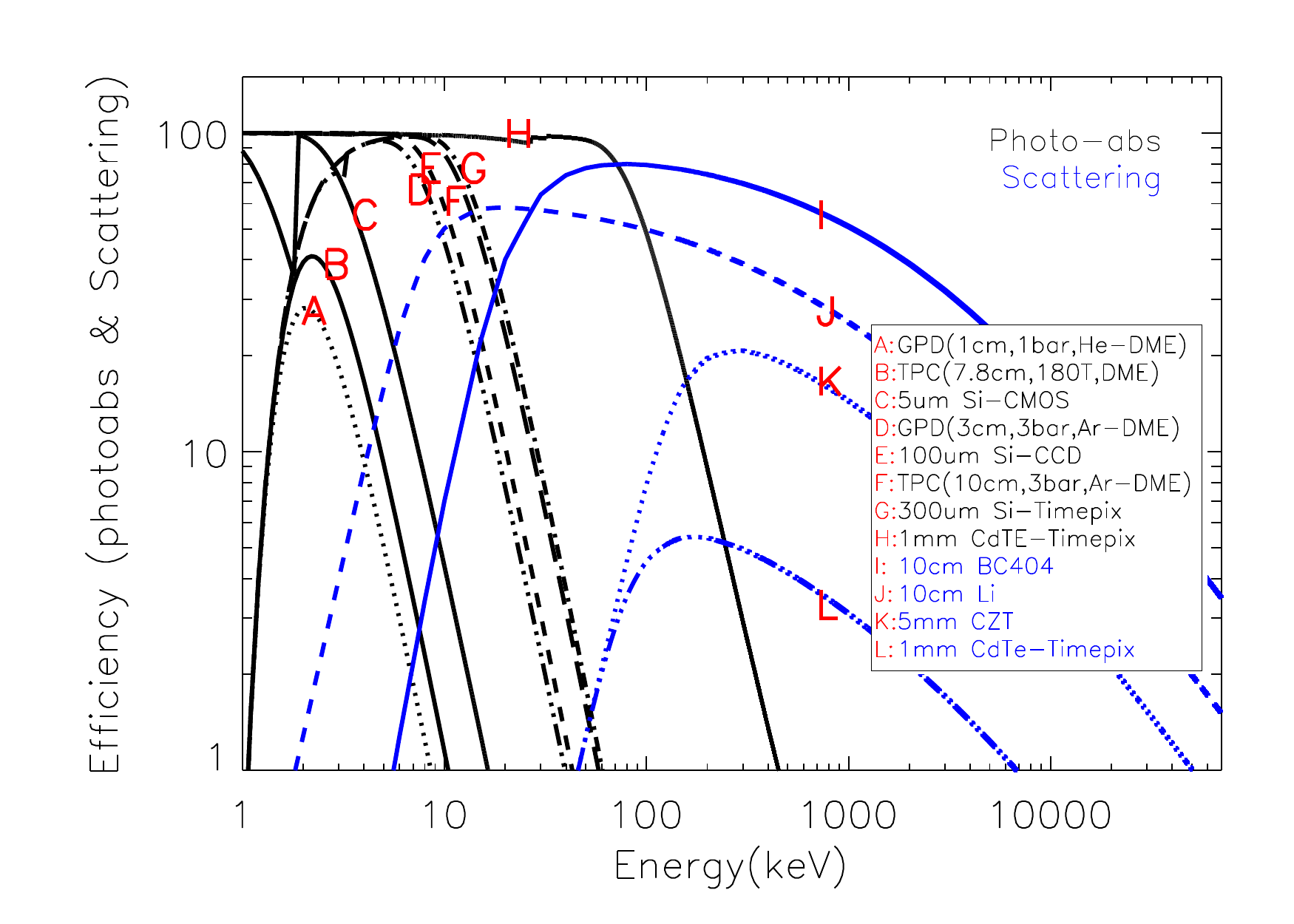}
\includegraphics[height=.3\textheight]{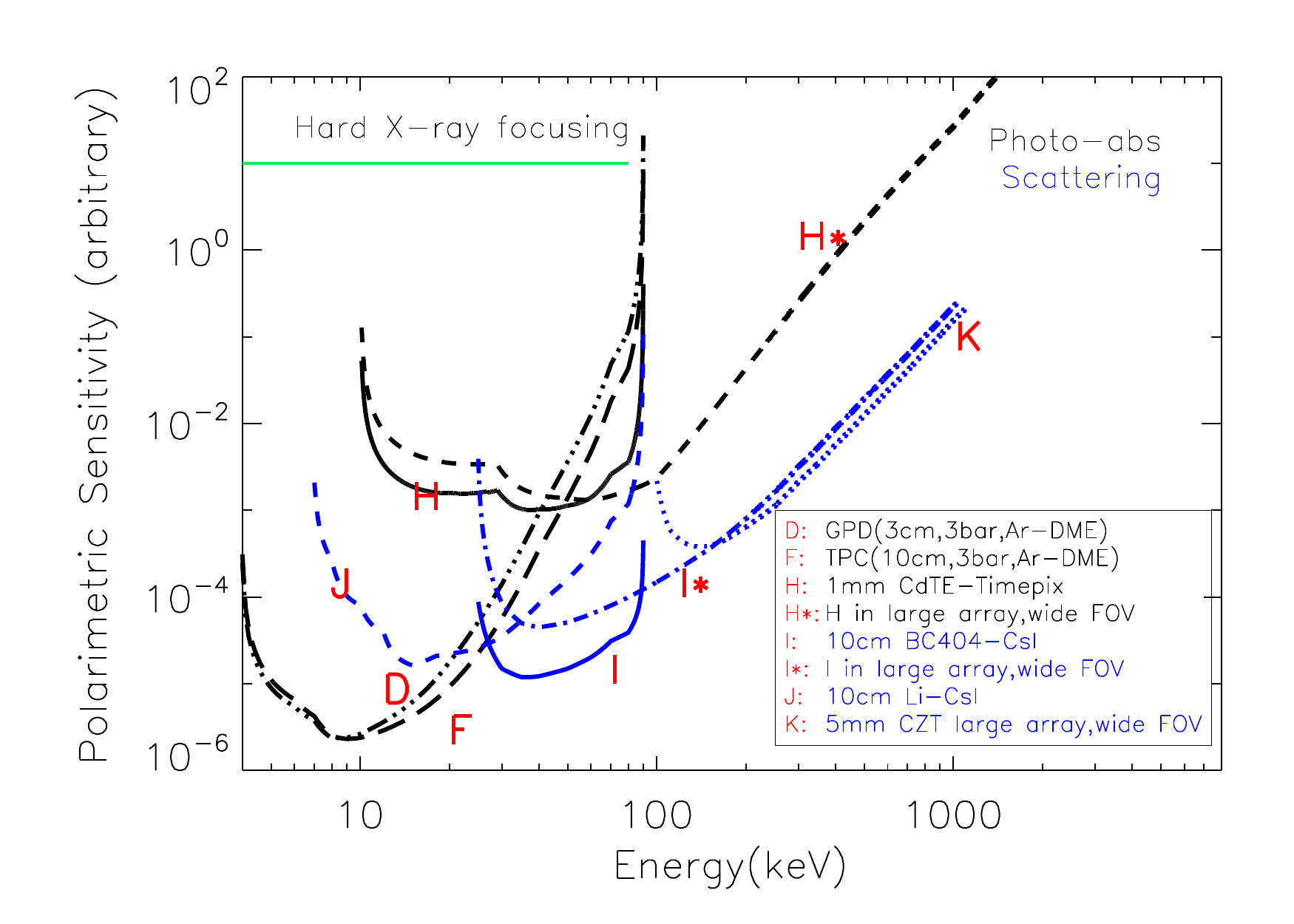}
\caption{top: A comparison of efficiencies in Photo-electric (black) and scattering method (blue) of polarimetry as a function of energy is made for various materials and thicknesses suitable for these techniques, bottom: A comparison in polarimetric sensitivity or MDP is made for various photo-electric and scattering polarimetry configurations as a function of energy. The MDP values are in arbitrary unit. See text for more details.}
\label{effy}
\end{figure*}

Top panel of Fig. \ref{effy} depicts a quantitative picture where a comparison between photo-electric and scattering efficiency has been made for materials suitable for these techniques.
For photoelectric efficiency (shown in black), we have considered two configurations $-$ 
\begin{enumerate}
    \item GEM based gas detectors in GPD and TPC configurations with standard and slightly optimistic gas depth, pressure and composition labelled as, 
    \begin{itemize}
        \item A $-$ GPD with 1 cm of He-DME at 1 bar
        \item B $-$ TPC with 7.8 cm of DME at 180 Torr
        \item D $-$ GPD with 3 cm of Ar-DME at 3 bar
        \item F $-$ TPC with 10 cm of Ar-DME at 3 bar.
    \end{itemize}
    \item State-of-the-art semiconductor detectors with available substrate material and thicknesses,
    \begin{itemize}
        \item E $-$ 100 $\mu$m Silicon CCD
        \item C $-$ 5 $\mu$m Silicon monolithic CMOS
        \item G $-$ 300 $\mu$m Silicon Timepix detector
        \item H $-$ 1 mm CdTe Timepix detector.
    \end{itemize}
\end{enumerate}
For scattering efficiency (shown in blue), two types of materials have been considered $-$ 
\begin{enumerate}
    \item low atomic number (Z) materials as scatterer of suitable depth and shape,
    \begin{itemize}
        \item I $-$ 10 cm thick BC404 bar
        \item J $-$ 10 cm thick lithium bar.
    \end{itemize}
    \item State-of-the-art high Z semiconductor detectors,
    \begin{itemize}
        \item K $-$ 5 mm thick CZT detectors
        \item L $-$ 1 mm thick CdTe Timepix detectors.
    \end{itemize}
\end{enumerate}
From the efficiency plots, it is evident that for energies above a few tens of keV, scattering technique is more suitable for polarimetry. 
The slightly optimistic configurations of GPD and TPC \citep[see `D' and `F',][]{fabiani18} have sensitivity ranging up to a few tens of keV. 
Photo-electron track imaging in pixelated semiconductors (`E', `G', `H') can also extend the measurements to a few tens of keV, in particular, 1 mm thick CdTe Timepix (see `H') possesses high detection efficiency in a broad energy range from a few keV to a few hundreds of keV.

In the bottom panel of Fig. \ref{effy}, we compare the polarimetric sensitivity of some of the photoelectric and scattering polarimetric configurations which show significant detection efficiency in hard X-rays $-$ D, F, H, I, J, K. Polarimetric sensitivity for instruments based on these materials will depend on the collecting area, background rates and multiple other factors. Therefore, the following assumptions were made on the collecting area and backgrounds while estimating and comparing the sensitivities among the various polarimetry configurations. 
\begin{itemize}
    \item Because of the availability of hard X-ray focusing optics in 3 $-$ 80 keV range (shown by a green line in the bottom panel of Fig. \ref{effy}), for materials sensitive up to a few tens of keV, we considered small area focal plane configurations. 
    
    For example, for `D' and `F', a compact 16 cm$^2$ geometric area of GPD and TPC is considered at the focal plane of a {\em NuSTAR} or {\em Hitomi} type of hard X-ray telescope.  
    \item For `H', `I', `J', because of the high photon detection efficiency in a broad energy range from a few tens to hundreds of keV, we considered both hard X-ray focal plane and a large area wide FOV configuration. 
    
    For example, for `H', a 16 cm$^2$ geometric area and 1 mm thick CdTe Timepix detector was assumed with state-of-the-art pixel size of 55 $\mu$m \citep{llopart07} at the focal plane of a hard X-ray telescope. For the non-focal configuration (labelled as `H*'), we assumed a large array of CdTe Timepix detectors with total collecting area of 900 cm$^2$ and FOV of 60$^\circ \times$ 60$^\circ$. 
    
    For `I' (or `J'), let us consider a 10 cm long active BC404 plastic scintillator (or a 10 cm long passive Lithium) surrounded by an array of CsI scintillators at the focal plane of a hard X-ray telescope. Clearly, while `I' because of the active scattering element, works as a Compton polarimeter, `J' with passive Lithium scatterer, works as a Rayleigh polarimeter. For the non-focal configuration (labelled as `I*'), a 900 cm$^2$ of geometric area is considered from an array of 10 cm long and 10 mm cross BC404 scintillators (in 6 $\times$ 6 array surrounded by a single array of CsI scintillators in 25 identical modules) with a FOV of 60$^\circ \times $60$^\circ$. 
    \item For materials sensitive beyond 100 keV which is inaccessible by focusing optics, only possible polarimetric configuration is a large area wide FOV detector array.   
    
    For example, for `K' or 5 mm thick CZT or any other high Z materials like Germanium,  we consider a 900 cm$^2$ CZT detector array with pixel pitch of 2.5 mm  \citep[same as CZTI instrument on board {\em AstroSat},][]{bhalerao16} with 60$^\circ$ $\times$ 60$^\circ$ FOV. A lower substrate thickness of CZT or CdTe like 1 mm CdTe in Timepix configuration would result in a relatively lower sensitivity due to lower scattering efficiency \citep{hahn16} and therefore has not been considered for sensitivity calculations. 
\end{itemize}

For sensitivity calculations (MDP, see Eq. \ref{clq4}), values of modulation factor are either taken from literature or calculated analytically in some cases \citep[for details of the analytical model, see][]{chattopadhyay13}. In case of analytical modeling, we ignored effects of K$_\alpha$ fluorescence photons, charge sharing in semiconductor detectors, charge diffusion in gas detectors and other complicated phenomena affecting the azimuthal modulation. However, the values quoted here should not be widely apart from the true values and suitable for the purpose of comparison among different techniques. We considered a Crab like spectrum for source flux calculations. For calculation of  background, two different contributions were considered $-$ 1. Cosmic X-ray background (CXB) within the FOV (CXB multiplied by either photoelectric efficiency of the photoelectric polarimeters or by scattering efficiency in case of scattering based polarimeters). The spectrum for hard X-ray cosmic background used in the calculations was taken from \citet{turler10} computed based on {\em INTEGRAL} observations, 2. A conservative number of 10 counts/cm$^2$ particle induced X-ray background was considered to estimate the random chance events between the scatterer and absorbers for Compton polarimeters assuming a coincidence time window of 10 $\mu$s. The values of MDP shown in the figure are in arbitrary units, meant to compare the different techniques as a function of energy. 

From the bottom panel of Fig. \ref{effy}, we see that CdTe Timepix in both focal and non-focal configurations yields lower sensitivity because of the large pixel size compared to the electron track length resulting in low modulation factors. In this regard, there have been a several attempts to make small pixel size CCDs \citep{tsunemi92,michel08,schmidt98,hill97} and CMOS \citep[2.5 $\mu$m pixel, see][]{asakura19} detectors. But the polarization analyzing power of these detectors is still on the lower side particularly at the lower energies. A thicker silicon or 5 micron CdTe instead of a 5 micron silicon on CMOS readout integrated chip (ROIC) can see a significantly higher polarization sensitivity in a broad energy range due to the enhanced photoelectric efficiency. Similarly, a much smaller pixel size in 1 mm CdTe-Timepix detectors will see a significant improvement in the overall sensitivity. 
This is particularly interesting because such a detector can provide sensitive timing, imaging, spectroscopic and polarization information simultaneously as a focal plane detector, while in the large area wide FOV mode, the instrument will provide spectroscopic, timing and polarimetry sensitivity in a broad energy range from a few keV to a few hundreds of keV.
However, there is no report on small pixel size CdTe Timepix detectors in literature. Also, one should keep in mind the effects of charge sharing on polarization for such smaller pixel size and thick CdTe detectors. 

Though at energies beyond 10 keV, scattering polarimetry outweighs photoelectric polarimeters, the optimistic configurations of GPD and TPC \citep[see `D' and `F',][]{fabiani18} can result in higher sensitivity up to $\sim$25 keV. However, in absence of any experimental demonstration of such thicker gas detectors, Lithium scatterer based Rayleigh polarimeters at the focal plane of a hard X-ray telescope, have the best sensitivity in 10 $-$ 25 keV range. Due to the enhanced background level in Rayleigh polarimeters, the sensitivity goes down significantly at higher energies. At those energies (25 $-$ 80 keV), a compact Compton polarimeter can provide the best polarimetric sensitivity because of the concentration of flux and lower background. 

Beyond 100 keV, an array of CZT detectors or a 2D array of low Z plastics and high Z CsI absorbers are expected to provide similar level of sensitivity. 
Though plastics have significantly higher scattering efficiency at these energies compared to CZTs and similar modulation factor values, the relatively higher detection volume in case of plastic-CsI array (10 cm depth compared to 5 mm depth in CZTs) introduces large chance background events. 
Plastic scatterer array has the advantage of lower energy threshold compared to the CZTs which make them sensitive below 100 keV as well. One should also keep in mind the relative complexities in realization of a large area detector plane in case of scintillators and CZTs, requirement of power in these configurations, readout electronics, availability of active coincidence and other critical factors to develop the payload. 

Consistent with this discussion, we have seen a lot of efforts in the last few years to develop Rayleigh polarimeters in 10 $-$ 30 keV range \citep{paul16} and Compton polarimeters at higher energies both as focal plane detector \citep{chattopadhyay13, chattopadhyay14,chattopadhyay15,krawczynski11} and large area of scintillators or CZTs or Germanium detectors \citep{mcconnell09,chattopadhyay14_2,vadawale15,orsi11,caroli12,caroli18,yang18}.
In the next section, we will review instrumentation of different scattering polarimetry configurations suitable in the three energy bands: 10 $-$ 25 keV, 25 $-$ 80 keV and beyond 80 keV.

\section{Scattering polarimetry configurations: Instrumentation in hard X-ray polarimetry}
\label{pol_instrument}
As we discuss the best possible scattering polarimetry configurations in detail in three hard X-ray energy bands, we will touch upon the actual instrument designs of the past, present and upcoming hard X-ray polarimetry experiments.
For details of the overall payload designs, readers are requested to refer to the references given here and Table \ref{ins_Table} where all the past, present and planned scattering polarimetry experiments have been summarized. 

\begin{table*}[ht!]
\begin{tiny}
\begin{threeparttable}
	\caption{Details of various hard X-ray polarimetry instruments (scattering polarimeters). See a review by \citet{fabiani18}}\label{ins_Table} 
\begin{tabular}{p{2cm}p{2.1cm}p{1.2cm}cp{4.2cm}p{1.4cm}p{2.2cm}}
\topline
\\
Instrument name & Configuration & Mission type & year & Material and geometry& Energy& Science\\
\\
\midline

{\bf Past}\\
SPR-N \citep{bogomolov03}&50 cm$^2$&Satellite&2001 $-$ 2005&Hexahedral prism of 5 cm thick Be with 3 pairs of surrounding scintillators (2 cm$\times$4 cm$\times$.3 cm)&20 $-$ 100 keV&Solar flares\\
PHENEX \citep{kishimoto07,gunji08}&$\sim$44 cm$^2$ (88 cm$^2$ in 2009 flight), collimated, FOV $\sim$4.8$^\circ$ FWHM&Balloon-borne&2006, 09&4 units (8 units in 2009 flight) of array of 36 plastic scintillators (5.5, 5.5, 40 mm) surrounded by 28 CsI(Tl) scintillators (5.5, 5.5, 40 mm)&40 $-$ 200 keV&Crab\\
PENGUIN-M \citep{dergachev09}&$\sim$78 cm$^2$&Satellite&2009 $-$ 10&4 P-terphenyl scintillators (5 cm dia, 3 cm thick) surrounded by 6 NaI(Tl) scintillators (7 cm dia, 5 mm thick) &20 $-$ 150 keV&Solar flares\\
GAP \citep{yonetoku06}&176 cm$^2$ area, wide FOV&{\em IKAROS} spececraft&2010 $-$ 11&dodecagon
plastic scintillator (140 mm gap between two opposite faces) surrounded by 12 CsI scintillators (6 cm long, 5 mm width)&50 $-$ 300 keV&GRBs\\
GRAPE \citep{bloser09,mcconnell09}&$\sim$144 cm$^2$ (216 cm$^2$ in 2014 flight), wide FOV&Balloon-borne&2011, 14&16 units (24 units in 2014 flight) of 6$\times$6 plastics (5, 5, 50 mm) surrounded by 28 CsI(Tl) scintillators (5, 5, 50 mm)&50 $-$ 500 keV&GRBs\\
PoGO+ \citep{chauvin16_pogo,friis18}&Large area ($\sim$400 cm$^2$), collimated, narrow FOV ($\sim$2$^\circ$)&Balloon-borne&2016&Array of 12 cm long and 3 cm cross 61 hexagonal plastic scintillators (EJ-204)&20 $-$ 180 keV&Crab, Cygnus X-1\\
SGD \citep{hitomi18}&large area (210 cm$^2$, narrow FOV Compton camera)&Satellite&2016&32 layers of Si ($\sim$25, 25, 0.6 mm, 3.2 mm pixel) and 8 layers of CdTe sensors ($\sim$25, 25, 0.75 mm, 3.2 mm pixel) surrounded by 2 layers of CdTe from all sides &50 $-$ 200 keV&Pulsars, BH XRBs, AGNs, supernova remnanats\\
COSI \citep{yang18}&Large area (256 cm$^2$), wide FOV Compton camera&Balloon-borne&2016&12 cross-strip Ge detectors (8, 8, 1.5 cm) in 2$\times$2$\times$3 array, 2 mm of strip pitch &0.2 $-$ 2 MeV&GRBs, galactic sources\\
POLAR \citep{sun16,produit18}&large area ($\sim$500 cm$^2$), wide FOV&Space lab in LE orbit&2016 $-$ 17&25 identical units of 64 plastic scintillator array (EJ-248M, 176 mm long, 5.8$\times$5.8 mm$^2$ cross-section)&50 $-$ 500 keV&GRBs\\
X-Calibur \citep{guo11_2,guo13,beilicke14,kislat19}&Focal plane&Balloon-borne&2016, 19&Plastic scintillator (13 mm dia, 14 cm long) with array of CZTs (2,2,0.2 cm)&20 $-$ 80 keV&BH XRBs, NS, AGNs\\
\hline
{\bf Active}\\
{\em INTEGRAL}-IBIS \citep{forot08}&large area (2600 cm$^2$ (CdTe array) \& 3100 cm$^2$ (CsI array)), 9$^\circ\times9^\circ$ FOV Compton camera&Satellite&2002-&Compton camera $-$ 1$^{st}$ layer of 128$\times$128 CdTe array (4, 4, 2 mm, 9.2 mm pixel pitch) and 2$^{nd}$ layer of 64$\times$64 CsI scintillator array (8.4, 8.4, 300 mm); gap of 94 mm between the two layers&250 $-$ 2000 keV&Crab, Cygnus X-1, transients\\
{\em INTEGRAL}-SPI \citep{chauvin13}&large area (500 cm$^2$), 16$^\circ$ FOV &Satellite&2002-&Array of 19 Germanium detectors (7 cm long, hexagonal shape)  &130 $-$ 1000 keV&Crab, Cygnus X-1, transients\\
{\em AstroSat}-CZTI \citep{chattopadhyay14,vadawale15}&Large area ($\sim$924 cm$^2$), wide FOV (above 100 keV)&Satellite&2015-&Array of 64 CZTs (4, 4, 0.5 cm, 256 pixels)&100 $-$ 350 keV&Crab, Cygnus X-1, transients\\
\hline
{\bf Planned}\\
{\em POLIX} \citep{paul10,paul16}&large area (640 cm$^2$), collimated (3$^\circ \times$3$^\circ$ FOV)&Satellite&2021&Li/Be scatterer with 4 position sensitive proportional counters (4$\times$1080 cm$^2$)&5 $-$ 30 keV&Pulsars, BH XRBs, AGNs, supernova remnanats\\
{\em Daksha$^1$}&Large area, wide FOV&Satellite&2023&Three separate detector planes of SDD, CZT and NaI detectors, polarimetry from CZT array&$\sim$100 $-$ 400  keV&galactic sources, transients\\
{\em POLAR-2} \citep{kole19}&large area ($\sim$2000 cm$^2$), wide FOV&Space lab in LE orbit&2024&100 units of 64 plastic scintillator array (EJ-248M, 178 mm long, 6$\times$6 mm$^2$ cross-section)&0.03 $-$ 1 MeV&GRBs\\
PING-P \citep{kotov16}&$\sim$30 cm$^2$ area&Satellite&2025&3 P-terphenyl detectors (3.6 mm dia, 3 cm thick) surrounded by 6 CsI scintillators (4.5 cm dia, 5 cm thick)&20 $-$ 150 keV&Solar flares\\
{\em PolariS} \citep{hayashida14}&Focal plane&Satellite&&8$\times$8 array of plastic scintillators (2.1, 2.1, 40 mm) surrounded by 4$\times$4 array of GSO scintillators (4.3, 4.3, 6 cm)&10 $-$ 80 keV&BH XRBs, NS, SNRs\\
SpHiNX \citep{pearce19}&large area (800 cm$^2$), wide FOV&Satellite&&6 hexagonically packed plastic (EJ-204) surrounded by 6 cm long, 24 GAGG scintillators (7 units, total 120 GAGGs and 42 plastics)&50 $-$ 600 keV&GRBs\\
COSI-2 \citep{tomsick19}&Large area (256 cm$^2$), wide FOV Compton camera&Satellite&&Same as COSI with an additional layer of Ge detector array&0.2 $-$ 2 MeV&GRBs, galactic sources\\
LEAP \citep{mcconnel16_leap}&large area (5000 cm$^2$), wide FOV&ISS&&Similar to GRAPE but with segmented scintillators with individual SiPM readout&50 $-$ 500 keV&GRBs\\
AMEGO \citep{amego19}&Large area, wide FOV&Satellite&&60 layers of 4 $\times$ 4 array DSSDs (95, 95, 0.5 mm, 500 $\mu$m strip pitch), an array 3D CZTs (8, 8, 40 mm) and  6 layers of CsI(Tl) crystal bars (15, 15, 380 mm)&200 keV $-$ 1 GeV&Pulsars, XRBs, AGNs, PWNs, GRBs\\
e-ASTROGAM \citep{de18}&Large area, wide FOV&Satellite&&layers of 5 $\times$ 5 array DSSDs (95, 95, 0.5 mm, 240 $\mu$m pitch) and an array of small scintillation crystals (33856 CsI(Tl) bars of 5, 5, 80 mm) readout by SDDs&300 keV $-$ 3 GeV&Pulsars, XRBs, AGNs, PWNs, GRBs\\
\hline
\end{tabular}
	\begin{tablenotes}
\item[1] \url{https://www.star-iitb.in/research/daksha}
\end{tablenotes}
\end{threeparttable}
\end{tiny}
\end{table*}
\subsection{10 $-$ 25 keV: Narrow FOV Rayleigh scattering polarimeter}
As discussed in Fig. \ref{effy}, a Rayleigh scattering based polarimeter is best suitable for polarization measurements in 10 $-$ 25 keV.
Since Rayleigh polarimeters do not work in active coincidence between the scatterer and surrounding absorbers, total background is dominated by the cosmic X-rays on the surrounding absorbers. 
A small contribution comes from the scattered cosmic X-rays within the FOV of the instrument onto the surrounding detectors. It is important to use proper active or passive shielding around the absorbers to keep the background level at its minimum.  
The overall sensitivity depends on the modulation factor and scattering efficiency of the scatterer. Particularly, the modulation factor critically depends on the shape of the scatterer. \citet{vadawale10} has explored various scatterer geometry (disk, hollow cone, hollow pyramid, slab) and scatterer material (Li, LiH, Be) and compared their sensitivities. There are multiple options for detectors and their configuration for absorbers surrounding the scatterer $-$ an array of CZT detectors in rectangular configuration on four sides of the scatterer, four rectangular shaped position sensitive multi-wire proportional counters, cylindrical array of high Z scintillators (e.g. CsI(Tl), NaI etc.). High Z scintillators and the CZT detectors have high detection efficiency to high energies making the polarimeters sensitive up to $\sim$100 keV, whereas the proportional counters are sensitive up to a few tens of keV. Proportional counters are however more preferred in the 10 $-$ 25 keV range because of their lower energy threshold compared to scintillators and CZTs. 
 
\begin{figure}[h]
\centering
\includegraphics[scale=.9]{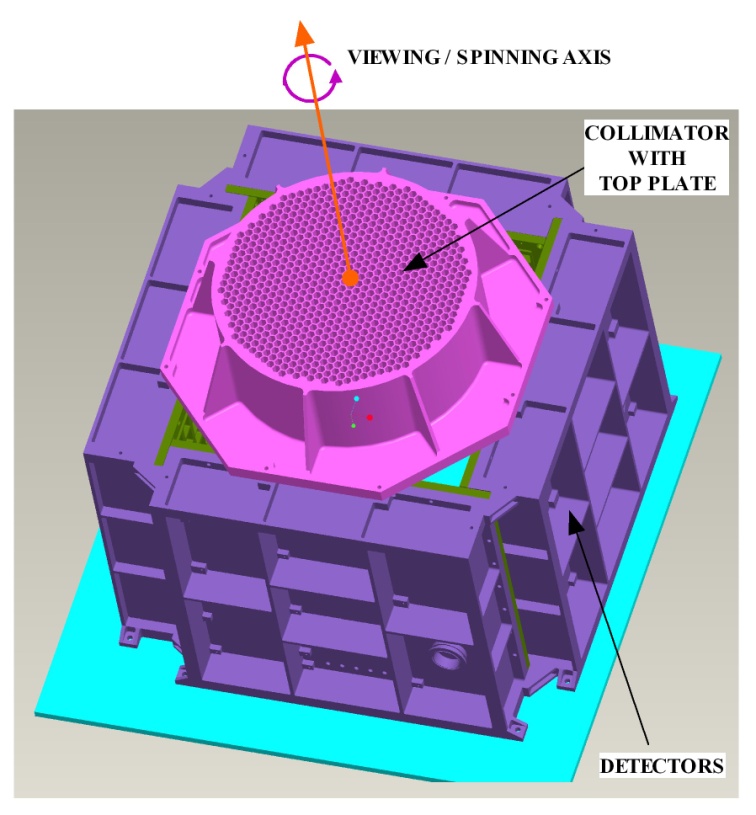}
\caption
{{\em POLIX}: a Rayleigh scattering based polarimeter sensitive in 5 $-$ 30 keV, scheduled for launch in 2021. Figure credit $-$ \citet{paul10}.}
\label{polix}
\end{figure}

{\em POLIX} \citep{paul10,paul16} is an example of Rayleigh scattering polarimeter sensitive in 5 $-$ 30 keV (see Fig. \ref{polix}). 
The instrument has been selected for a small satellite mission for Indian Space Research Organization (ISRO), scheduled for launch in 2021 in a low Earth and low inclination equatorial orbit. The polarimeter consists of a Be/Li disk shaped scatterer of 640 cm$^2$ collecting area surrounded by 
proportional counters (4$\times$1080 cm$^2$), placed on four sides of the scatterer. A collimator is used to 
restrict the FOV of the instrument to $3^\circ \times 3^\circ$ with a flat top response to overcome the satellite pointing inaccuracies. To reduce systematic, the payload spins around the roll axis by 0.5 $-$ 5 rpm. The instrument is expected to provide an MDP level of 2 $-$ 3 \% for a 50 mCrab source in 10$^6$ s of exposure.

SPR-N \citep{bogomolov03} on board solar mission {\em CORONAS-F} exploits Rayleigh scattering on a 5 cm thick and 50 cm$^2$ collecting area Beryllium packed inside a hexahedral prism
with 3 pairs of scintillation detectors surrounding the scatterer. Because of the use of scintillators, SPR-N polarimeter is sensitive in 20 $-$ 100 keV range. As discussed in the previous sections, the instrument has been successful in providing polarization measurements for 25 solar flares \citep{zhitnik06}.

\subsection{25 $-$ 80 keV: Focal plane Compton polarimeter}
\begin{figure}[h]
\centering
\includegraphics[scale=.6]{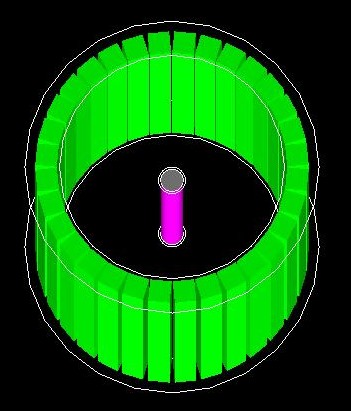}
\caption
{An example of focal plane Compton polarimeter used in CXPOL experiment \citep{chattopadhyay13,chattopadhyay15}. It uses a 
a long narrow plastic scatterer surrounded by a cylindrical array of long CsI scintillators.}
\label{cxpol_geom}
\end{figure}

Above $\sim$25 keV, Compton polarimeters at the focal plane of hard X-ray optics can be the most sensitive polarimetry instruments. Focal plane configuration ensures ideal scattering geometry (high modulation factor) which along with concentration of flux in hard X-rays and low background due to the small detector area and active co-incidence between the scatterer and absorbers makes a multi-fold enhancement in the sensitivity. 
In recent years, multiple groups across the globe have started investigating possible implementation of 
Compton scattering based X-ray polarimeter 
coupled with {\em NuSTAR} type of hard X-ray optics, e.g. X-Calibur \citep{guo13,beilicke14}, CXPOL \citep{chattopadhyay13,chattopadhyay14_2,chattopadhyay15}, {\em PolariS} \citep{hayashida14}.  X-Calibur had two balloon flights in 2016 and 2019 with InFOC$\mu$S (International Focusing Optics Collaboration for $\mu-$Crab Sensitivity) focusing optics. {\em PolariS} uses a modified {\em Hitomi} focusing telescope. Intial launch date of {\em PolariS} under JAXA was set in 2020. CXPOL is a focal plane Compton polarimetry instrument development program in India which uses a {\em NuSTAR} type of optic (under development) and is under active consideration for future X-ray astronomy mission under ISRO (private communication).   
All these experiments use a narrow long scatterer (X-Calibur$-$13 mm diameter and 14 cm long plastic, CXPOL$-$ 5 mm diameter and 10 cm long long plastic, {\em PolariS}$-$ 2.1$\times$2.1$\times$40 mm$^3$ plastic in 8$\times$8 array) surrounded
by either a cylindrical or rectangular array of detectors (see Fig. \ref{cxpol_geom} as an example of focal plane Compton polarimeter). 
The long and narrow cylindrical scatterer geometry yields high scattering efficiency and low probability of multiple interactions in the scattering volume.  

This type of instruments can achieve $\sim$1 $\%$ MDP level in 1 Ms exposure for 100 mCrab sources assuming {\em NuSTAR} type of focusing area. Polarimetric sensitivity is supposed to be critically dependent on the lower energy threshold of the plastic scatterer which will depend on many factors $-$ light output of the plastic, light collection efficiency, position of interaction, readout method and the readout noise. 
There are variety of different plastic scintillator manufacturer like ELJEN technology \footnote{\url{https://eljentechnology.com/products/plastic-scintillators}}, SAINT-GOBAIN \footnote{\url{https://www.crystals.saint-gobain.com/products/bc-408-bc-412-bc-416}}, CRYOS-BETA \footnote{\url{http://www.cryos-beta.kharkov.ua/organic.php}} which provide sensitive plastic scintillators. Scintillation properties of most of these scintillators (ELJEN and BICRON) are similar with around 1.36$\times$10$^4$ photons/MeV light output, 0.7 $-$ 1 ns rise time and 1.8 $-$ 3 ns of decay time and spectral emission peaking in 410 $-$ 430 nm range. CRYOS-BETA manufactured p-terphenyl provides higher light output 2.74$\times$10$^4$ and slightly higher density, however longer decay time constant of 3.7 ns. 
Proper reflection wrapping material to contain the scintillation photons inside the detector volume is another critical parameter to improve the lower energy threshold. 
All these properties should be carefully inspected along with the photo-multiplier tube characteristics (e.g. quantum efficiency spectrum) to design a sensitive Compton polarimeter.   
Because of all these statistical processes involved in the detection of X-rays in scintillators, plastics are expected to have a decreasing detection probability at lower energies, proper understanding of which is extremely important. 
\citet{fabiani13} and \citet{chattopadhyay14_2} described a method based on Compton scattering to estimate the behavior of plastics at lower energies where 
Compton scattered X-rays from the plastic rod is collected by an independent detector at different known polar scattering angles and known incident energies. Since the plastic and the surrounding detector are in active coincidence, at lower scattering angles the deposited energy being low, below the detection threshold, there will be no signal from the other detector. \citet{chattopadhyay14_2}, for a 5 mm diameter and 10 cm long BC404 plastic coupled to a photo-multiplier tube (PMT $-$ Hamamatsu R6095), found
$\sim$6 $\%$ detection probability at energies down to $\sim$0.5
keV, with a linear increase in the probability up to 3 keV. Detection probability reaches maximum (100 $\%$) at around 7 keV.
These parameters set the energy threshold of the polarimeter at around 20 keV. Better sensitivity requires better detection probability in the plastic which can be obtained by further optimization in the use of wrapping material (e.g. TETRATEX, teflon, VM2000, BC-620 paint) or better scatterers \citep{fabiani13}.

There are several options in the choice of detectors for the surrounding absorber array. X-Calibur uses an array of 2$\times$2$\times$0.2 cm$^3$ CZT detectors (64 pixels) on four sides of the plastic scatterer. CXPOL \citep{chattopadhyay15} and {\em PolariS} \citep{hayashida14}, on the other hand, use long CsI(Tl) and GSO scintillator bars respectively to detect the scattered X-rays. Use of CZT detector array provides 2D positional information of the scattered photons which can be used to extract the polar scattering angle (if plastic is vertically position sensitive) and therefore can be used to further filter out the chance background events. CZT detectors also provide better energy resolution which enable sensitive spectroscopic measurements from the Compton events ($E=E_{plastic}+E_{CZT}$). Advantage of scintillators is simpler electronics and better uniformity. They can be used in cylindrical array which does not suffer any intrinsic azimuthal asymmetry as in case of rectangular configuration and therefore can be operated without the need of payload spin. Particularly, with the development of MAPMTs \footnote{\url{https://www.hamamatsu.com/us/en/product/optical-sensors/pmt/index.html}} and Si photomultipliers (SiPM) \footnote{\url{https://www.hamamatsu.com/us/en/product/optical-sensors/mppc/index.html}}, because of their small and compact size, low weight, scintillators can have the compactness needed to develop focal plane instruments. {\em PolariS} uses MAPMT to readout the GSO scintillator array from the flat bottom side. MAPMTs still employs high HV bias ($\sim$1000 V) between the anode and cathode and crosstalk between the anode pixels reduces the sensitivity of the detectors. SiPM units \citep{buzhan02,buzhan03,otte06}, on the other hand, operates with low voltage ($\sim$31 V, above breakdown voltage) and do not suffer from crosstalk. The extremely small size ($\sim$3 mm$\times$3 mm), light weight, robust nature provides required compactness. The wide spectral range matches with the emission spectra of most scintillators. It is also possible to read a scintillator bar from both sides which will provide vertical position sensitivity as in case of CZTs. 

Though SiPMs have been mostly used at higher energies \citep{bloser13,sanaei15}, lower energy (20 $-$ 80 keV) application of SiPMs was demonstrated only recently during CXPOL experiment \citep{chattopadhyay15} to readout CsI(Tl) scintillators with SiPM PM3350 from KETEK \footnote{\url{https://www.ketek.net/sipm/}}. One major issue of SiPMs is the high background level ($\sim$500 kHz/mm$^2$), therefore use of SiPMs at lower energies requires very good light collection at the SiPM which will depend on the scintillator light output and decay time constant. A smaller decay time constant ensures simultaneous arrival of the optical photons to the SiPM which will yield higher signal to noise ratio. Therefore, NaI(Tl), LaBr$_3$(Ce), CeBr$_3$ with small decay constants make them good candidates for SiPM readout \citep{goyal18} (see Table \ref{Table_scintillator}). However, some of these detectors suffer from intrinsic background in the MeV region \citep{iyudin09,Cebrian12} which can be a disadvantage for the polarimeter.
\begin{table*}[htb]
\begin{tiny}
\begin{threeparttable}
	\caption{Possible inorganic scintillators as absorber array and their properties}\label{Table_scintillator} 
\begin{tabular}{p{1.8cm}p{2.1cm}p{1.4cm}cp{1.4cm}p{1.4cm}p{1.2cm}}
\topline
\\
Scintillator name & Light Yield (/keV) & Decay time constant (ns) & Density (gm/cc) & Hygroscopic& Maximum emission (nm) & Hardness (Mho)\\
\midline
\\
GSO(Ce)$^1$ &8 $-$ 10&30 $-$ 60&6.7&No&430&5.7\\
CeBr$_3^2$ &60&19&5.1&Yes&380&\\
NaI(Tl)$^3$&38&250&3.67&Yes&415&2\\
CsI(Tl)$^4$ &54&1000&4.51&Slight&550&2\\
CsI(Na)$^4$ &41&630&4.51&Yes&420&2\\
GAGG(Ce)$^5$ &40 $-$ 60&50 $-$ 150&6.63&No&520&8\\
\hline
\end{tabular}
\begin{tablenotes}
\item[1] \url{https://www.advatech-uk.co.uk/gso_ce.html}
\item[2] \url{https://www.advatech-uk.co.uk/cebr3.html}
	\item[3] \url{https://www.crystals.saint-gobain.com/products/nai-sodium-iodide}
		\item[4] \url{https://www.crystals.saint-gobain.com/products/csitl-cesium-iodide-thallium}
			\item[5] \url{https://www.advatech-uk.co.uk/gagg_ce.html}
\end{tablenotes}
\end{threeparttable}
\end{tiny}
\end{table*}
New generation SiPMs with lower background level and faster post-avalanche recovery of the micro-pixels are the key to achieve lower energy threshold of 20 keV of the scintillators and Compton polarimeters. Cooling down the SiPMs to lower temperature should also improve the overall background and sensitivity of the instruments.

PENGUIN-M \citep{dergachev09} on board the CORONAS-PHOTON solar mission, launched in 2009 and PING-P \citep{kotov16} in the 2025 upcoming solar mission use similar Compton polarimetry geometry but with non-focal configuration to measure polarization of solar flares in 20 $-$ 150 keV. Both the instruments use p-terphenyl active scatterer with NaI(Tl) and CsI scintillators as absorbers respectively read by individual PMTs (see Table \ref{ins_Table}). 

\subsection{$>$80 keV: large detector array configuration}
As discussed in section \ref{pol_measurement} and shown in Fig. \ref{effy}, at energies above 80 keV with no focusing optics in hard X-rays, large array of high Z detectors like pixelated CZTs or Germanium detectors or an array of plastic and inorganic scintillators make sensitive Compton polarimeters.   
They can be utilized both in narrow FOV (when collimated) for pointed observations or in wide FOV configuration to detect transients like GRBs or solar flares. Use of collimators restricts the energy range to a few hundreds of keV, but possesses higher sensitivity due to lower background. However, one should consider the fact that a large FOV instrument can observe multiple hard X-ray objects simultaneously and therefore the effective exposure in this case is T/N rather than T, where T is the exposure and N is the number of sources in the FOV.  
Most of the Compton polarimetry experiments seen in the last decade e.g. {\em AstroSat}-CZTI, {\em INTEGRAL}-SPI, POLAR, GRAPE, GAP, PHENEX, PoGO+ utilize one of these configurations.  

\subsubsection{Array of high Z detectors}
Any spatially segmented spectrometer should intrinsically provide Compton polarimetry capability if the readout preserves the 2-pixel or multi-pixel events. CZT detector based spectrometers can be used in this configuration because of the significant Compton scattering cross-section of thick CZTs (as shown in Fig. \ref{effy}).  
CZT detectors have already been used extensively in developing sensitive spectrometers over the years in SWIFT \citep{barthelmy05}, {\em INTEGRAL} \citep{ubertini03},
{\em NuSTAR} \citep{harrison13}, {\em AstroSat} \citep{singh14}, {\em Hitomi} \citep{kunieda10}, EXIST \citep{hong09},
RT-2 experiment on board CORONAS-PHOTON \citep{nandi11}  exploiting 
its high detection efficiency
in a broad energy band from a few keV to a few hundreds of keV, fine spectral energy resolution at near room temperature and sufficient radiation hardness. 
CZT spectrometers can also provide simultaneous polarimetry measurements at energies beyond 100 keV utilizing the spatial segmentation (see bottom panel of Fig. \ref{effy}). 

\begin{figure}[h]
\centering
\includegraphics[scale=.45]{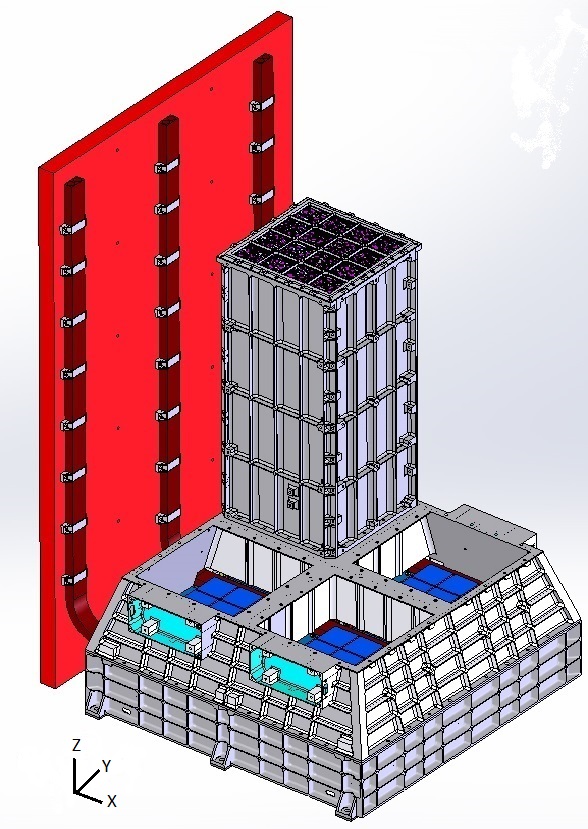}
\caption
{Schematic of CZT Imager on board {\em AstroSat}. The 64 CZT detector array is shown in blue along with CZTI bottom housing (grey), collimator and supporting structures (grey towers) and radiator plate (red). Picture credit $-$ \citet{bhalerao16} (reprinted by permission from Springer Nature Customer Service Centre GmbH).}
\label{czt_astrosat}
\end{figure}

Multiple groups across the globe are actively involved in developing CZT detectors optimized for polarimetry and experimentally demonstrating their polarization measurement capability \citep[see an overview on  the ongoing efforts by][]{caroli18}. CZT Imager \citep[CZTI,][]{bhalerao16,rao16} on board {\em AstroSat} \citep{agrawal06,singh14} is a large area coded mask hard X-ray spectrometer. It has been successfully utilized as a hard X-ray Compton polarimeter in 100 $-$ 350 keV \citep{chattopadhyay14,vadawale15}. 
{\em AstroSat}-CZTI employs a total of 
64 CZT detector modules (each module is 40$\times$40$\times$5 mm$^3$ in size with 16$\times$16 pixel array of 2.5 mm pitch) with integrated readout ASIC providing a collecting area of 1024 cm$^2$ for hard X-ray imaging and spectroscopy in 10 keV to 150 keV range (see Fig. \ref{czt_astrosat}). 
The readout provides a time resolution of 20 $\mu$s which makes it possible to use the instrument as a sensitive polarimeter in 100 $-$ 300 keV \citep{vadawale17}. The collimators, coded mask and the supporting structure of the payload become increasingly transparent at higher energies enabling the instrument to detect GRBs and measure their polarization \citep{chattopadhyay19}. 

\begin{figure}[h]
\centering
\includegraphics[scale=.7]{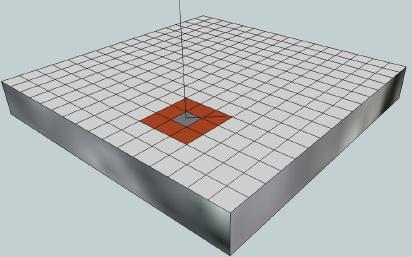}
\caption
{Compton polarimetry in pixelated CZT detectors. The pixel with low energy deposition shown in grey or the 1$^{st}$ event and one of the adjacent 8 pixels with high energy deposition shown in red or the 2$^{nd}$ event provide the azimuthal Compton scattering angle for polarization analysis.}
\label{CZT_pol}
\end{figure}

\begin{figure}[h]
\centering
\includegraphics[scale=.52]{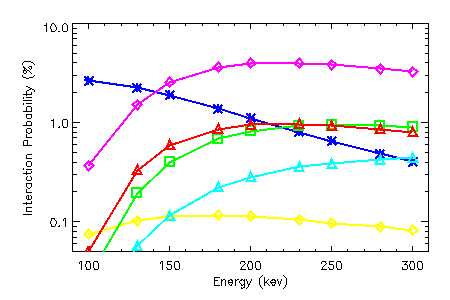}
\caption
{Different processes generating 2-pixel events in 2.5 mm pixel and 5 mm thick CZT detectors as a function of energies: photo-electric $-$ photo-electric
(blue, asterisks), 1-compton $-$ photo-electric (pink, diamond), 1-compton $-$ 1-photo $-$ electric (red, triangle),
photo-electric $-$ multi-Compton (green, square), photo-electric $-$ bremsstrahlung (yellow, diamond),
photo-electric $-$ bremsstrahlung-Compton (light blue, triangle). Figure credit $-$ \citet{chattopadhyay14} (reprinted by permission from Springer Nature Customer Service Centre GmbH).}
\label{czt_evt}
\end{figure}

Unlike other polarimetry configurations where the low Z scatterer(s) and high Z absorbers are physically isolated, in CZT detectors, a similar distinction of scatterer and absorbers is not possible which can make the identification of the 1$^{st}$ or the scattering event and 2$^{nd}$ or the photo-absorption event slightly complicated. At lower energies (below 260 keV), this identification can be done from the respective energy depositions as the electron recoil energy is always lower than the scattered photon energy (see Fig. \ref{CZT_pol}).
However, at energies beyond 260 keV, energy deposited in the electron recoil can be larger than the scattered photon energy for some range of scattering angles which makes the extraction of azimuthal angle complicated.  
Unlike the low Z plastic scatterers, CZT detectors (Z $\sim$50) have escape lines around 30 keV than can produce fake 2-pixel events (around 1 \% probability for the CZT detectors used in CZTI) that can reduce the sensitivity of the instrument. On the other hand at higher energies (above 200 keV), bremsstrahlung photons emitted from the high energy photo-electrons can register as 2-pixel events diluting the modulation patterns (see Fig. \ref{czt_evt}). 
Therefore, it is important to accurately extract the true Compton events based on the energy depositions and other Compton kinematics criteria. It is possible to filter out most of the 2-pixel hits arising from the photo-electric escape events by setting the energy threshold of the CZT pixels above 30 keV (in the analysis software), however at the expense of higher low energy cut ($\sim$150 keV) for polarization analysis resulting in slightly lower sensitivity.

Because of the asymmetry in the scattering geometry due to the square pixels, these detectors can possess inherent modulation in the azimuthal angle distribution.
Modulation factors in CZT detectors also depend on the polarization angle. For polarization along the corner pixels, modulation factors are significantly higher than that for the edge pixels. This is because the electric field vector when aligned along the corner pixels favors polar scattering angle close to 90$^\circ$ for the Compton scattering events which yields high modulation factors as shown in Fig. \ref{comp_cross} \citep[for more details, see][]{chattopadhyay14}. All these effects should be carefully evaluated using detailed laboratory experiments and simulations for correct interpretation of the data from this type of polarimetry instruments.  

A thick CZT detector ($\sim$5 mm) provides sufficient scattering probability, however limits the pixel size to a couple of millimeter in order to minimize charge sharing between the pixels \citep{kalemci02,kim11,veale14,chattopadhyay16}. This in turn limits the number of bins to only eight as most of the two pixel events are contained within a 9-pixel island. Accurate measurement of polarization angle requires a larger number of azimuthal bins. Reduction in the substrate thickness will allow for smaller pixel sizes which in turn might provide larger number of azimuthal bins. Optimization in pixel size and thickness of the detectors and pixel geometry (e.g. square vs. hexagonal) can play an important role in developing sensitive CZT based Compton polarimeter. 

\begin{figure}
\centering
\includegraphics[height=.3\textheight]{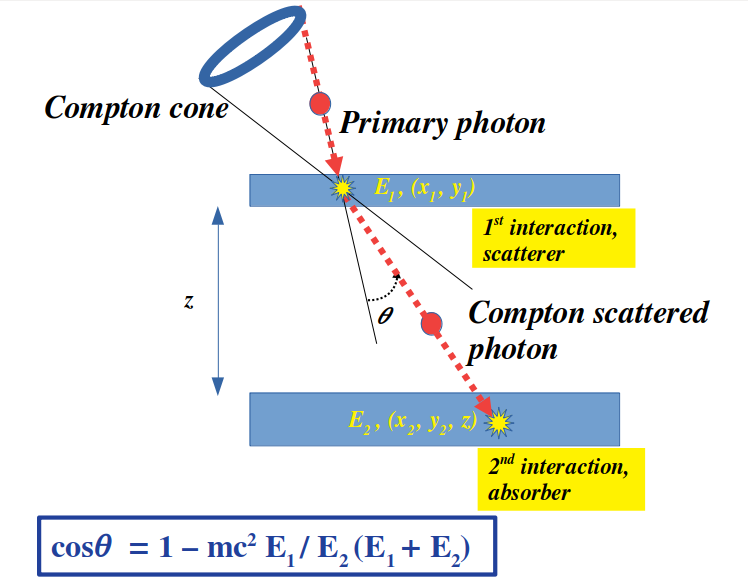}
\caption{Working principle of a Compton camera. From the two known energy depositions, polar scattering angle, $\theta$, can be inferred which along with the known x, y, z-locations of the events can be used to obtain the source direction within the circumference of the Compton cone.}
\label{comp_camera}
\end{figure}

\begin{figure}
\centering
\includegraphics[height=.25\textheight]{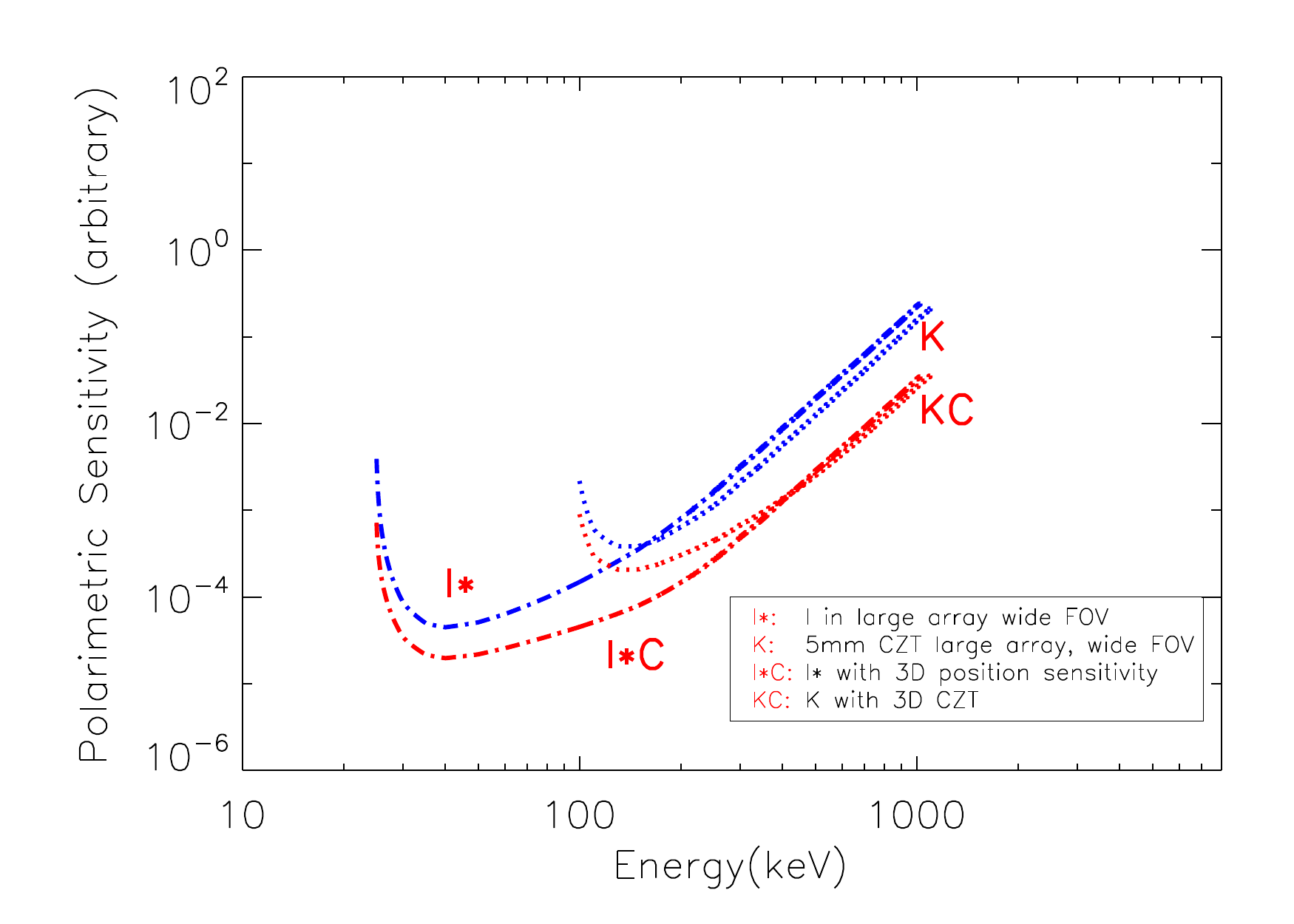}
\caption{Same as bottom panel of Fig. \ref{effy} with sensitivity of a 3D CZT detector array (KC) and 3D sensitive scintillator detector array (I*C) as a function of photon energy. We see a significant improvement in sensitivity compared to the respective 2D configurations (K and I*) with the availability of 3D position of each event enabling reconstruction of source photon direction.}
\label{effy_camera}
\end{figure}

With the increasing success of inferring photon interaction site along the depth of the detector \citep{zhang12_3DCZT,kuvvetli14_3dCZT}, CZTs can now be utilized as Compton camera and therefore can constrain the source direction for each 2-pixel event which will reduce the chance background events significantly. Working principle of a Compton camera has been demonstrated in Fig. \ref{comp_camera}. 
From the two known energy depositions and x, y, z-locations of the events, the source direction can be constrained within the circumference of a Compton cone.
The interaction depth (z) is calculated from the ratio of charges in the segmented cathode to the charge in strip anode in CZT drift detectors \citep{kuvvetli14_3dCZT} or by calculating the electron cloud drift time based on the time difference between the
signals from the cathode and the anode \citep{kim19_3dCZT}. 
Compton polarimetry based on 3D CZT detectors has already been evaluated and proposed \citep{caroli12}. See \citet{caroli18} for an overview of 3D CZT detector development for polarimetry experiments. 
The uncertainty in determining the Compton scattering angle depends on the energy resolution of the detector for E$_1$ (1$^{st}$ interaction) and E$_2$ (2$^{nd}$ interaction) and E, total energy (see Eq. \ref{eqcompang}) and position resolution in x, y, z direction,
\begin{equation}
\label{eqcompang}
\cos\theta=1-\dfrac{mc^{2}E_1}{E_2(E_1+E_2)}.
\end{equation}
CZT detectors are expected to provide better than $\sim$5 \% energy resolution around 100 keV at room temperatures. 
The 2D spatial resolution in the x, y-direction is limited by the pixel size (2.5 mm pixel size assuming CZTI type of detectors). A $<$1 mm resolution in the z direction for a 5 mm thick CZT has been reported \citep{kim19_3dCZT,amego19}. Based on these parameters, comparison in polarization sensitivity between a 900 cm$^2$ planar array of 2D CZT and 3D CZT detectors is made in Fig. \ref{effy_camera}.  
The sensitivity improves significantly for 3D CZTs due to the restricted FOV along the circumference of the Compton scattering cone. However, one should keep in mind the complexity due to large volume of data flow (full temporal profile with a few tens of $\mu$s timing resolution for both anode and cathode for each event) and realization of a large detection plane using this type of detector.

Germanium detector based hard X-ray spectrometers \citep[e.g SPI,][]{vedrenne03} on board {\em INTEGRAL}, can also be utilized for Compton polarimetry where Compton scattered events are identified from two events in two individual detector bars. Advantage of Germanium detectors is that they can be made thicker $\sim$ a few cm, which makes them sensitive in a broad energy range with high Compton scattering efficiency. Unlike CZT detectors, Germanium detectors require cooling to very low temperatures to minimize the leakage current. SPI is a coded mask instrument utilizing 19 hexagonal high purity Germanium detector bars (each with thickness of 7 cm) closely packed in a hexagonal shape. Events depositing energies in two detectors within a coincidence window of 350 ns are used to extract polarization information in 200 $-$ 1000 keV \citep{chauvin18}. SPI was not optimized for polarimetry and never tested for polarization before the launch of {\em INTEGRAL}. The instrument in 18 years of operation lost 4 detectors with a $\sim$53 \% reduction in the multi-detector effective area and significant degrade in the polarimetric sensitivity. Germanium detector array with proper optimization for Compton polarimetry can serve as a sensitive spectro-polarimeter in a wide energy band.

\subsubsection{Array of low Z and high Z scintillators}
An obvious extension of the focal plane configuration with plastic scatterer based Compton polarimeter in $>$80 keV is a square array of plastic scintillators for the requirement of large collecting area surrounded by an array of high Z inorganic scintillators on four sides. Availability of MAPMTs makes this type of configuration possible where a single MAPMT (e.g. 8$\times$8 pixels) can read the full array of plastic and inorganic scintillators (total 8$\times$8 array of scintillators). One advantage of this configuration compared to the high Z detector array is that the instruments are also sensitive at lower energies $\sim$50 keV because of higher Compton scattering efficiency of the low Z plastics at lower energies. The lower energy limit can be further improved with the improvement in the lower energy threshold in the plastic scintillators (discussed in the previous section). This type of configuration, however, suffers from light leak between the adjacent plastic scintillators and between the anode pixels in the MAPMTs which reduce the modulation factors of the instruments and enhance the intrinsic chance background events. The upcoming polarimetry designs under this configuration take care most of these shortcomings by employing new detection technologies (see below).

\begin{figure}[h]
\centering
\includegraphics[scale=.4]{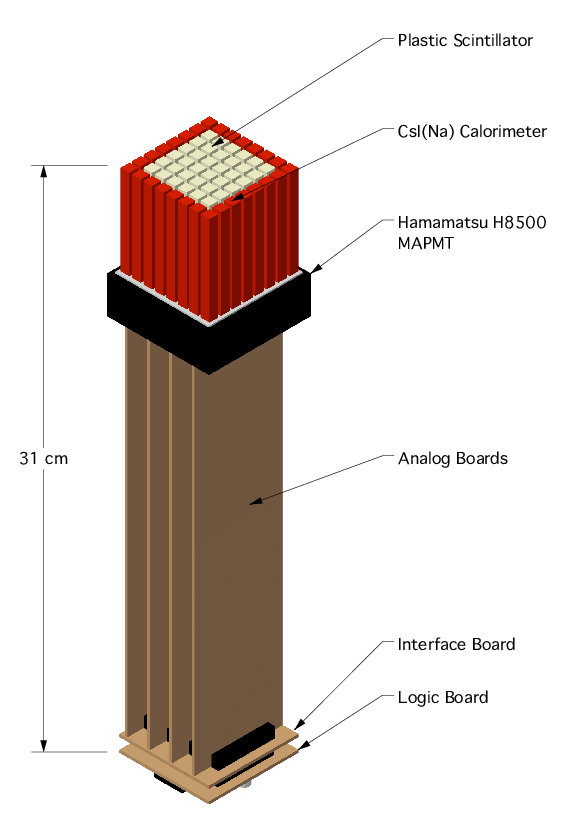}
\caption
{An example of scintillator array configuration for Compton polarimetry as used in GRAPE experiment. See text for more details. Picture courtesy $-$ \citet{mcconnell09} (reprinted by permission from authors and SPIE).}
\label{grape_geom}
\end{figure}

GRAPE \citep{bloser09,mcconnell09}, a balloon-borne wide FOV hard X-ray GRB polarimeter and PHENEX \citep{kishimoto07,gunji08}, another balloon-borne but narrow FOV hard X-ray polarimeter employ this type of geometry. 
Both PHENEX and GRAPE use long (40 and 50 mm respectively), square shaped (5.5$\times$5.5 mm$^2$ and 5$\times$5 mm$^2$ respectively) plastic scintillators in 6$\times$6 array at the center surrounded by 28 CsI (Tl) scintillators of same dimensions to detect the plastic scattered X-rays (see Fig. \ref{grape_geom}). All the scintillators are read by a single MAPMT.    
Molybdenum collimators in PHENEX modules restrict the FOV to 4.8$^\circ$ (FWHM) in an energy range of 20 $-$ 200 to perform
pointed observations. The balloon flights of GRAPE and PHENEX did not yield any significant measurements either due to small effective exposure because of problems in attitude control system or higher than expected background level resulting in low sensitivity. 
One major issue with the MAPMT readout is the cross-talk between the pixels reducing the polarimetric sensitivity of the instruments. SiPM readout can solve the cross-talk problem. An array of SiPMs along the length of the scintillators can provide significant improvement in the sensitivity by lowering down the energy threshold in the plastic scintillators, also enabling Compton imaging from 3D readout of the events which in turn reduces the background events. 
Fig. \ref{effy_camera} compares sensitivity between 2D (same as in Fig. \ref{effy}) and 3D scintillator array. Even though the lower energy threshold is kept same in both 2D and 3D plastics, we see a significant improvement in the sensitivity for the 3D scintillator array due to reduced background.  
Updated design of GRAPE employs a large number of small, scintillator cubes, each of which is readout by its own SiPM \citep{mcconnel18_grape}. LEAP \citep{mcconnel16_leap}, recently selected for phase A study for NASA's future astronomy mission is based on the same design. In a tightly packed scintillator array, light leak between the adjacent scintillators is another concern, minimization of which requires an efficient wrapping material making them optically isolated. This also promotes better light collecting efficiency and thereby improving the sensitivity.  
SPHiNX \citep{pearce19}, a newly proposed Swedish GRB polarimetry mission (sensitive in 50 $-$ 500 keV) uses SiPM readout for the surrounding GAGG scintillators along its width and length. It uses normal PMT readout for the plastic scatterers.  
The design uses 7 units of 6 hexagonically packed triangle-shaped EJ-204 plastic scintillators each read by its own PMT surrounded by 24 GAGG scitillator bars read by SiPM array. The shared walls of the 7 hexagons are read by the common SiPMs in order to reduce the number of readout channels.

Design of GRB polarimeter, GAP \citep{yonetoku06}, on board 2010 {\em IKAROS} solar power sail mission, is based on a simple Compton scattering geometry. GAP uses a single but large 12 sided plastic scintillator (gap between two opposite sides $\sim$ 140 mm and height of 60 mm) read by a single large 51 mm diameter PMT surrounded by 12 CsI scintillators (6 cm long and 5 mm width) read by individual small subminiature PMTs. The large plastic provides a collecting area of  $\sim$176 cm$^2$ with a large FOV. 

\subsubsection{Array of low Z scintillators}
This type of configuration uses only an array of low Z plastic scintillators and therefore play a dual role of both scatterer and absorber. The advantage of this configuration is that the instrument can be scaled up easily in collecting area by increasing the number plastic scintillators.
The recent polarimetry experiments, POLAR and PoGO+, utilized this configuration in developing the instruments.

POLAR \citep{sun16,produit18} on board Chinese space laboratory TG-2, is a wide field GRB polarimeter, sensitive in 50 $-$ 500 keV. One unit of POLAR (total of 25 units)
comprises 64 plastic scintillator bars (EJ-248M, 17.6 cm long and 5.8 mm sides) readout by one MAPMT. As mentioned earlier, MAPMT suffers from optical cross-talk between the adjacent pixels which limits the sensitivity of the instrument. It also requires high voltage for operation, failure of which caused an untimely end of the POLAR mission. These shortcoming will be taken care of in the successor mission,
POLAR-2 \citep{kole19}, planned for launch in 2024, by employing SiPMs to readout the plastic bars from the back side instead of MAPMTs. POLAR-2 uses similar configuration of array of 64 plastics of similar depth and cross-section, however with a total of 100 units, thereby enhancing the effective area by an order of magnitude. To overcome the high background of SiPMs, POLAR-2 will passively cool the SiPMs to -10$^\circ$C which ensures a lower background level and therefore better energy threshold of the plastics ($\sim$2 keV). This improves the effective area performance of POLAR-2 at lower energies. POLAR-2 with its planned sensitivity and energy range of 30 $-$ 1000 keV is expected to provide accurate polarization measurements for $\sim$50 GRBs/year with polarization level of $\sim$10 \%. 

\begin{figure}[h]
\centering
\includegraphics[scale=.42]{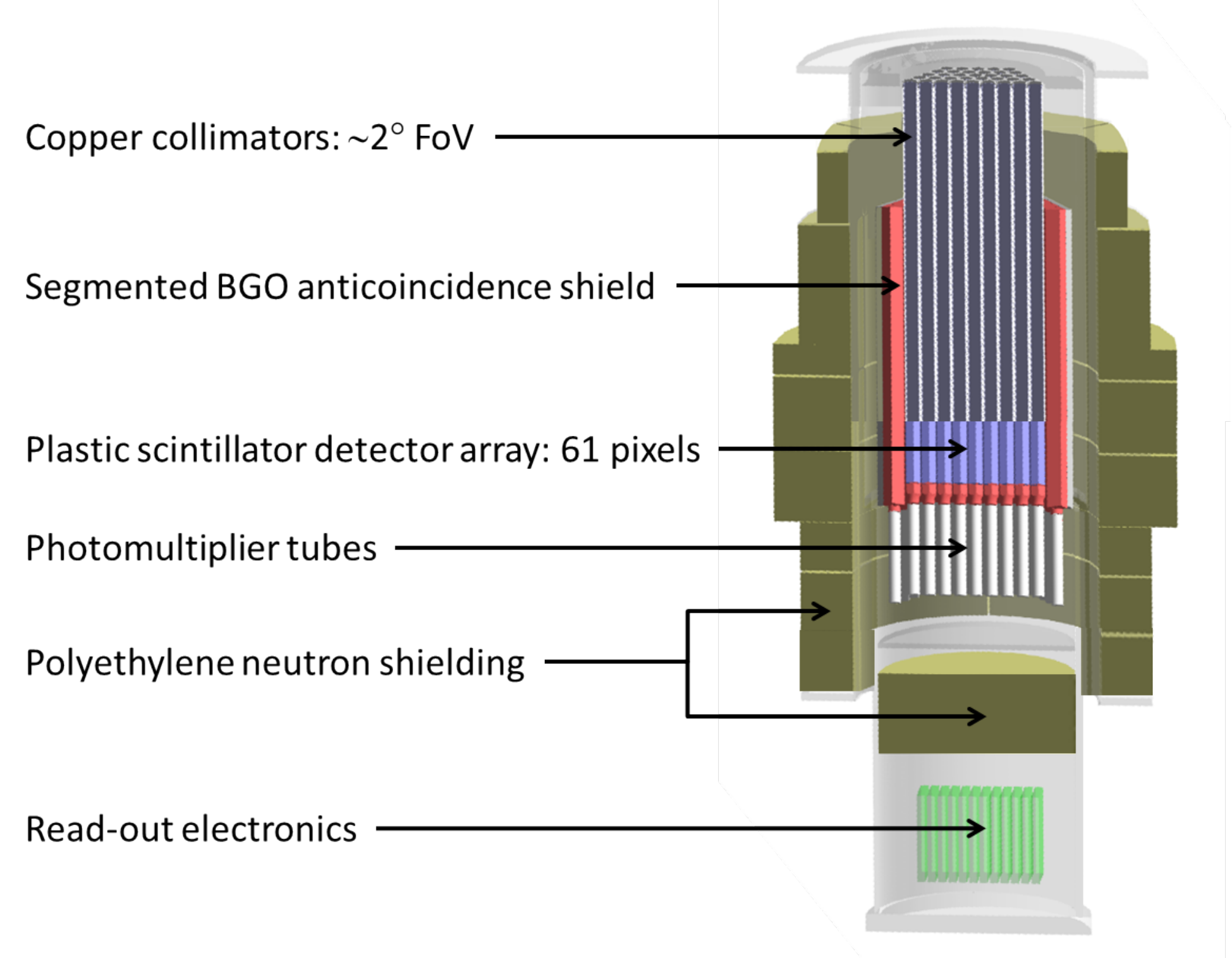}
\caption
{Array of plastic scintillators for Compton polarimetry as used in PoGO+ experiment. Picture courtesy $-$ \citet{friis18} (reprinted based on MDPI open access license policy).}
\label{pogo_geom}
\end{figure}

PoGO+ \citep{chauvin16_pogo,friis18} is a balloon-borne polarimetry mission, flown in 2016. It was optimized for pointing observation of Crab and Cygnus X-1. PoGO+ utilizes Compton scattering within an array of 61 plastic scintillators (EJ-204), each 12 cm long and 3 cm cross in hexagonal shape, read by individual PMTs (see Fig. \ref{pogo_geom}). 
The design is a matured version of the 
pathfinder balloon-born experiment, PoGOLite, flown in 2013 \citep{chauvin16_pogolite}. 
The graded shield collimators (graded shield of Copper, Tin and Lead layers) restrict the FOV to $\sim$2$^\circ$ for pointed observations in 20 $-$ 180 keV. While the collimators suppress most of the off-axis background events from front side, active BGO shielding on the sides and back of the scintillators vetoes most of the background events from other directions.
In order to reduce the light leak between the tightly packed adjacent scintillator rods, a special wrapping material is used. PoGO+ type of instrument design is expected to provide a sensitivity level of 11.1 \% for Crab like sources for a 5-day balloon flight mission.

\subsubsection{Compton camera configuration}
This configuration uses two separate layers of high Z detector planes with Compton scattering in the first layer and subsequent absorption of the scattered photon by the second layer. Therefore the instruments provide 3D positional information of each event (2D position in x and y-direction from individual layers and position in z from the known gap between the two layers) which along with the deposited energy information can be used to reconstruct the original photon direction. Illustration of the basic technique is shown in Fig .\ref{comp_camera}. This technique efficiently filters out most of the background events which yields a high signal to noise ratio. From the location of 1$^{st}$ and 2$^{nd}$ event from layer 1 and layer 2 respectively, obtained for each of the Compton scattering event, can be used to extract the azimuthal angle of scattering. 
Though Compton imager provides a better background rejection due to accurate determination of the source direction (better positional resolution in z-direction), because the geometry allows only forward scattering, polarimetric efficiency and modulation factor are on the lower side (see Fig. \ref{comp_cross}).  
IBIS \citep{ubertini03} on board {\em INTEGRAL}, COSI \citep{yang18}, SGD \citep{tajima10} on board {\em Hitomi} use this type of configuration to carry out  spectro-polarimetry study in hard X-rays. IBIS consists of a top layer of 128$\times$128 CdTe detector array (4$\times$4$\times$2 mm$^3$) with pixel pitch of 9.2 mm and a bottom layer of 64$\times$64 CsI detector bars (each 8.4$\times$8.4$\times$300 mm$^3$). The two layers are separated by 94 mm. Though primarily developed as a spectro-imager, the same events which deposit energy in both the layers within a coincidence window of 3.8 $\mu$s for reconstructing the source position, can be used in extracting the azimuthal scattering angle distribution. 

COSI, on the other hand, is a dedicated balloon borne Compton polarimeter and spectro-imager sensitive in 200 keV to several MeV. It uses 12 cooled high-purity Germanium double-sided strip
detectors (each 8$\times$8$\times$1.5 cm$^3$) with a 2 mm strip pitch in 2$\times$2$\times$3 array to detect and reconstruct the Compton events. With 37 orthogonal strips per side, the instrument provides an excellent 2D and depth resolution (0.5 mm FWHM) which along with excellent spectral
resolution makes this instrument a sensitive spectro-polarimeter. COSI-2 \citep{tomsick19}, a similar design with an additional layer of Germanium detector, has been selected for phase A study for NASA SMEX mission. COSI-2 is expected to detect 20 GRBs a year with polarization capability of 50 \% MDP for 4$\times$10$^6$ erg/cm$^2$ fluence, apart from polarization measurements for bright galactic X-ray sources.      

SGD Compton camera configuration consists of 32 layers of 0.6 mm thick Silicon sensors and 8 layers of 0.75 mm thick CdTe detectors. These layers are surrounded by 2 more layers of same CdTe detectors from all sides. Each detector is pixelated with a pixel size of 3.2 mm. Total eight such units (210 cm$^2$ geometric area) are collimated separately to a narrow FOV. Excellent energy and positional resolution and optimized scattering geometry of SGD allow sensitive polarimetry measurements in 50 $-$ 200 keV with MDP level of a few percent for 100 mCrab sources in 100 ks exposure.  

\section{Summary and conclusion}

X-ray polarization measurements provide two independent parameters $-$ degree and 
angle of polarization characterizing the incoming radiation from any 
X-ray source. Measurement of these parameters can provide a unique opportunity to 
study the behavior of matter and radiation under extreme magnetic 
and gravitational fields. Importance of X-ray polarimetry in astronomy has been emphasized over the years, however its wealth has never really been utilized until recently when a number of polarimetry and non-polarimetry instruments provided some interesting polarization measurements in hard X-rays for Crab, Cygnus X-1, V404-Cygni and a number of GRBs and solar flares. While some of the instruments e.g. {\em AstroSat}, {\em INTEGRAL} will continue to provide new measurements or improve upon the existing results for years to come, there is a large number of confirmed and proposed dedicated hard X-ray polarimetry experiments which ensures an extremely exciting time ahead in the field of hard X-ray polarimetry. The primary reason behind these new measurements and the urge of conducting new experiments at these energies is the higher level of polarization that is expected in hard X-rays compared to that in soft X-rays where the dominance of thermal emission dilutes the polarization level to only a few percent. Consequently, a moderately sensitive polarimeter can also provide interesting measurements for some of the bright sources. 
Hard X-rays being generated close to the compact object and from fast cooling of the highest energy particles, carry some unique insights of acceleration and cooling mechanisms, signatures of high gravity and magnetic field close to the black hole jets and low altitude magnetosphere of magnetars.  
We have already seen a glimpse of such interesting science cases from {\em AstroSat} and {\em Integral} measurements. With a significant advancement in the detection technology in the last couple of decades in the field of scintillator readout methods, high-speed readout Silicon sensors, 3D CZT and Germanium detectors and above of the all, development of hard X-ray focusing optics, developing a sensitive compact hard X-ray focal plane Compton polarimeter has come well within our reach. 
All these new activities demand more focus in the field of hard X-ray polarimetry to improve and develop sensitive polarimetry configurations using the new detection and readout technologies, and develop robust methods to analyze the existing and upcoming polarization data.

In this article, we reviewed the scientific potential of hard X-ray polarimetry and some of the recent findings from various hard X-ray polarimetry experiments. We discussed the existing polarimetry techniques and compared them in hard X-rays. While Compton scattering is best suited in the hard X-ray regime, photo-electron track imagers can also be explored in hard X-rays in the future, in particular, the TPC detectors with higher gas depth and density and the CdTe-Timepix detectors with smaller pixel sizes. 
Within scattering based polarimetry technique, we found that a compact Compton scattering polarimeter at the focal plane of a hard X-ray telescope can provide an order of magnitude higher sensitivity in 25 $-$ 80 keV range.  At the lower energies, {\em POLIX} \citep{paul16} type Rayleigh polarimeters are still the best option for polarimetry whereas at the most difficult photon starved 100 $-$ 1000 keV region is probably best served by a polarimetry-optimized large area wide FOV Compton camera. 

{With the growing realization of scientific potential of X-ray polarimetry, we expect many more new approvals for dedicated X-ray polarimetry missions like {\em IXPE}, the NASA mission and {\em POLIX}, the Indian next space mission, both set to launch in 2021. 
In the hard X-rays, we have seen some proposals for dedicated focal plane Compton polarimeters in 20 $-$ 80 keV range. The conventional focal plane polarimeters employ plastic scintillator rod as scatterer and sensitivity of the instruments is primarily limited by the lower energy threshold of the plastic. In the future, we expect to see improvement in the detection threshold with the advent of better readout technologies. Improvisation in the scatterer configurations, e.g. replacing the plastic by a hybrid passive (Be/Li) and active plastic scatterers in a single cylindrical bundle might improve the overall energy threshold of the polarimeters by utilizing the Rayleigh scattered X-rays at lower energies from the passive scatterers.  
Thanks to the advancement in the detector technology like in Double Sided Silicon Strip Detector \citep[DSSD,][]{ichiro08_DSSD} or in the Silicon based active pixel sensors, e.g. fast readout X-ray Hybrid CMOS detectors \citep{bai08,hull17,chattopadhyay18_HCDoverview}, in particular the SPEEDSTER-EXDs \citep{griffith16}, and Timepix detectors \citep{llopart07}, it might also be possible to replace the plastic scatterer by a position sensitive Silicon sensor to enable simultaneous spectroscopic, timing, imaging and polarimetry \citep[see][for a conceptual design]{vadawale12}.  

Unavailability of focusing optics in $>$80 keV warrants a large area detector plane to be 
the obvious solution to build a sensitive polarimeter.
Emergence of Laue lens based on diffraction in crystals allowing concentration of X-rays above 100 keV \citep{barriere11} or the newly proposed Stacked Prism Lenses \citep[SPL,][]{Mi19_spl} will provide an excellent solution in future to develop sensitive detectors in the sub-MeV region \citep{roques12}. However, in absence of any real implementation of these techniques at present, an effective approach might be to use a large area detector in wide FOV configuration. The recently developed 3D CZT detectors can provide excellent polarimetry and spectroscopic sensitivity in Compton camera mode in this regime. Alternatively, a Compton camera made of a top layer of 2D CZT detectors (like in CZTI) and a bottom layer of CeBr$_3$ crystal with 2D SiPM readout \citep[recently shown by][]{goyal18} might be an effective way to tackle this photon starved region.    

With all these new developments in the hardware and upcoming dedicated polarimetry missions, we expect many more new polarization measurements in hard X-rays in the coming years.}   


\appendix




\section*{Acknowledgements}
I would like to thank Prof. A. R. Rao, Prof. Santosh Vadawale and Prof. David Burrows for the useful discussions while writing this article. 
Dr. Moszi Kiss and Prof. Maxime Chauvin of PoGO+ collaboration provided some important inputs about the PoGO+ payload and future hard X-ray polarimetry techniques. I am thankful to Prof. Mark McConnell for sharing with me some useful contents about GRAPE and LEAP polarimetry designs. 
My wife, Suravi, helped me with some of the figures used in the article; a big thank to her. I would also like to acknowledge Journal of Astrophysics and Astronomy and in particular, Prof. Tushar P. Prabhu for inviting me to write this article. Some of the figures and pictures used in this article are either adapted or taken from published articles. I am grateful to the authors and the publishers of the articles for providing me the necessary copyright permissions for the reuse of the materials.    
\vspace{-1em}




\begin{thebibliography}{}
\expandafter\ifx\csname natexlab\endcsname\relax\def\natexlab#1{#1}\fi

\bibitem[{{Agrawal}(2006)}]{agrawal06}
{Agrawal}, P.~C. 2006, Advances in Space Research, 38, 2989

\bibitem[{Aharonian {$et~al$.}(2018)Aharonian, Akamatsu, Akimoto, Allen,
  Angelini, Audard, Awaki, Axelsson, Bamba, \& et~al.}]{hitomi18}
Aharonian, F., Akamatsu, H., Akimoto, F., {$et~al$.} 2018, Publications of the
  Astronomical Society of Japan, 70, doi:10.1093/pasj/psy118

\bibitem[{{Angel} {$et~al$.}(1969){Angel}, {Novick}, {vanden Bout}, \&
  {Wolff}}]{angel69}
{Angel}, J.~R., {Novick}, R., {vanden Bout}, P., \& {Wolff}, R. 1969, Physical
  Review Letters, 22, 861

\bibitem[{{Angel} \& {Weisskopf}(1970)}]{angel70}
{Angel}, J.~R.~P., \& {Weisskopf}, M.~C. 1970, The Astronomical Journal, 75,
  231

\bibitem[{Asakura {$et~al$.}(2019)Asakura, Hayashida, Hanasaka, Kawabata,
  Yoneyama, Okazaki, Ide, Noda, Matsumoto, Tsunemi, Awaki, \&
  Nakajima}]{asakura19}
Asakura, K., Hayashida, K., Hanasaka, T., {$et~al$.} 2019, X-ray imaging
  polarimetry with a 2.5 micrometer pixel CMOS sensor for visible light at
  room temperature, arXiv:1906.00012

\bibitem[{{Bai} \& {Ramaty}(1978)}]{bai78}
{Bai}, T., \& {Ramaty}, R. 1978, \apj, 219, 705

\bibitem[{{Bai} {$et~al$.}(2008){Bai}, {Bajaj}, {Beletic}, {Farris}, {Joshi},
  {Lauxtermann}, {Petersen}, \& {Williams}}]{bai08}
{Bai}, Y., {Bajaj}, J., {Beletic}, J.~W., {$et~al$.} 2008, 7021, 702102

\bibitem[{Baring \& Harding(2006)}]{baring06}
Baring, M.~G., \& Harding, A.~K. 2006, Resonant Compton Upscattering in
  Anomalous X-ray Pulsars, arXiv:astro-ph/0610382

\bibitem[{Barrière {$et~al$.}(2011)Barrière, Natalucci, \&
  Ubertini}]{barriere11}
Barrière, N.~M., Natalucci, L., \& Ubertini, P. 2011, Hard X / soft gamma ray
  polarimetry using a Laue lens, arXiv:1109.1313

\bibitem[{{Barthelmy} {$et~al$.}(2005){Barthelmy}, {Barbier}, {Cummings},
  {Fenimore}, {Gehrels}, {Hullinger}, {Krimm}, {Markwardt}, {Palmer},
  {Parsons}, {Sato}, {Suzuki}, {Takahashi}, {Tashiro}, \&
  {Tueller}}]{barthelmy05}
{Barthelmy}, S.~D., {Barbier}, L.~M., {Cummings}, J.~R., {$et~al$.} 2005, Space
  Science Reviews, 120, 143

\bibitem[{{Basak} \& {Rao}(2015)}]{basak15}
{Basak}, R., \& {Rao}, A.~R. 2015, \apj, 807, 34

\bibitem[{{Beilicke} {$et~al$.}(2014){Beilicke}, {Kislat}, {Zajczyk}, {Guo},
  {Endsley}, {Stork}, {Cowsik}, {Dowkontt}, {Barthelmy}, {Hams}, {Okajima},
  {Sasaki}, {Zeiger}, {de Geronimo}, {Baring}, \& {Krawczynski}}]{beilicke14}
{Beilicke}, M., {Kislat}, F., {Zajczyk}, A., {$et~al$.} 2014, Journal of
  Astronomical Instrumentation, 3, 40008

\bibitem[{{Bellazzini} {$et~al$.}(2006){Bellazzini}, {Angelini}, {Baldini},
  {Bitti}, {Brez}, {Cavalca}, {Del Prete}, {Kuss}, {Latronico}, {Omodei},
  {Pinchera}, {Massai}, {Minuti}, {Razzano}, {Sgro}, {Spandre}, {Tenze},
  {Costa}, \& {Soffitta}}]{bellazzini06}
{Bellazzini}, R., {Angelini}, F., {Baldini}, L., {$et~al$.} 2006, Nuclear
  Instruments and Methods in Physics Research A, 560, 425

\bibitem[{{Bellazzini} {$et~al$.}(2007){Bellazzini}, {Spandre}, {Minuti},
  {Baldini}, {Brez}, {Latronico}, {Omodei}, {Razzano}, {Massai},
  {Pesce-Rollins}, {Sgr{\'o}}, {Costa}, {Soffitta}, {Sipila}, \&
  {Lempinen}}]{bellazzini07}
{Bellazzini}, R., {Spandre}, G., {Minuti}, M., {$et~al$.} 2007, Nuclear
  Instruments and Methods in Physics Research A, 579, 853

\bibitem[{Beloborodov(2012)}]{Beloborodov12}
Beloborodov, A.~M. 2012, The Astrophysical Journal, 762, 13

\bibitem[{{Bhalerao} {$et~al$.}(2017){Bhalerao}, {Bhattacharya}, {Vibhute},
  {Pawar}, {Rao}, {Hingar}, {Khanna}, {Kutty}, {Malkar}, {Patil}, {Arora},
  {Sinha}, {Priya}, {Samuel}, {Sreekumar}, {Vinod}, {Mithun}, {Vadawale},
  {Vagshette}, {Navalgund}, {Sarma}, {Pandiyan}, {Seetha}, \&
  {Subbarao}}]{bhalerao16}
{Bhalerao}, V., {Bhattacharya}, D., {Vibhute}, A., {$et~al$.} 2017, Journal of
  Astrophysics and Astronomy, 38, 31

\bibitem[{{Black} {$et~al$.}(2007){Black}, {Baker}, {Deines-Jones}, {Hill}, \&
  {Jahoda}}]{black07}
{Black}, J.~K., {Baker}, R.~G., {Deines-Jones}, P., {Hill}, J.~E., \& {Jahoda},
  K. 2007, Nuclear Instruments and Methods in Physics Research A, 581, 755

\bibitem[{{Bloser} {$et~al$.}(2013){Bloser}, {Legere}, {Bancroft}, {McConnell},
  {Ryan}, \& {Schwadron}}]{bloser13}
{Bloser}, P.~F., {Legere}, J., {Bancroft}, C., {$et~al$.} 2013, in Society of
  Photo-Optical Instrumentation Engineers (SPIE) Conference Series, Vol. 8859,
  Society of Photo-Optical Instrumentation Engineers (SPIE) Conference Series,
  0

\bibitem[{{Bloser} {$et~al$.}(2009){Bloser}, {Legere}, {McConnell}, {Macri},
  {Bancroft}, {Connor}, \& {Ryan}}]{bloser09}
{Bloser}, P.~F., {Legere}, J.~S., {McConnell}, M.~L., {$et~al$.} 2009, Nuclear
  Instruments and Methods in Physics Research A, 600, 424

\bibitem[{{Boggs} {$et~al$.}(2006){Boggs}, {Coburn}, \& {Kalemci}}]{boggs06}
{Boggs}, S.~E., {Coburn}, W., \& {Kalemci}, E. 2006, \apj, 638, 1129

\bibitem[{{Bogomolov} {$et~al$.}(2003){Bogomolov}, {Denisov}, {Kuznetsov},
  {Lisin}, {Logachev}, {Morozov}, {Myagkova}, {Svertilov}, {Zhitnik},
  {Ignat'ev}, {Oparin}, {Pertsov}, {Stepanov}, \& {Tindo}}]{bogomolov03}
{Bogomolov}, A.~V., {Denisov}, Y.~I., {Kuznetsov}, S.~N., {$et~al$.} 2003,
  Solar System Research, 37, 112

\bibitem[{{Buzhan} {$et~al$.}(2002){Buzhan}, {Dolgoshein}, {Ilyin},
  {Kantserov}, {Kaplin}, {Karakash}, {Pleshko}, {Popova}, {Smirnov}, {Volkov},
  {Filatov}, {Klemin}, \& {Kayumov}}]{buzhan02}
{Buzhan}, P., {Dolgoshein}, B., {Ilyin}, A., {$et~al$.} 2002, in Advanced
  Technology - Particle Physics, ed. M.~{Barone}, E.~{Borchi}, J.~{Huston},
  C.~{Leroy}, P.~G. {Rancoita}, P.~{Riboni}, \& R.~{Ruchti}, 717--728

\bibitem[{{Buzhan} {$et~al$.}(2003){Buzhan}, {Dolgoshein}, {Filatov}, {Ilyin},
  {Kantzerov}, {Kaplin}, {Karakash}, {Kayumov}, {Klemin}, {Popova}, \&
  {Smirnov}}]{buzhan03}
{Buzhan}, P., {Dolgoshein}, B., {Filatov}, L., {$et~al$.} 2003, Nuclear
  Instruments and Methods in Physics Research A, 504, 48

\bibitem[{{Bykov} {$et~al$.}(2009){Bykov}, {Uvarov}, {Bloemen}, {den Herder},
  \& {Kaastra}}]{bykov09}
{Bykov}, A.~M., {Uvarov}, Y.~A., {Bloemen}, J.~B.~G.~M., {den Herder}, J.~W.,
  \& {Kaastra}, J.~S. 2009, \mnras, 399, 1119

\bibitem[{{Caroli} {$et~al$.}(2018){Caroli}, {Moita}, {da Silva}, {Del Sordo},
  {de Cesare}, {Maia}, \& {P{\`a}scoa}}]{caroli18}
{Caroli}, E., {Moita}, M., {da Silva}, R., {$et~al$.} 2018, Galaxies, 6, 69

\bibitem[{{Caroli} {$et~al$.}(2012){Caroli}, {Alvarez}, {Auricchio},
  {Budtz-J{\o}rgensen}, {Curado da Silva}, {Del Sordo}, {Ferrando}, {Laurent},
  {Limousin}, {Galv{\`e}z}, {Gloster}, {Hernanz}, {Isern}, {Kuvvetli}, {Maia},
  {Meuris}, {Stephen}, \& {Zappettini}}]{caroli12}
{Caroli}, E., {Alvarez}, J.~M., {Auricchio}, N., {$et~al$.} 2012, in Society of
  Photo-Optical Instrumentation Engineers (SPIE) Conference Series, Vol. 8443,
  Society of Photo-Optical Instrumentation Engineers (SPIE) Conference Series

\bibitem[{Cebrián {$et~al$.}(2012)Cebrián, Cuesta, Amaré, Borjabad,
  Fortuño, García, Ginestra, Gómez, Martínez, Oliván, Ortigoza, {Ortiz de
  Solórzano}, Pobes, Puimedón, Sarsa, \& Villar}]{Cebrian12}
Cebrián, S., Cuesta, C., Amaré, J., {$et~al$.} 2012, Astroparticle Physics,
  37, 60

\bibitem[{{Celotti} \& {Matt}(1994)}]{celotti94}
{Celotti}, A., \& {Matt}, G. 1994, \mnras, 268, 451

\bibitem[{{Chand} {$et~al$.}(2019){Chand}, {Chattopadhyay}, {Oganesyan}, {Rao},
  {Vadawale}, {Bhattacharya}, {Bhalerao}, \& {Misra}}]{chand18b}
{Chand}, V., {Chattopadhyay}, T., {Oganesyan}, G., {$et~al$.} 2019, \apj, 874,
  70

\bibitem[{{Chand} {$et~al$.}(2018){Chand}, {Chattopadhyay}, {Iyyani}, {Basak},
  {Aarthy}, {Rao}, {Vadawale}, {Bhattacharya}, \& {Bhalerao}}]{chand18a}
{Chand}, V., {Chattopadhyay}, T., {Iyyani}, S., {$et~al$.} 2018, \apj, 862, 154

\bibitem[{{Chattopadhyay} {$et~al$.}(2013){Chattopadhyay}, {Vadawale}, \&
  {Pendharkar}}]{chattopadhyay13}
{Chattopadhyay}, T., {Vadawale}, S.~V., \& {Pendharkar}, J. 2013, Experimental
  Astronomy, 35, 391

\bibitem[{{Chattopadhyay} {$et~al$.}(2016){Chattopadhyay}, {Vadawale}, {Rao},
  {Bhattacharya}, {Mithun}, \& {Bhalerao}}]{chattopadhyay16}
{Chattopadhyay}, T., {Vadawale}, S.~V., {Rao}, A.~R., {$et~al$.} 2016, in
  \procspie, Vol. 9905, Society of Photo-Optical Instrumentation Engineers
  (SPIE) Conference Series, 99054D

\bibitem[{{Chattopadhyay} {$et~al$.}(2014{\natexlab{a}}){Chattopadhyay},
  {Vadawale}, {Rao}, {Sreekumar}, \& {Bhattacharya}}]{chattopadhyay14}
{Chattopadhyay}, T., {Vadawale}, S.~V., {Rao}, A.~R., {Sreekumar}, S., \&
  {Bhattacharya}, D. 2014{\natexlab{a}}, Experimental Astronomy, 37, 555

\bibitem[{{Chattopadhyay} {$et~al$.}(2014{\natexlab{b}}){Chattopadhyay},
  {Vadawale}, {Shanmugam}, \& {Goyal}}]{chattopadhyay14_2}
{Chattopadhyay}, T., {Vadawale}, S.~V., {Shanmugam}, M., \& {Goyal}, S.~K.
  2014{\natexlab{b}}, Astrophysical Journal Supplement, 212, 12

\bibitem[{Chattopadhyay {$et~al$.}(2015)Chattopadhyay, Vadawale, Goyal, P.~S.,
  Patel, Shukla, Ladiya, Shanmugam, Patel, \& Ubale}]{chattopadhyay15}
Chattopadhyay, T., Vadawale, S., Goyal, S., {$et~al$.} 2015, Experimental
  Astronomy, 1

\bibitem[{Chattopadhyay {$et~al$.}(2018)Chattopadhyay, Falcone, Burrows, Hull,
  Bray, Wages, McQuaide, Buntic, Crum, O'Dell, \&
  Anderson}]{chattopadhyay18_HCDoverview}
Chattopadhyay, T., Falcone, A.~D., Burrows, D.~N., {$et~al$.} 2018, in Space
  Telescopes and Instrumentation 2018: Ultraviolet to Gamma Ray, Vol. 10709

\bibitem[{{Chattopadhyay} {$et~al$.}(2019){Chattopadhyay}, {Vadawale},
  {Aarthy}, {Mithun}, {Chand}, {Ratheesh}, {Basak}, {Rao}, {Bhalerao}, {Mate},
  {Arvind}, {Sharma}, \& {Bhattacharya}}]{chattopadhyay19}
{Chattopadhyay}, T., {Vadawale}, S.~V., {Aarthy}, E., {$et~al$.} 2019, \apj,
  884, 123

\bibitem[{{Chauvin} {$et~al$.}(2013){Chauvin}, {Roques}, {Clark}, \&
  {Jourdain}}]{chauvin13}
{Chauvin}, M., {Roques}, J.~P., {Clark}, D.~J., \& {Jourdain}, E. 2013, \apj,
  769, 137

\bibitem[{{Chauvin} {$et~al$.}(2016){Chauvin}, {Jackson}, {Kawano}, {Kiss},
  {Kole}, {Mikhalev}, {Moretti}, {Takahashi}, \& {Pearce}}]{chauvin16_pogo}
{Chauvin}, M., {Jackson}, M., {Kawano}, T., {$et~al$.} 2016, Astroparticle
  Physics, 82, 99

\bibitem[{Chauvin {$et~al$.}(2016)Chauvin, Jackson, Kawano, Kiss, Kole,
  Mikhalev, Moretti, Takahashi, \& Pearce}]{chauvin16_pogolite}
Chauvin, M., Jackson, M., Kawano, T., {$et~al$.} 2016, Astroparticle Physics,
  72, 1

\bibitem[{{Chauvin} {$et~al$.}(2018{\natexlab{a}}){Chauvin}, {Flor{\'e}n},
  {Friis}, {Jackson}, {Kamae}, {Kataoka}, {Kawano}, {Kiss}, {Mikhalev},
  {Mizuno}, {Ohashi}, {Stana}, {Tajima}, {Takahashi}, {Uchida}, \&
  {Pearce}}]{chauvin18a}
{Chauvin}, M., {Flor{\'e}n}, H.~G., {Friis}, M., {$et~al$.} 2018{\natexlab{a}},
  Nature Astronomy, 2, 652

\bibitem[{{Chauvin} {$et~al$.}(2018{\natexlab{b}}){Chauvin}, {Flor{\'e}n},
  {Friis}, {Jackson}, {Kamae}, {Kataoka}, {Kawano}, {Kiss}, {Mikhalev},
  {Mizuno}, {Tajima}, {Takahashi}, {Uchida}, \& {Pearce}}]{chauvin18bcrab}
---. 2018{\natexlab{b}}, \mnras, 477, L45

\bibitem[{{Chauvin} {$et~al$.}(2018{\natexlab{c}}){Chauvin}, {Flor{\'e}n},
  {Friis}, {Jackson}, {Kamae}, {Kataoka}, {Kawano}, {Kiss}, {Mikhalev},
  {Mizuno}, {Tajima}, {Takahashi}, {Uchida}, \& {Pearce}}]{chauvin18}
---. 2018{\natexlab{c}}, \mnras, 477, L45

\bibitem[{{Chauvin} {$et~al$.}(2019){Chauvin}, {Flor{\'e}n}, {Jackson},
  {Kamae}, {Kataoka}, {Kiss}, {Mikhalev}, {Mizuno}, {Takahashi}, {Uchida}, \&
  {Pearce}}]{chauvin18b}
{Chauvin}, M., {Flor{\'e}n}, H.-G., {Jackson}, M., {$et~al$.} 2019, \mnras,
  483, L138

\bibitem[{{Cheng} {$et~al$.}(2000){Cheng}, {Ruderman}, \& {Zhang}}]{cheng00}
{Cheng}, K.~S., {Ruderman}, M., \& {Zhang}, L. 2000, Astrophysical Journal,
  537, 964

\bibitem[{{Coburn} \& {Boggs}(2003)}]{coburn03}
{Coburn}, W., \& {Boggs}, S.~E. 2003, Nature, 423, 415

\bibitem[{{Costa} {$et~al$.}(2001){Costa}, {Soffitta}, {Bellazzini}, {Brez},
  {Lumb}, \& {Spandre}}]{costa01}
{Costa}, E., {Soffitta}, P., {Bellazzini}, R., {$et~al$.} 2001, Nature, 411,
  662

\bibitem[{{Covino} \& {Gotz}(2016)}]{covino16}
{Covino}, S., \& {Gotz}, D. 2016, Astronomical and Astrophysical Transactions,
  29, 205

\bibitem[{{Daugherty} \& {Harding}(1982)}]{daugherty82}
{Daugherty}, J.~K., \& {Harding}, A.~K. 1982, Astrophysical Journal, 252, 337

\bibitem[{De~Angelis {$et~al$.}(2018)De~Angelis, Tatischeff, Grenier, McEnery,
  Mallamaci, Tavani, Oberlack, Hanlon, Walter, Argan, \& et~al.}]{de18}
De~Angelis, A., Tatischeff, V., Grenier, I., {$et~al$.} 2018, Journal of High
  Energy Astrophysics, 19, 1–106

\bibitem[{{Dean} {$et~al$.}(2008){Dean}, {Clark}, {Stephen}, {McBride},
  {Bassani}, {Bazzano}, {Bird}, {Hill}, {Shaw}, \& {Ubertini}}]{dean08}
{Dean}, A.~J., {Clark}, D.~J., {Stephen}, J.~B., {$et~al$.} 2008, Science, 321,
  1183

\bibitem[{den Hartog {$et~al$.}(2008)den Hartog, Kuiper, Hermsen, Kaspi, Dib,
  Knoedlseder, \& Gavriil}]{hartog08}
den Hartog, P.~R., Kuiper, L., Hermsen, W., {$et~al$.} 2008, Detailed
  high-energy characteristics of AXP 4U 0142+61 - Multi-year observations with
  INTEGRAL, RXTE, XMM-Newton and ASCA, arXiv:0804.1640

\bibitem[{{Dergachev} {$et~al$.}(2009){Dergachev}, {Matveev}, {Kruglov},
  {Lazutkov}, {Savchenko}, {Skorodumov}, {Pyatigorsky}, {Chichikaluk},
  {Shishov}, {Khmylko}, {Vasiliev}, {Dranevich}, {Krut'Kov}, {Stepanov},
  {Kotov}, {Yurov}, {Glyanenko}, {Arkhangelsky}, {Gorelyi}, \&
  {Rubtsov}}]{dergachev09}
{Dergachev}, V.~A., {Matveev}, G.~A., {Kruglov}, E.~M., {$et~al$.} 2009,
  Bulletin of the Russian Academy of Sciences, Physics, 73, 419

\bibitem[{{Dov{\v c}iak} {$et~al$.}(2008){Dov{\v c}iak}, {Muleri}, {Goosmann},
  {Karas}, \& {Matt}}]{dovciak08}
{Dov{\v c}iak}, M., {Muleri}, F., {Goosmann}, R.~W., {Karas}, V., \& {Matt}, G.
  2008, \mnras, 391, 32

\bibitem[{{Duncan} \& {Thompson}(1992)}]{duncan92}
{Duncan}, R.~C., \& {Thompson}, C. 1992, Astrophysical Journal Letter, 392, L9

\bibitem[{{Dyks} \& {Rudak}(2003)}]{dyks03}
{Dyks}, J., \& {Rudak}, B. 2003, Astrophysical Journal, 598, 1201

\bibitem[{{Elsner} {$et~al$.}(1997){Elsner}, {Ramsey}, {O'dell}, {Sulkanen},
  {Tennant}, {Weisskopf}, {Gunji}, {Minamitani}, {Austin}, {Kolodziejczak},
  {Swartz}, {Garmire}, {Meszaros}, \& {Pavlov}}]{elsner97}
{Elsner}, R.~F., {Ramsey}, B.~D., {O'dell}, S.~L., {$et~al$.} 1997, in Bulletin
  of the American Astronomical Society, Vol.~29, American Astronomical Society
  Meeting Abstracts \#190, 790

\bibitem[{{Emslie} \& {Brown}(1980)}]{emslie80}
{Emslie}, A.~G., \& {Brown}, J.~C. 1980, Astrophysical Journal, 237, 1015

\bibitem[{Enoto {$et~al$.}(2017)Enoto, Shibata, Kitaguchi, Suwa, Uchide,
  Nishioka, Kisaka, Nakano, Murakami, \& Makishima}]{enoto17}
Enoto, T., Shibata, S., Kitaguchi, T., {$et~al$.} 2017, Magnetar Broadband
  X-ray Spectra Correlated with Magnetic Fields: Suzaku Archive of SGRs and
  AXPs Combined with NuSTAR, Swift, and RXTE, arXiv:1704.07018

\bibitem[{Esposito {$et~al$.}(2018)Esposito, Rea, \& Israel}]{esposito18}
Esposito, P., Rea, N., \& Israel, G.~L. 2018, Magnetars: a short review and
  some sparse considerations, arXiv:1803.05716

\bibitem[{{Fabiani}(2018)}]{fabiani18}
{Fabiani}, S. 2018, Galaxies, 6, 54

\bibitem[{{Fabiani} {$et~al$.}(2013){Fabiani}, {Campana}, {Costa}, {Del Monte},
  {Muleri}, {Rubini}, \& {Soffitta}}]{fabiani13}
{Fabiani}, S., {Campana}, R., {Costa}, E., {$et~al$.} 2013, Astroparticle
  Physics, 44, 91

\bibitem[{{Feng} {$et~al$.}(2020){Feng}, {Li}, {Long}, {Bellazzini}, {Costa},
  {Wu}, {Huang}, {Jiang}, {Minuti}, {Wang}, {Xu}, {Yang}, {Baldini}, {Citraro},
  {Nasimi}, {Soffitta}, {Muleri}, {Jung}, {Yu}, {Jin}, {Zeng}, {An}, {Brez},
  {Latronico}, {Sgro}, {Spandre}, \& {Pinchera}}]{feng20}
{Feng}, H., {Li}, H., {Long}, X., {$et~al$.} 2020, Nature Astronomy, 4, 511

\bibitem[{{Forot} {$et~al$.}(2008){Forot}, {Laurent}, {Grenier},
  {Gouiff{\`e}s}, \& {Lebrun}}]{forot08}
{Forot}, M., {Laurent}, P., {Grenier}, I.~A., {Gouiff{\`e}s}, C., \& {Lebrun},
  F. 2008, Astrophysical Journal Letter, 688, L29

\bibitem[{{Friis} {$et~al$.}(2018){Friis}, {Kiss}, {Mikhalev}, {Pearce}, \&
  {Takahashi}}]{friis18}
{Friis}, M., {Kiss}, M., {Mikhalev}, V., {Pearce}, M., \& {Takahashi}, H. 2018,
  Galaxies, 6, 30

\bibitem[{{Gehrels} \& {M{\'e}sz{\'a}ros}(2012)}]{gehrels12}
{Gehrels}, N., \& {M{\'e}sz{\'a}ros}, P. 2012, Science, 337, 932

\bibitem[{{Gehrels} {$et~al$.}(2004){Gehrels}, {Chincarini}, {Giommi}, {Mason},
  {Nousek}, {Wells}, {White}, {Barthelmy}, {Burrows}, {Cominsky}, {Hurley},
  {Marshall}, {M{\'e}sz{\'a}ros}, {Roming}, {Angelini}, {Barbier}, {Belloni},
  {Campana}, {Caraveo}, {Chester}, {Citterio}, {Cline}, {Cropper}, {Cummings},
  {Dean}, {Feigelson}, {Fenimore}, {Frail}, {Fruchter}, {Garmire}, {Gendreau},
  {Ghisellini}, {Greiner}, {Hill}, {Hunsberger}, {Krimm}, {Kulkarni}, {Kumar},
  {Lebrun}, {Lloyd-Ronning}, {Markwardt}, {Mattson}, {Mushotzky}, {Norris},
  {Osborne}, {Paczynski}, {Palmer}, {Park}, {Parsons}, {Paul}, {Rees},
  {Reynolds}, {Rhoads}, {Sasseen}, {Schaefer}, {Short}, {Smale}, {Smith},
  {Stella}, {Tagliaferri}, {Takahashi}, {Tashiro}, {Townsley}, {Tueller},
  {Turner}, {Vietri}, {Voges}, {Ward}, {Willingale}, {Zerbi}, \&
  {Zhang}}]{gehrels04}
{Gehrels}, N., {Chincarini}, G., {Giommi}, P., {$et~al$.} 2004, \apj, 611, 1005

\bibitem[{{Ghisellini} \& {Celotti}(1999)}]{ghisellini99}
{Ghisellini}, G., \& {Celotti}, A. 1999, \apjl, 511, L93

\bibitem[{Ghisellini \& Lazzati(1999)}]{ghisellini_lazzati99}
Ghisellini, G., \& Lazzati, D. 1999, Monthly Notices of the Royal Astronomical
  Society, 309, L7

\bibitem[{{Ghisellini} {$et~al$.}(2000){Ghisellini}, {Lazzati}, {Celotti}, \&
  {Rees}}]{ghisellini00}
{Ghisellini}, G., {Lazzati}, D., {Celotti}, A., \& {Rees}, M.~J. 2000, \mnras,
  316, L45

\bibitem[{{Gill} {$et~al$.}(2018){Gill}, {Granot}, \& {Kumar}}]{gill18}
{Gill}, R., {Granot}, J., \& {Kumar}, P. 2018, arXiv e-prints, arXiv:1811.11555

\bibitem[{{Goosmann} \& {Matt}(2011)}]{goosmann11}
{Goosmann}, R.~W., \& {Matt}, G. 2011, \mnras, 415, 3119

\bibitem[{{G{\"o}tz} {$et~al$.}(2013){G{\"o}tz}, {Covino},
  {Fern{\'a}ndez-Soto}, {Laurent}, \& {Bo{\v s}njak}}]{gotz13}
{G{\"o}tz}, D., {Covino}, S., {Fern{\'a}ndez-Soto}, A., {Laurent}, P., \&
  {Bo{\v s}njak}, {\v Z}. 2013, \mnras, 431, 3550

\bibitem[{{G{\"o}tz} {$et~al$.}(2014){G{\"o}tz}, {Laurent}, {Antier}, {Covino},
  {D'Avanzo}, {D'Elia}, \& {Melandri}}]{gotz14}
{G{\"o}tz}, D., {Laurent}, P., {Antier}, S., {$et~al$.} 2014, \mnras, 444, 2776

\bibitem[{{G{\"o}tz} {$et~al$.}(2009){G{\"o}tz}, {Laurent}, {Lebrun}, {Daigne},
  \& {Bo{\v s}njak}}]{gotz09}
{G{\"o}tz}, D., {Laurent}, P., {Lebrun}, F., {Daigne}, F., \& {Bo{\v s}njak},
  {\v Z}. 2009, Astrophysical Journal Letter, 695, L208

\bibitem[{{Gowen} {$et~al$.}(1977){Gowen}, {Cooke}, {Griffiths}, \&
  {Ricketts}}]{gowen77}
{Gowen}, R.~A., {Cooke}, B.~A., {Griffiths}, R.~E., \& {Ricketts}, M.~J. 1977,
  \mnras, 179, 303

\bibitem[{Goyal {$et~al$.}(2018)Goyal, Naik, Mithun, Vadawale, Nagrani, Madhvi,
  Tiwari, Ladiya, Patel, Adalja, Chattopadhyay, Shanmugam, Auknoor, Sharma, \&
  Patel}]{goyal18}
Goyal, S.~K., Naik, A.~P., Mithun, N. P.~S., {$et~al$.} 2018, Journal of
  Astronomical Telescopes, Instruments, and Systems, 5, 1

\bibitem[{{Granot} \& {K{\"o}nigl}(2003)}]{granot03}
{Granot}, J., \& {K{\"o}nigl}, A. 2003, Astrophysical Journal Letter, 594, L83

\bibitem[{{Griffith} {$et~al$.}(2016){Griffith}, {Falcone}, {Prieskorn}, \&
  {Burrows}}]{griffith16}
{Griffith}, C.~V., {Falcone}, A.~D., {Prieskorn}, Z.~R., \& {Burrows}, D.~N.
  2016, Journal of Astronomical Telescopes, Instruments, and Systems, 2, 016001

\bibitem[{{Griffiths} {$et~al$.}(1976){Griffiths}, {Ricketts}, \&
  {Cooke}}]{griffiths76}
{Griffiths}, R.~E., {Ricketts}, M.~J., \& {Cooke}, B.~A. 1976, \mnras, 177, 429

\bibitem[{Gruner {$et~al$.}(2002)Gruner, Tate, \& Eikenberry}]{gruner02_ccd}
Gruner, S.~M., Tate, M.~W., \& Eikenberry, E.~F. 2002, Review of Scientific
  Instruments, 73, 2815

\bibitem[{{Gunji} {$et~al$.}(2008){Gunji}, {Kishimoto}, {Sakurai}, {Tokanai},
  {Kanno}, {Ishikawa}, {Hayashida}, {Anabuke}, {Tsunemi}, {Mihara}, {Kohama},
  {Suzuki}, \& {Saito}}]{gunji08}
{Gunji}, S., {Kishimoto}, Y., {Sakurai}, H., {$et~al$.} 2008, in Polarimetry
  days in Rome: Crab status, theory and prospects, 5

\bibitem[{{Guo} {$et~al$.}(2013){Guo}, {Beilicke}, {Garson}, {Kislat},
  {Fleming}, \& {Krawczynski}}]{guo13}
{Guo}, Q., {Beilicke}, M., {Garson}, A., {$et~al$.} 2013, Astroparticle
  Physics, 41, 63

\bibitem[{{Guo} {$et~al$.}(2011){Guo}, {Garson}, {Beilicke}, {Martin}, {Lee},
  \& {Krawczynski}}]{guo11_2}
{Guo}, Q., {Garson}, A., {Beilicke}, M., {$et~al$.} 2011, ArXiv e-prints,
  arXiv:1101.0595

\bibitem[{{Hahn} {$et~al$.}(2016){Hahn}, {Weber}, {M{\"a}rtin}, {H{\"o}fer},
  {K{\"a}mpfer}, \& {St{\"o}hlker}}]{hahn16}
{Hahn}, C., {Weber}, G., {M{\"a}rtin}, R., {$et~al$.} 2016, Review of
  Scientific Instruments, 87, 043106

\bibitem[{Harding(2019)}]{Harding19_book}
Harding, A.~K. 2019, Multi-Wavelength Polarimetry of Isolated Pulsars, ed.
  R.~Mignani, A.~Shearer, A.~S{\l}owikowska, \& S.~Zane (Cham: Springer
  International Publishing), 277--299

\bibitem[{{Harrison} {$et~al$.}(2013){Harrison}, {Craig}, {Christensen},
  {Hailey}, {Zhang}, {Boggs}, {Stern}, {Cook}, {Forster}, {Giommi},
  {Grefenstette}, {Kim}, {Kitaguchi}, {Koglin}, {Madsen}, {Mao}, {Miyasaka},
  {Mori}, {Perri}, {Pivovaroff}, {Puccetti}, {Rana}, {Westergaard}, {Willis},
  {Zoglauer}, {An}, {Bachetti}, {Barri{\`e}re}, {Bellm}, {Bhalerao},
  {Brejnholt}, {Fuerst}, {Liebe}, {Markwardt}, {Nynka}, {Vogel}, {Walton},
  {Wik}, {Alexander}, {Cominsky}, {Hornschemeier}, {Hornstrup}, {Kaspi},
  {Madejski}, {Matt}, {Molendi}, {Smith}, {Tomsick}, {Ajello}, {Ballantyne},
  {Balokovi{\'c}}, {Barret}, {Bauer}, {Blandford}, {Brandt}, {Brenneman},
  {Chiang}, {Chakrabarty}, {Chenevez}, {Comastri}, {Dufour}, {Elvis}, {Fabian},
  {Farrah}, {Fryer}, {Gotthelf}, {Grindlay}, {Helfand}, {Krivonos}, {Meier},
  {Miller}, {Natalucci}, {Ogle}, {Ofek}, {Ptak}, {Reynolds}, {Rigby},
  {Tagliaferri}, {Thorsett}, {Treister}, \& {Urry}}]{harrison13}
{Harrison}, F.~A., {Craig}, W.~W., {Christensen}, F.~E., {$et~al$.} 2013,
  Astrophysical Journal, 770, 103

\bibitem[{{Hayashida} {$et~al$.}(2014){Hayashida}, {Yonetoku}, {Gunji},
  {Tamagawa}, {Mihara}, {Mizuno}, {Takahashi}, {Dotani}, {Kubo}, {Yatsu},
  {Tokanai}, {Nakamori}, {Shibata}, {Hayato}, {Furuzawa}, {Kishimoto},
  {Kitamoto}, {Toma}, {Sadamoto}, {Yoshinaga}, {Kim}, {Ide}, {Kamitsukasa},
  {Anabuki}, {Tsunemi}, {Katagiri}, \& {Sugimoto}}]{hayashida14}
{Hayashida}, K., {Yonetoku}, D., {Gunji}, S., {$et~al$.} 2014, in Society of
  Photo-Optical Instrumentation Engineers (SPIE) Conference Series, Vol. 9144,
  Society of Photo-Optical Instrumentation Engineers (SPIE) Conference Series,
  0

\bibitem[{{Heitler}(1954)}]{heitler54}
{Heitler}, W. 1954, in International Series of Monographs on Physics, Oxford:
  Clarendon, 1954, 3rd ed.

\bibitem[{{Heyl} \& {Shaviv}(2002)}]{heyl02}
{Heyl}, J.~S., \& {Shaviv}, N.~J. 2002, Physical Review D, 66, 023002

\bibitem[{{Hill} {$et~al$.}(1997){Hill}, {Holland}, {Castelli}, {Short},
  {Turner}, \& {Burt}}]{hill97}
{Hill}, J.~E., {Holland}, A.~D., {Castelli}, C.~M., {$et~al$.} 1997, in Society
  of Photo-Optical Instrumentation Engineers (SPIE) Conference Series, Vol.
  3114, EUV, X-Ray, and Gamma-Ray Instrumentation for Astronomy VIII, ed. O.~H.
  {Siegmund} \& M.~A. {Gummin}, 241--249

\bibitem[{{Hong} {$et~al$.}(2009){Hong}, {Allen}, {Grindlay}, {Chammas},
  {Barthelemy}, {Baker}, {Gehrels}, {Nelson}, {Labov}, {Collins}, {Cook},
  {McLean}, \& {Harrison}}]{hong09}
{Hong}, J., {Allen}, B., {Grindlay}, J., {$et~al$.} 2009, Nuclear Instruments
  and Methods in Physics Research A, 605, 364

\bibitem[{{Hughes} {$et~al$.}(1984){Hughes}, {Long}, \& {Novick}}]{hughes84}
{Hughes}, J.~P., {Long}, K.~S., \& {Novick}, R. 1984, Astrophysical Journal,
  280, 255

\bibitem[{{Hull} {$et~al$.}(2017){Hull}, {Falcone}, {Burrows}, {Wages},
  {Chattopadhyay}, {McQuaide}, {Bray}, \& {Kern}}]{hull17}
{Hull}, S.~V., {Falcone}, A.~D., {Burrows}, D.~N., {$et~al$.} 2017, 10397,
  1039704

\bibitem[{ichiro {$et~al$.}(2008)ichiro, Takahashi, Watanabe, Tajima, Tanaka,
  Nakazawa, \& Fukazawa}]{ichiro08_DSSD}
ichiro, S., Takahashi, T., Watanabe, S., {$et~al$.} 2008, Spie Newsroom

\bibitem[{Iyudin {$et~al$.}(2009)Iyudin, Bogomolov, Svertilov, Yashin, Klassen,
  Shmurak, \& Orlov}]{iyudin09}
Iyudin, A., Bogomolov, V., Svertilov, S., {$et~al$.} 2009, Instruments and
  Experimental Techniques - INSTRUM EXP TECH-ENGL TR, 52, 774

\bibitem[{{Iyyani} {$et~al$.}(2015){Iyyani}, {Ryde}, {Ahlgren}, {Burgess},
  {Larsson}, {Pe'er}, {Lundman}, {Axelsson}, \& {McGlynn}}]{iyyani15}
{Iyyani}, S., {Ryde}, F., {Ahlgren}, B., {$et~al$.} 2015, \mnras, 450, 1651

\bibitem[{{Jahoda}(2010)}]{jahoda10}
{Jahoda}, K. 2010, in Society of Photo-Optical Instrumentation Engineers (SPIE)
  Conference Series, Vol. 7732, Society of Photo-Optical Instrumentation
  Engineers (SPIE) Conference Series

\bibitem[{{Jeffrey} \& {Kontar}(2011)}]{jeffrey11}
{Jeffrey}, N.~L.~S., \& {Kontar}, E.~P. 2011, Astronomy and Astrophysics, 536,
  A93

\bibitem[{Jourdain \& Roques(2019)}]{Jourdain19}
Jourdain, E., \& Roques, J.-P. 2019, The Astrophysical Journal, 882, 129

\bibitem[{{Jourdain} {$et~al$.}(2012){Jourdain}, {Roques}, {Chauvin}, \&
  {Clark}}]{jourdain12}
{Jourdain}, E., {Roques}, J.~P., {Chauvin}, M., \& {Clark}, D.~J. 2012,
  Astrophysical Journal, 761, 27

\bibitem[{{Kaaret}(2014)}]{kaaret14}
{Kaaret}, P. 2014, arXiv:1408.5899

\bibitem[{{Kaaret} {$et~al$.}(1994){Kaaret}, {Schwartz}, {Soffitta}, {Dwyer},
  {Shaw}, {Hanany}, {Novick}, {Sunyaev}, {Lapshov}, {Silver}, {Ziock},
  {Weisskopf}, {Elsner}, {Ramsey}, {Costa}, {Rubini}, {Feroci}, {Piro},
  {Manzo}, {Giarrusso}, {Santangelo}, {Scarsi}, {Perola}, {Massaro}, \&
  {Matt}}]{kaaret93}
{Kaaret}, P.~E., {Schwartz}, J., {Soffitta}, P., {$et~al$.} 1994, in Society of
  Photo-Optical Instrumentation Engineers (SPIE) Conference Series, Vol. 2010,
  X-Ray and Ultraviolet Polarimetry, ed. S.~{Fineschi}, 22--27

\bibitem[{{Kalemci} {$et~al$.}(2007){Kalemci}, {Boggs}, {Kouveliotou},
  {Finger}, \& {Baring}}]{kalemci07}
{Kalemci}, E., {Boggs}, S.~E., {Kouveliotou}, C., {Finger}, M., \& {Baring},
  M.~G. 2007, Astrophysical Journals, 169, 75

\bibitem[{{Kalemci} \& {Matteson}(2002)}]{kalemci02}
{Kalemci}, E., \& {Matteson}, J.~L. 2002, Nuclear Instruments and Methods in
  Physics Research A, 478, 527

\bibitem[{Kantzas {$et~al$.}(2020)Kantzas, Markoff, Beuchert, Lucchini,
  Chhotray, Ceccobello, Tetarenko, Miller-Jones, Bremer, Garcia, \&
  et~al.}]{Kantzas20}
Kantzas, D., Markoff, S., Beuchert, T., {$et~al$.} 2020, Monthly Notices of the
  Royal Astronomical Society, 500, 2112–2126

\bibitem[{Kaspi \& Beloborodov(2017)}]{Kaspi17}
Kaspi, V.~M., \& Beloborodov, A.~M. 2017, Annual Review of Astronomy and
  Astrophysics, 55, 261–301

\bibitem[{{Kim} {$et~al$.}(2011){Kim}, {Anderson}, {Kaye}, {Zhang}, {Zhu},
  {Kaye}, \& {He}}]{kim11}
{Kim}, J.~C., {Anderson}, S.~E., {Kaye}, W., {$et~al$.} 2011, Nuclear
  Instruments and Methods in Physics Research A, 654, 233

\bibitem[{Kim {$et~al$.}(2019)Kim, Lee, \& Lee}]{kim19_3dCZT}
Kim, Y., Lee, T., \& Lee, W. 2019, Nuclear Engineering and Technology, 51, 1417

\bibitem[{{Kirk} {$et~al$.}(2002){Kirk}, {Skj{\ae}raasen}, \&
  {Gallant}}]{kirk02}
{Kirk}, J.~G., {Skj{\ae}raasen}, O., \& {Gallant}, Y.~A. 2002, \aap, 388, L29

\bibitem[{{Kishimoto} {$et~al$.}(2007){Kishimoto}, {Gunji}, {Ishigaki},
  {Kanno}, {Murayama}, {Ito}, {Tokanai}, {Suzuki}, {Sakurai}, {Mihara},
  {Kohama}, {Suzuki}, {Hayato}, {Hayashida}, {Anabuki}, {Morimoto}, {Tsunemi},
  {Saito}, {Yamagami}, \& {Kishimoto}}]{kishimoto07}
{Kishimoto}, Y., {Gunji}, S., {Ishigaki}, Y., {$et~al$.} 2007, IEEE
  Transactions on Nuclear Science, 54, 561

\bibitem[{{Kislat} {$et~al$.}(2019){Kislat}, {Abarr}, {Bose}, {Braun}, {De
  Geronimo}, {Dowkontt}, {Errando}, {Gadson}, {Guarino}, {Heatwole}, {Iyer},
  {Kiss}, {Kitaguchi}, {Krawczynski}, {Kushwah}, {Lanzi}, {Li}, {Lisalda},
  {Okajima}, {Pearce}, {Peterson}, {Rauch}, {Stuchlik}, {Takahashi}, {Uchida},
  \& {West}}]{kislat19}
{Kislat}, F., {Abarr}, Q., {Bose}, R., {$et~al$.} 2019, in AAS/High Energy
  Astrophysics Division, AAS/High Energy Astrophysics Division, 109.78

\bibitem[{{Kole}(2019)}]{kole19}
{Kole}, M. 2019, in International Cosmic Ray Conference, Vol.~36, 36th
  International Cosmic Ray Conference (ICRC2019), 572

\bibitem[{Kole {$et~al$.}(2020)Kole, De~Angelis, Berlato, Burgess, Gauvin,
  Greiner, Hajdas, Li, Li, Pollo, \& et~al.}]{Kole20polar_catalog}
Kole, M., De~Angelis, N., Berlato, F., {$et~al$.} 2020, Astronomy and
  Astrophysics, 644, A124

\bibitem[{{Kotov} {$et~al$.}(2016){Kotov}, {Yurov}, {Glyanenko}, {Lupar},
  {Kochemasov}, {Trofimov}, {Zakharov}, {Faradzhaev}, {Tyshkevich}, {Rubtsov},
  {Dergachev}, {Kruglov}, {Lazutkov}, {Savchenko}, \& {Skorodumov}}]{kotov16}
{Kotov}, Y.~D., {Yurov}, V.~N., {Glyanenko}, A.~S., {$et~al$.} 2016, Advances
  in Space Research, 58, 635

\bibitem[{{Kotthaus} {$et~al$.}(1998){Kotthaus}, {Buschhorn}, {Rzepka},
  {Schmidt}, \& {Weinmann}}]{schmidt98}
{Kotthaus}, R., {Buschhorn}, G., {Rzepka}, M., {Schmidt}, K.~H., \& {Weinmann},
  P.~M. 1998, in Society of Photo-Optical Instrumentation Engineers (SPIE)
  Conference Series, Vol. 3443, X-Ray and Ultraviolet Spectroscopy and
  Polarimetry II, ed. S.~{Fineschi}, 105--116

\bibitem[{{Krawczynski} {$et~al$.}(2011){Krawczynski}, {Garson}, {Guo},
  {Baring}, {Ghosh}, {Beilicke}, \& {Lee}}]{krawczynski11}
{Krawczynski}, H., {Garson}, A., {Guo}, Q., {$et~al$.} 2011, Astroparticle
  Physics, 34, 550

\bibitem[{{Kumar} \& {Zhang}(2015)}]{kumar15}
{Kumar}, P., \& {Zhang}, B. 2015, \physrep, 561, 1

\bibitem[{{Kunieda} {$et~al$.}(2010){Kunieda}, {Awaki}, {Furuzawa}, {Haba},
  {Iizuka}, {Ishibashi}, {Ishida}, {Itoh}, {Kosaka}, {Maeda}, {Matsumoto},
  {Miyazawa}, {Mori}, {Namba}, {Ogasaka}, {Ogi}, {Okajima}, {Suzuki}, {Tamura},
  {Tawara}, {Uesugi}, {Yamashita}, \& {Yamauchi}}]{kunieda10}
{Kunieda}, H., {Awaki}, H., {Furuzawa}, A., {$et~al$.} 2010, in Society of
  Photo-Optical Instrumentation Engineers (SPIE) Conference Series, Vol. 7732,
  Society of Photo-Optical Instrumentation Engineers (SPIE) Conference Series

\bibitem[{Kuvvetli {$et~al$.}(2014)Kuvvetli, Budtz-Jørgensen, Zappettini,
  Zambelli, Benassi, Kalemci, Caroli, Stephen, \& Auricchio}]{kuvvetli14_3dCZT}
Kuvvetli, I., Budtz-Jørgensen, C., Zappettini, A., {$et~al$.} 2014, in High
  Energy, Optical, and Infrared Detectors for Astronomy VI, ed. A.~D. Holland
  \& J.~Beletic, Vol. 9154, International Society for Optics and Photonics
  (SPIE), 272 -- 281

\bibitem[{Laurent {$et~al$.}(2017)Laurent, Gouiffes, Rodriguez, \&
  Chambouleyron}]{laurent17}
Laurent, P., Gouiffes, C., Rodriguez, J., \& Chambouleyron, V. 2017, PoS,
  INTEGRAL2016, 022

\bibitem[{{Laurent} {$et~al$.}(2011){Laurent}, {Rodriguez}, {Wilms}, {Cadolle
  Bel}, {Pottschmidt}, \& {Grinberg}}]{laurent11}
{Laurent}, P., {Rodriguez}, J., {Wilms}, J., {$et~al$.} 2011, Science, 332, 438

\bibitem[{{Lazzati} {$et~al$.}(2004){Lazzati}, {Rossi}, {Ghisellini}, \&
  {Rees}}]{lazzati04}
{Lazzati}, D., {Rossi}, E., {Ghisellini}, G., \& {Rees}, M.~J. 2004, \mnras,
  347, L1

\bibitem[{{Leach} {$et~al$.}(1985){Leach}, {Emslie}, \& {Petrosian}}]{leach83}
{Leach}, J., {Emslie}, A.~G., \& {Petrosian}, V. 1985, \solphys, 96, 331

\bibitem[{{Lei} {$et~al$.}(1997){Lei}, {Dean}, \& {Hills}}]{lei97}
{Lei}, F., {Dean}, A.~J., \& {Hills}, G.~L. 1997, Space Science Reviews, 82,
  309

\bibitem[{Lesser(2015)}]{Lesser15_ccd}
Lesser, M. 2015, Publications of the Astronomical Society of the Pacific, 127,
  1097

\bibitem[{{Llopart} {$et~al$.}(2007){Llopart}, {Ballabriga}, {Campbell},
  {Tlustos}, \& {Wong}}]{llopart07}
{Llopart}, X., {Ballabriga}, R., {Campbell}, M., {Tlustos}, L., \& {Wong}, W.
  2007, Nuclear Instruments and Methods in Physics Research A, 581, 485

\bibitem[{{Long} {$et~al$.}(1980){Long}, {Chanan}, \& {Novick}}]{long80}
{Long}, K.~S., {Chanan}, G.~A., \& {Novick}, R. 1980, \apj, 238, 710

\bibitem[{{Lowell} {$et~al$.}(2017){Lowell}, {Boggs}, {Chiu}, {Kierans},
  {Sleator}, {Tomsick}, {Zoglauer}, {Chang}, {Tseng}, {Yang}, {Jean}, {von
  Ballmoos}, {Lin}, \& {Amman}}]{lowell17}
{Lowell}, A.~W., {Boggs}, S.~E., {Chiu}, C.~L., {$et~al$.} 2017, \apj, 848, 120

\bibitem[{{Lumb} {$et~al$.}(1991){Lumb}, {Berthiaume}, {Burrows}, {Garmire}, \&
  {Nousek}}]{lumb91_ccd}
{Lumb}, D.~H., {Berthiaume}, G.~D., {Burrows}, D.~N., {Garmire}, G.~P., \&
  {Nousek}, J.~A. 1991, Experimental Astronomy, 2, 179

\bibitem[{{Lyutikov} {$et~al$.}(2003){Lyutikov}, {Pariev}, \&
  {Blandford}}]{lyutikov03}
{Lyutikov}, M., {Pariev}, V.~I., \& {Blandford}, R.~D. 2003, Astrophysical
  Journal, 597, 998

\bibitem[{{Madsen} {$et~al$.}(2015){Madsen}, {Reynolds}, {Harrison}, {An},
  {Boggs}, {Christensen}, {Craig}, {Fryer}, {Grefenstette}, {Hailey},
  {Markwardt}, {Nynka}, {Stern}, {Zoglauer}, \& {Zhang}}]{madsen15}
{Madsen}, K.~K., {Reynolds}, S., {Harrison}, F., {$et~al$.} 2015, \apj, 801, 66

\bibitem[{{Marin}(2018)}]{marin18}
{Marin}, F. 2018, Galaxies, 6, 38

\bibitem[{{Markoff} {$et~al$.}(2001){Markoff}, {Falcke}, \&
  {Fender}}]{markoff01}
{Markoff}, S., {Falcke}, H., \& {Fender}, R. 2001, Astronomy and Astrophysics,
  372, L25

\bibitem[{Marshall {$et~al$.}(2010)Marshall, Heilmann, Schulz, \&
  Murphy}]{marshall10}
Marshall, H.~L., Heilmann, R.~K., Schulz, N.~S., \& Murphy, K.~D. 2010, in
  Space Telescopes and Instrumentation 2010: Ultraviolet to Gamma Ray, ed.
  M.~Arnaud, S.~S. Murray, \& T.~Takahashi, Vol. 7732, International Society
  for Optics and Photonics (SPIE), 103 -- 112

\bibitem[{{Marshall} {$et~al$.}(2003){Marshall}, {Murray}, {Chappell},
  {Schnopper}, {Silver}, \& {Weisskopf}}]{marshall03}
{Marshall}, H.~L., {Murray}, S.~S., {Chappell}, J.~H., {$et~al$.} 2003, in
  Society of Photo-Optical Instrumentation Engineers (SPIE) Conference Series,
  Vol. 4843, Polarimetry in Astronomy, ed. S.~{Fineschi}, 360--371

\bibitem[{Marshall {$et~al$.}(2018)Marshall, Günther, Heilmann, Schulz, Egan,
  Hellickson, Heine, Windt, Gullikson, Ramsey, Tagliaferri, \&
  Pareschi}]{marshall18}
Marshall, H.~L., Günther, H.~M., Heilmann, R.~K., {$et~al$.} 2018, Journal of
  Astronomical Telescopes, Instruments, and Systems, 4, 1

\bibitem[{{McConnell} {$et~al$.}(2019){McConnell}, {Ajello}, {Baring},
  {Bloser}, {Chattopadhyay}, {da Silva}, {Guirec}, {Hartmann}, {Li}, {Lowell},
  {Prescod-Weinstein}, {Rani}, {Tatischeff}, {Tomsick}, {van der Horst},
  {Vestrand}, {Wadiasingh}, {Zane}, {Zhang}, \& {Zhang}}]{mcconnell19}
{McConnell}, M., {Ajello}, M., {Baring}, M., {$et~al$.} 2019, \baas, 51, 100

\bibitem[{{McConnell}(2016)}]{mcconnell16}
{McConnell}, M.~L. 2016, ArXiv e-prints, arXiv:1611.06579

\bibitem[{{McConnell} {$et~al$.}(2009){McConnell}, {Bancroft}, {Bloser},
  {Connor}, {Legere}, \& {Ryan}}]{mcconnell09}
{McConnell}, M.~L., {Bancroft}, C., {Bloser}, P.~F., {$et~al$.} 2009, in
  Society of Photo-Optical Instrumentation Engineers (SPIE) Conference Series,
  Vol. 7435, Society of Photo-Optical Instrumentation Engineers (SPIE)
  Conference Series, 0

\bibitem[{McConnell {$et~al$.}(2018)McConnell, Bloser, Legere, Ryan, Hanlon,
  McBreen, \& Uliyanov}]{mcconnel18_grape}
McConnell, M.~L., Bloser, P.~F., Legere, J.~S., {$et~al$.} 2018, in Space
  Telescopes and Instrumentation 2018: Ultraviolet to Gamma Ray, ed. J.-W.~A.
  den Herder, S.~Nikzad, \& K.~Nakazawa, Vol. 10699, International Society for
  Optics and Photonics (SPIE), 659 -- 671

\bibitem[{{McConnell} \& {LEAP Collaboration}(2016)}]{mcconnel16_leap}
{McConnell}, M.~L., \& {LEAP Collaboration}. 2016, in Eighth Huntsville
  Gamma-Ray Burst Symposium, Vol. 1962, 4051

\bibitem[{{McConnell} {$et~al$.}(2007){McConnell}, {Ryan}, {Smith}, {Hurford},
  {Fivian}, {Lin}, {Emslie}, \& {Hajdas}}]{mcconnel07}
{McConnell}, M.~L., {Ryan}, J.~M., {Smith}, D.~M., {$et~al$.} 2007, in American
  Astronomical Society Meeting Abstracts, Vol. 210, American Astronomical
  Society Meeting Abstracts \#210, 93.01

\bibitem[{{McConnell} {$et~al$.}(2002){McConnell}, {Ryan}, {Smith}, {Lin}, \&
  {Emslie}}]{mcconnell02}
{McConnell}, M.~L., {Ryan}, J.~M., {Smith}, D.~M., {Lin}, R.~P., \& {Emslie},
  A.~G. 2002, Solar Physics, 210, 125

\bibitem[{McEnery {$et~al$.}(2019)McEnery, Barrio, Agudo, Ajello, Álvarez,
  Ansoldi, Anton, Auricchio, Stephen, Baldini, Bambi, Baring, Barres, Bastieri,
  Beacom, Beckmann, Bednarek, Bernard, Bissaldi, Bloser, Blumer, Boettcher,
  Boggs, Bolotnikov, Bottacini, Bozhilov, Bozzo, Briggs, Buckley, Burns, Buson,
  Campana, Caputo, Cardillo, Caroli, Castro, Cenko, Charles, Chen, Cheung,
  Ciprini, Coppi, da~Silva, Cutini, D'Ammando, Angelis, Becker, Nolfo, Sordo,
  Mauro, Venere, Dietrich, Digel, Dominguez, Doro, Ferrara, Fields, Finke,
  Foffano, Fryer, Fukazawa, Funk, Gasparrini, Gelfand, Georganopoulos,
  Giordano, Giuliani, Gouiffes, Grefenstette, Grenier, Griffin, Grove, Guiriec,
  Harding, Harding, Hartmann, Hays, Hernanz, Hewitt, Holder, Hui, Inglis,
  Johnson, Jones, Kanbach, Kargaltsev, Kaufmann, Kerr, Kierans, Kislat,
  Klimenko, Knodlseder, Kocveski, Kopp, Krawczynsiki, Krizmanic, Kubo, Neilson,
  Laurent, Lenain, Li, Lien, Linden, Lommler, Longo, Lovellette, López,
  Manousakis, Marcotulli, Marcowith, Martinez, McConnell, Metcalfe, Meyer,
  Meyer, Mignani, Mitchell, Mizuno, Moiseev, Morcuende, Moskalenko, Moss,
  Nakazawa, Mazziotta, Oberlack, Ohno, Oikonomou, Ojha, Omodei, Orlando, Otte,
  Paliya, Parker, Patricelli, Perkins, Petropoulou, Pittori, Pohl, Porter,
  Prandini, Prescod-Weinstein, Racusin, Rando, Rani, Ribó, Rodi,
  Sanchez-Conde, Parkinson, Schirato, Shawhan, Shrader, Smith, Smith, Stamerra,
  Stawarz, Strong, Stumke, Tajima, Takahashi, Tanaka, Tatischeff, The,
  Thompson, Tibaldo, Tomsick, Uhm, Venters, Vestrand, Vianello, Wadiasingh,
  Walter, Wang, Williams, Wilson-Hodge, Wood, Woolf, Wulf, Younes, Zampieri,
  Zane, Zhang, Zhang, Zimmer, Zoglauer, \& van~der Horst}]{amego19}
McEnery, J., Barrio, J.~A., Agudo, I., {$et~al$.} 2019, All-sky Medium Energy
  Gamma-ray Observatory: Exploring the Extreme Multimessenger Universe,
  arXiv:1907.07558

\bibitem[{{McGlynn} {$et~al$.}(2007{\natexlab{a}}){McGlynn}, {Clark}, {Dean},
  {Hanlon}, {McBreen}, {Willis}, {McBreen}, {Bird}, \& {Foley}}]{mcgltnn07}
{McGlynn}, S., {Clark}, D.~J., {Dean}, A.~J., {$et~al$.} 2007{\natexlab{a}},
  Astronomy and Astrophysics, 466, 895

\bibitem[{{McGlynn} {$et~al$.}(2007{\natexlab{b}}){McGlynn}, {Clark}, {Dean},
  {Hanlon}, {McBreen}, {Willis}, {McBreen}, {Bird}, \& {Foley}}]{mcglynn07}
---. 2007{\natexlab{b}}, Astronomy and Astrophysics, 466, 895

\bibitem[{{McGlynn} {$et~al$.}(2009){McGlynn}, {Foley}, {McBreen}, {Hanlon},
  {McBreen}, {Clark}, {Dean}, {Martin-Carrillo}, \& {O'Connor}}]{mcglynn09}
{McGlynn}, S., {Foley}, S., {McBreen}, B., {$et~al$.} 2009, Astronomy and
  Astrophysics, 499, 465

\bibitem[{{McNamara} {$et~al$.}(2009){McNamara}, {Kuncic}, \&
  {Wu}}]{mcnamara09}
{McNamara}, A.~L., {Kuncic}, Z., \& {Wu}, K. 2009, \mnras, 395, 1507

\bibitem[{{Medvedev}(2007)}]{medvedev07}
{Medvedev}, M.~V. 2007, \apss, 307, 245

\bibitem[{{Medvedev} \& {Loeb}(1999)}]{medvedev99}
{Medvedev}, M.~V., \& {Loeb}, A. 1999, The Astrophysical Journal, 526, 697

\bibitem[{{Meegan} {$et~al$.}(2009){Meegan}, {Lichti}, {Bhat}, {Bissaldi},
  {Briggs}, {Connaughton}, {Diehl}, {Fishman}, {Greiner}, {Hoover}, {van der
  Horst}, {von Kienlin}, {Kippen}, {Kouveliotou}, {McBreen}, {Paciesas},
  {Preece}, {Steinle}, {Wallace}, {Wilson}, \& {Wilson-Hodge}}]{meegan09}
{Meegan}, C., {Lichti}, G., {Bhat}, P.~N., {$et~al$.} 2009, \apj, 702, 791

\bibitem[{Mereghetti {$et~al$.}(2015)Mereghetti, Pons, \&
  Melatos}]{Mereghetti15}
Mereghetti, S., Pons, J.~A., \& Melatos, A. 2015, Space Science Reviews, 191,
  315–338

\bibitem[{{Meszaros} {$et~al$.}(1988){Meszaros}, {Novick}, {Szentgyorgyi},
  {Chanan}, \& {Weisskopf}}]{meszaros88}
{Meszaros}, P., {Novick}, R., {Szentgyorgyi}, A., {Chanan}, G.~A., \&
  {Weisskopf}, M.~C. 1988, Astrophysical Journal, 324, 1056

\bibitem[{{Meszaros} \& {Rees}(1993)}]{meszaros93}
{Meszaros}, P., \& {Rees}, M.~J. 1993, \apj, 405, 278

\bibitem[{Mi {$et~al$.}(2019)Mi, Nillius, Pearce, \& Danielsson}]{Mi19_spl}
Mi, W., Nillius, P., Pearce, M., \& Danielsson, M. 2019, Nature Astronomy, 3,
  867–872

\bibitem[{Michel \& Durst(2008)}]{michel08}
Michel, T., \& Durst, J. 2008, Nuclear Instruments and Methods in Physics
  Research Section A: Accelerators, Spectrometers, Detectors and Associated
  Equipment, 594, 188

\bibitem[{{Moran} {$et~al$.}(2013){Moran}, {Shearer}, {Gouiffes}, \&
  {Laurent}}]{moran13}
{Moran}, P., {Shearer}, A., {Gouiffes}, C., \& {Laurent}, P. 2013, ArXiv
  e-prints, arXiv:1302.3622

\bibitem[{Nakamura \& Shibata(2007)}]{nakamura07}
Nakamura, Y., \& Shibata, S. 2007, Monthly Notices of the Royal Astronomical
  Society, 381, 1489

\bibitem[{{Nakar} {$et~al$.}(2003){Nakar}, {Piran}, \& {Waxman}}]{nakar03}
{Nakar}, E., {Piran}, T., \& {Waxman}, E. 2003, Journal of Cosmology and
  Astroparticle Physics, 10, 005

\bibitem[{{Nandi} {$et~al$.}(2011){Nandi}, {Palit}, {Debnath}, {Chakrabarti},
  {Kotoch}, {Sarkar}, {Yadav}, {Girish}, {Rao}, \& {Bhattacharya}}]{nandi11}
{Nandi}, A., {Palit}, S., {Debnath}, D., {$et~al$.} 2011, Experimental
  Astronomy, 29, 55

\bibitem[{{Novick}(1975)}]{novick75}
{Novick}, R. 1975, Space Science Reviews, 18, 389

\bibitem[{{Novick} {$et~al$.}(1972){Novick}, {Weisskopf}, {Berthelsdorf},
  {Linke}, \& {Wolff}}]{novick72}
{Novick}, R., {Weisskopf}, M.~C., {Berthelsdorf}, R., {Linke}, R., \& {Wolff},
  R.~S. 1972, Astrophysical Journal Letter, 174, L1

\bibitem[{{Orsi} \& {Polar Collaboration}(2011)}]{orsi11}
{Orsi}, S., \& {Polar Collaboration}. 2011, Astrophysics and Space Sciences
  Transactions, 7, 43

\bibitem[{{Otte}(2006)}]{otte06}
{Otte}, N. 2006, in IX International Symposium on Detectors for Particle,
  Astroparticle and Synchrotron Radiation Experiments, SNIC Symposium,
  Stanford, California, 1--9

\bibitem[{{Paul} {$et~al$.}(2016){Paul}, {Gopala Krishna}, \& {Puthiya
  Veetil}}]{paul16}
{Paul}, B., {Gopala Krishna}, M.~R., \& {Puthiya Veetil}, R. 2016, in 41st
  COSPAR Scientific Assembly, Vol.~41, E1.15--8--16

\bibitem[{{Paul} {$et~al$.}(2010){Paul}, {Rishin}, {Maitra}, {Gopalakrishna},
  {Duraichelvan}, {Ateequlla}, {Cowsik}, {Devasia}, \& {Marykutty}}]{paul10}
{Paul}, B., {Rishin}, P.~V., {Maitra}, C., {$et~al$.} 2010, in The First Year
  of MAXI: Monitoring Variable X-ray Sources, 68P

\bibitem[{{Pavlov} \& {Zavlin}(2000)}]{pavlov00}
{Pavlov}, G.~G., \& {Zavlin}, V.~E. 2000, \apj, 529, 1011

\bibitem[{{Pearce} {$et~al$.}(2019){Pearce}, {Eliasson}, {Kumar Iyer}, {Kiss},
  {Kushwah}, {Larsson}, {Lundman}, {Mikhalev}, {Ryde}, {Stana}, {Takahashi}, \&
  {Xie}}]{pearce19}
{Pearce}, M., {Eliasson}, L., {Kumar Iyer}, N., {$et~al$.} 2019, Astroparticle
  Physics, 104, 54

\bibitem[{{Pe'er} \& {Ryde}(2011)}]{peer11}
{Pe'er}, A., \& {Ryde}, F. 2011, \apj, 732, 49

\bibitem[{{P{\'e}tri} \& {Kirk}(2005)}]{petri05}
{P{\'e}tri}, J., \& {Kirk}, J.~G. 2005, Astrophysical Journal Letter, 627, L37

\bibitem[{Produit {$et~al$.}(2018)Produit, Bao, Batsch, Bernasconi, Britvich,
  Cadoux, Cernuda, Chai, Dong, Gauvin, \& et~al.}]{produit18}
Produit, N., Bao, T., Batsch, T., {$et~al$.} 2018, Nuclear Instruments and
  Methods in Physics Research Section A: Accelerators, Spectrometers, Detectors
  and Associated Equipment, 877, 259–268

\bibitem[{{Rao} {$et~al$.}(2016){Rao}, {Chand}, {Hingar}, {Iyyani}, {Khanna},
  {Kutty}, {Malkar}, {Paul}, {Bhalerao}, {Bhattacharya}, {Dewangan}, {Pawar},
  {Vibhute}, {Chattopadhyay}, {Mithun}, {Vadawale}, {Vagshette}, {Basak},
  {Pradeep}, {Samuel}, {Sreekumar}, {Vinod}, {Navalgund}, {Pandiyan}, {Sarma},
  {Seetha}, \& {Subbarao}}]{rao16}
{Rao}, A.~R., {Chand}, V., {Hingar}, M.~K., {$et~al$.} 2016, \apj, 833, 86

\bibitem[{Rea \& Esposito(2010)}]{Rea10}
Rea, N., \& Esposito, P. 2010, Astrophysics and Space Science Proceedings,
  247–273

\bibitem[{{Romani} {$et~al$.}(2001){Romani}, {Miller}, {Cabrera}, {Nam}, \&
  {Martinis}}]{romani01}
{Romani}, R.~W., {Miller}, A.~J., {Cabrera}, B., {Nam}, S.~W., \& {Martinis},
  J.~M. 2001, Astrophysical Journal, 563, 221

\bibitem[{{Romero} {$et~al$.}(2014){Romero}, {Vieyro}, \& {Chaty}}]{romero14}
{Romero}, G.~E., {Vieyro}, F.~L., \& {Chaty}, S. 2014, Astronomy and
  Astrophysics, 562, L7

\bibitem[{{Roques} {$et~al$.}(2012){Roques}, {Jourdain}, {Bassani}, {Bazzano},
  {Belmont}, {Bird}, {Caroli}, {Chauvin}, {Clark}, {Gehrels}, {Goerlach},
  {Harrisson}, {Laurent}, {Malzac}, {Medina}, {Merloni}, {Paltani}, {Stephen},
  {Ubertini}, \& {Wilms}}]{roques12}
{Roques}, J.-P., {Jourdain}, E., {Bassani}, L., {$et~al$.} 2012, Experimental
  Astronomy, 34, 489

\bibitem[{{Rutledge} \& {Fox}(2004)}]{rutledge04}
{Rutledge}, R.~E., \& {Fox}, D.~B. 2004, \mnras, 350, 1288

\bibitem[{{Ryde}(2004)}]{ryde04}
{Ryde}, F. 2004, \apj, 614, 827

\bibitem[{{Sanaei} {$et~al$.}(2015){Sanaei}, {Baei}, \&
  {Sayyed-Alangi}}]{sanaei15}
{Sanaei}, B., {Baei}, M.~T., \& {Sayyed-Alangi}, S.~Z. 2015, Journal of Modern
  Physics, 6, 425

\bibitem[{{Sari}(1999)}]{sari99}
{Sari}, R. 1999, \apjl, 524, L43

\bibitem[{{Sauli}(1997)}]{sauli97}
{Sauli}, F. 1997, Nuclear Instruments and Methods in Physics Research A, 386,
  531

\bibitem[{{Sazonov} \& {Sunyaev}(2001)}]{sazonov01}
{Sazonov}, S.~Y., \& {Sunyaev}, R.~A. 2001, Astronomy and Astrophysics, 373,
  241

\bibitem[{{Schnittman} \& {Krolik}(2009)}]{schnittman09}
{Schnittman}, J.~D., \& {Krolik}, J.~H. 2009, \apj, 701, 1175

\bibitem[{{Schnittman} \& {Krolik}(2010)}]{schnittman10}
---. 2010, Astrophysical Journal, 712, 908

\bibitem[{{Schnopper} \& {Kalata}(1969)}]{schnopper69}
{Schnopper}, H.~W., \& {Kalata}, K. 1969, \aj, 74, 854

\bibitem[{Shahbaz {$et~al$.}(2016)Shahbaz, Russell, Covino, Mooley, Fender, \&
  Rumsey}]{shahbaz16_v404}
Shahbaz, T., Russell, D.~M., Covino, S., {$et~al$.} 2016, Monthly Notices of
  the Royal Astronomical Society, 463, 1822

\bibitem[{{Sharma} {$et~al$.}(2019){Sharma}, {Iyyani}, {Bhattacharya},
  {Chattopadhyay}, {Rao}, {Aarthy}, {Vadawale}, {Mithun}, {Bhalerao}, {Ryde},
  \& {Peer}}]{vidushi19}
{Sharma}, V., {Iyyani}, S., {Bhattacharya}, D., {$et~al$.} 2019, arXiv
  e-prints, arXiv:1908.10885

\bibitem[{{Shaviv} \& {Dar}(1995)}]{shaviv95}
{Shaviv}, N.~J., \& {Dar}, A. 1995, \apj, 447, 863

\bibitem[{{Silver} {$et~al$.}(1990){Silver}, {Holley}, {Ziock}, {Novick},
  {Kaaret}, {Weisskopf}, \& {Elsner}}]{silver90}
{Silver}, E., {Holley}, J., {Ziock}, K., {$et~al$.} 1990, Optical Engineering,
  29, 759

\bibitem[{{Silver} {$et~al$.}(1979){Silver}, {Weisskopf}, {Kestenbaum}, {Long},
  {Novick}, \& {Wolff}}]{silver79}
{Silver}, E.~H., {Weisskopf}, M.~C., {Kestenbaum}, H.~L., {$et~al$.} 1979,
  Astrophysical Journal, 232, 248

\bibitem[{{Singh} {$et~al$.}(2014){Singh}, {Tandon}, {Agrawal}, {Antia},
  {Manchanda}, {Yadav}, {Seetha}, {Ramadevi}, {Rao}, {Bhattacharya}, {Paul},
  {Sreekumar}, {Bhattacharyya}, {Stewart}, {Hutchings}, {Annapurni}, {Ghosh},
  {Murthy}, {Pati}, {Rao}, {Stalin}, {Girish}, {Sankarasubramanian},
  {Vadawale}, {Bhalerao}, {Dewangan}, {Dedhia}, {Hingar}, {Katoch}, {Kothare},
  {Mirza}, {Mukerjee}, {Shah}, {Shah}, {Mohan}, {Sangal}, {Nagabhusana},
  {Sriram}, {Malkar}, {Sreekumar}, {Abbey}, {Hansford}, {Beardmore}, {Sharma},
  {Murthy}, {Kulkarni}, {Meena}, {Babu}, \& {Postma}}]{singh14}
{Singh}, K.~P., {Tandon}, S.~N., {Agrawal}, P.~C., {$et~al$.} 2014, in Society
  of Photo-Optical Instrumentation Engineers (SPIE) Conference Series, Vol.
  9144, Society of Photo-Optical Instrumentation Engineers (SPIE) Conference
  Series

\bibitem[{{S{\l}owikowska} {$et~al$.}(2009){S{\l}owikowska}, {Kanbach},
  {Kramer}, \& {Stefanescu}}]{slowikowska09}
{S{\l}owikowska}, A., {Kanbach}, G., {Kramer}, M., \& {Stefanescu}, A. 2009,
  \mnras, 397, 103

\bibitem[{{Smith} {$et~al$.}(1988){Smith}, {Jones}, {Dick}, \&
  {Pike}}]{smith88}
{Smith}, F.~G., {Jones}, D.~H.~P., {Dick}, J.~S.~B., \& {Pike}, C.~D. 1988,
  \mnras, 233, 305

\bibitem[{{Soffitta}(1997)}]{soffitta97}
{Soffitta}, P. 1997, in Ital. Phys. Soc. Conf. Ser. 57: Frontier Objects in
  Astrophysics and Particle Physics, ed. F.~{Giovannelli} \& G.~{Mannocchi},
  561

\bibitem[{{Soffitta}(2017)}]{soffitta17}
{Soffitta}, P. 2017, in Society of Photo-Optical Instrumentation Engineers
  (SPIE) Conference Series, Vol. 10397, Society of Photo-Optical
  Instrumentation Engineers (SPIE) Conference Series, 103970I

\bibitem[{{St{\c{e}}{\'s}licki} {$et~al$.}(2016){St{\c{e}}{\'s}licki},
  {Sylwester}, {P{\l}ocieniak}, {Baka{\l}a}, {Szaforz}, {{\'S}cis{\l}owski},
  {Kowali{\'n}ski}, {Hernandez}, {Kuzin}, \& {Shestov}}]{steslicki16}
{St{\c{e}}{\'s}licki}, M., {Sylwester}, J., {P{\l}ocieniak}, S., {$et~al$.}
  2016, in IAU Symposium, Vol. 320, Solar and Stellar Flares and their Effects
  on Planets, ed. A.~G. {Kosovichev}, S.~L. {Hawley}, \& P.~{Heinzel}, 450--455

\bibitem[{{Suarez-Garcia} {$et~al$.}(2006){Suarez-Garcia}, {Hajdas}, {Wigger},
  {Arzner}, {G{\"u}del}, {Zehnder}, \& {Grigis}}]{garcia06}
{Suarez-Garcia}, E., {Hajdas}, W., {Wigger}, C., {$et~al$.} 2006, \solphys,
  239, 149

\bibitem[{Sun {$et~al$.}(2016)Sun, Wu, Bao, Batsch, Bernasconi, Britvitch,
  Cadoux, Cernuda, Chai, Dong, Gauvin, Hajdas, He, Kole, Kong, Kong,
  Lechanoine-Leluc, Li, Liu, Liu, Marcinkowski, Orsi, Pohl, Produit, Rapin,
  Rutczynska, Rybka, Shi, Song, Szabelski, Wang, Wen, Xiao, Xiong, Xu, Xu,
  Zhang, Zhang, Zhang, Zhang, Zhang, \& Zwolinska}]{sun16}
Sun, J.~C., Wu, B.~B., Bao, T.~W., {$et~al$.} 2016, in , 99052P--99052P--13

\bibitem[{{Tajima} \& {et al.}(2010)}]{tajima10}
{Tajima}, H., \& {et al.} 2010, in X-ray Polarimetry: A New Window in
  Astrophysics by Ronaldo Bellazzini, Enrico Costa, Giorgio Matt and Gianpiero
  Tagliaferri.~Cambridge University Press, 2010.~ ISBN: 9780521191845, p.~275,
  275

\bibitem[{{Tavecchio} {$et~al$.}(2018){Tavecchio}, {Landoni}, {Sironi}, \&
  {Coppi}}]{tavecchio18}
{Tavecchio}, F., {Landoni}, M., {Sironi}, L., \& {Coppi}, P. 2018, \mnras, 480,
  2872

\bibitem[{{Taverna} {$et~al$.}(2014){Taverna}, {Muleri}, {Turolla}, {Soffitta},
  {Fabiani}, \& {Nobili}}]{taverna14}
{Taverna}, R., {Muleri}, F., {Turolla}, R., {$et~al$.} 2014, /mnras, 438, 1686

\bibitem[{{Toma} {$et~al$.}(2009){Toma}, {Sakamoto}, {Zhang}, {Hill},
  {McConnell}, {Bloser}, {Yamazaki}, {Ioka}, \& {Nakamura}}]{toma08}
{Toma}, K., {Sakamoto}, T., {Zhang}, B., {$et~al$.} 2009, Astrophysical
  Journal, 698, 1042

\bibitem[{Tomsick {$et~al$.}(2019)Tomsick, Zoglauer, Sleator, Lazar, Beechert,
  Boggs, Roberts, Siegert, Lowell, Wulf, Grove, Phlips, Brandt, Smale, Kierans,
  Burns, Hartmann, Leising, Ajello, Fryer, Amman, Chang, Jean, \& von
  Ballmoos}]{tomsick19}
Tomsick, J.~A., Zoglauer, A., Sleator, C., {$et~al$.} 2019, The Compton
  Spectrometer and Imager, arXiv:1908.04334

\bibitem[{{Toraskar}(1975)}]{toraskar75}
{Toraskar}, J.~R. 1975, Applied Optics, 14, 1727

\bibitem[{{Tsunemi} {$et~al$.}(1992){Tsunemi}, {Hayashida}, {Tamura}, {Nomoto},
  {Wada}, {Hirano}, \& {Miyata}}]{tsunemi92}
{Tsunemi}, H., {Hayashida}, K., {Tamura}, K., {$et~al$.} 1992, Nuclear
  Instruments and Methods in Physics Research A, 321, 629

\bibitem[{{T{\"u}rler} {$et~al$.}(2010){T{\"u}rler}, {Chernyakova},
  {Courvoisier}, {Lubi{\'n}ski}, {Neronov}, {Produit}, \& {Walter}}]{turler10}
{T{\"u}rler}, M., {Chernyakova}, M., {Courvoisier}, T.~J.-L., {$et~al$.} 2010,
  Astronomy and Astrophysics, 512, A49

\bibitem[{Turolla {$et~al$.}(2015)Turolla, Zane, \& Watts}]{Turolla15}
Turolla, R., Zane, S., \& Watts, A.~L. 2015, Reports on Progress in Physics,
  78, 116901

\bibitem[{{Ubertini} {$et~al$.}(2003){Ubertini}, {Lebrun}, {Di Cocco},
  {Bazzano}, {Bird}, {Broenstad}, {Goldwurm}, {La Rosa}, {Labanti}, {Laurent},
  {Mirabel}, {Quadrini}, {Ramsey}, {Reglero}, {Sabau}, {Sacco}, {Staubert},
  {Vigroux}, {Weisskopf}, \& {Zdziarski}}]{ubertini03}
{Ubertini}, P., {Lebrun}, F., {Di Cocco}, G., {$et~al$.} 2003, Astronomy and
  Astrophysics, 411, L131

\bibitem[{{Uttley} {$et~al$.}(2014){Uttley}, {Cackett}, {Fabian}, {Kara}, \&
  {Wilkins}}]{uttley14}
{Uttley}, P., {Cackett}, E.~M., {Fabian}, A.~C., {Kara}, E., \& {Wilkins},
  D.~R. 2014, \aapr, 22, 72

\bibitem[{{Vadawale} {$et~al$.}(2012){Vadawale}, {Chattopadhyay}, \&
  {Pendharkar}}]{vadawale12}
{Vadawale}, S.~V., {Chattopadhyay}, T., \& {Pendharkar}, J. 2012, in Society of
  Photo-Optical Instrumentation Engineers (SPIE) Conference Series, Vol. 8443,
  Society of Photo-Optical Instrumentation Engineers (SPIE) Conference Series

\bibitem[{{Vadawale} {$et~al$.}(2015){Vadawale}, {Chattopadhyay}, {Rao},
  {Bhattacharya}, {Bhalerao}, {Vagshette}, {Pawar}, \&
  {Sreekumar}}]{vadawale15}
{Vadawale}, S.~V., {Chattopadhyay}, T., {Rao}, A.~R., {$et~al$.} 2015,
  Astronomy and Astrophysics, 578, 73

\bibitem[{{Vadawale} {$et~al$.}(2010){Vadawale}, {Paul}, {Pendharkar}, \&
  {Naik}}]{vadawale10}
{Vadawale}, S.~V., {Paul}, B., {Pendharkar}, J., \& {Naik}, S. 2010, Nuclear
  Instruments and Methods in Physics Research A, 618, 182

\bibitem[{{Vadawale} {$et~al$.}(2001){Vadawale}, {Rao}, \&
  {Chakrabarti}}]{vadawale01}
{Vadawale}, S.~V., {Rao}, A.~R., \& {Chakrabarti}, S.~K. 2001, Astronomy and
  Astrophysics, 372, 793

\bibitem[{{Vadawale} {$et~al$.}(2003){Vadawale}, {Rao}, {Naik}, {Yadav},
  {Ishwara-Chandra}, {Pramesh Rao}, \& {Pooley}}]{vadawale03}
{Vadawale}, S.~V., {Rao}, A.~R., {Naik}, S., {$et~al$.} 2003, Astrophysical
  Journal, 597, 1023

\bibitem[{{Vadawale} {$et~al$.}(2018){Vadawale}, {Chattopadhyay}, {Mithun},
  {Rao}, {Bhattacharya}, {Vibhute}, {Bhalerao}, {Dewangan}, {Misra}, {Paul},
  {Basu}, {Joshi}, {Sreekumar}, {Samuel}, {Priya}, {Vinod}, \&
  {Seetha}}]{vadawale17}
{Vadawale}, S.~V., {Chattopadhyay}, T., {Mithun}, N.~P.~S., {$et~al$.} 2018,
  Nature Astronomy, 2, 50

\bibitem[{{Veale} {$et~al$.}(2014){Veale}, {Bell}, {Duarte}, {Schneider},
  {Seller}, {Wilson}, \& {Iniewski}}]{veale14}
{Veale}, M.~C., {Bell}, S.~J., {Duarte}, D.~D., {$et~al$.} 2014, Nuclear
  Instruments and Methods in Physics Research A, 767, 218

\bibitem[{{Vedrenne} {$et~al$.}(2003){Vedrenne}, {Roques}, {Sch{\"o}nfelder},
  {Mandrou}, {Lichti}, {von Kienlin}, {Cordier}, {Schanne}, {Kn{\"o}dlseder},
  {Skinner}, {Jean}, {Sanchez}, {Caraveo}, {Teegarden}, {von Ballmoos},
  {Bouchet}, {Paul}, {Matteson}, {Boggs}, {Wunderer}, {Leleux},
  {Weidenspointner}, {Durouchoux}, {Diehl}, {Strong}, {Cass{\'e}}, {Clair}, \&
  {Andr{\'e}}}]{vedrenne03}
{Vedrenne}, G., {Roques}, J.-P., {Sch{\"o}nfelder}, V., {$et~al$.} 2003, \aap,
  411, L63

\bibitem[{{Viironen} \& {Poutanen}(2004)}]{viironen04}
{Viironen}, K., \& {Poutanen}, J. 2004, Astronomy and Astrophysics, 426, 985

\bibitem[{Wadiasingh {$et~al$.}(2017)Wadiasingh, Baring, Gonthier, \&
  Harding}]{wadiasingh17}
Wadiasingh, Z., Baring, M.~G., Gonthier, P.~L., \& Harding, A.~K. 2017,
  Resonant Inverse Compton Scattering Spectra from Highly-magnetized Neutron
  Stars, arXiv:1712.09643

\bibitem[{Wadiasingh {$et~al$.}(2019)Wadiasingh, Younes, Baring, Harding,
  Gonthier, Hu, van~der Horst, Zane, Kouveliotou, Beloborodov,
  Prescod-Weinstein, Chattopadhyay, Chandra, Kalapotharakos, Parfrey, Blumer,
  \& Kazanas}]{wadiasingh19_white}
Wadiasingh, Z., Younes, G., Baring, M.~G., {$et~al$.} 2019, Magnetars as
  Astrophysical Laboratories of Extreme Quantum Electrodynamics: The Case for a
  Compton Telescope, arXiv:1903.05648

\bibitem[{{Waxman}(2003)}]{waxman03}
{Waxman}, E. 2003, Nature, 423, 388

\bibitem[{{Weisskopf}(2018)}]{weisskopf18}
{Weisskopf}, M. 2018, Galaxies, 6, 33

\bibitem[{{Weisskopf} {$et~al$.}(1976){Weisskopf}, {Cohen}, {Kestenbaum},
  {Long}, {Novick}, \& {Wolff}}]{weisskopf76}
{Weisskopf}, M.~C., {Cohen}, G.~G., {Kestenbaum}, H.~L., {$et~al$.} 1976,
  Astrophysical Journal Letter, 208, L125

\bibitem[{{Weisskopf} {$et~al$.}(2009){Weisskopf}, {Elsner}, {Kaspi}, {O'Dell},
  {Pavlov}, \& {Ramsey}}]{weisskopf09}
{Weisskopf}, M.~C., {Elsner}, D., {Kaspi}, V.~M., {$et~al$.} 2009, X-Ray
  Polarimetry and Its Potential Use for Understanding Neutron Stars, ed.
  I.~B.~W. (eds) (Berlin, Heidelberg: Springer)

\bibitem[{{Weisskopf} {$et~al$.}(1978){Weisskopf}, {Silver}, {Kestenbaum},
  {Long}, \& {Novick}}]{weisskopf78}
{Weisskopf}, M.~C., {Silver}, E.~H., {Kestenbaum}, H.~L., {Long}, K.~S., \&
  {Novick}, R. 1978, Astrophysical Journal Letter, 220, L117

\bibitem[{{Weisskopf} {$et~al$.}(2016){Weisskopf}, {Ramsey}, {O'Dell},
  {Tennant}, {Elsner}, {Soffitta}, {Bellazzini}, {Costa}, {Kolodziejczak},
  {Kaspi}, {Muleri}, {Marshall}, {Matt}, \& {Romani}}]{weisskopf16}
{Weisskopf}, M.~C., {Ramsey}, B., {O'Dell}, S., {$et~al$.} 2016, in Society of
  Photo-Optical Instrumentation Engineers (SPIE) Conference Series, Vol. 9905,
  Space Telescopes and Instrumentation 2016: Ultraviolet to Gamma Ray, 990517

\bibitem[{{Wigger} {$et~al$.}(2004){Wigger}, {Hajdas}, {Arzner}, {G{\"u}del},
  \& {Zehnder}}]{wigger04}
{Wigger}, C., {Hajdas}, W., {Arzner}, K., {G{\"u}del}, M., \& {Zehnder}, A.
  2004, Astrophysical Journal, 613, 1088

\bibitem[{{Willis} {$et~al$.}(2005){Willis}, {Barlow}, {Bird}, {Clark}, {Dean},
  {McConnell}, {Moran}, {Shaw}, \& {Sguera}}]{willis05}
{Willis}, D.~R., {Barlow}, E.~J., {Bird}, A.~J., {$et~al$.} 2005, \aap, 439,
  245

\bibitem[{{Winkler} {$et~al$.}(2003){Winkler}, {Courvoisier}, {Di Cocco},
  {Gehrels}, {Gim{\'e}nez}, {Grebenev}, {Hermsen}, {Mas-Hesse}, {Lebrun},
  {Lund}, {Palumbo}, {Paul}, {Roques}, {Schnopper}, {Sch{\"o}nfelder},
  {Sunyaev}, {Teegarden}, {Ubertini}, {Vedrenne}, \& {Dean}}]{winkler03}
{Winkler}, C., {Courvoisier}, T.~J.~L., {Di Cocco}, G., {$et~al$.} 2003, \aap,
  411, L1

\bibitem[{{Wolff} {$et~al$.}(1970){Wolff}, {Angel}, {Novick}, \& {vanden
  Bout}}]{wolff70}
{Wolff}, R.~S., {Angel}, J.~R.~P., {Novick}, R., \& {vanden Bout}, P. 1970,
  Astrophysical Journal Letter, 160, L21

\bibitem[{Wu {$et~al$.}(2010)Wu, McNamara, \& Kuncic}]{Wu09}
Wu, K., McNamara, A., \& Kuncic, Z. 2010, X-Ray Polarimetry, 187–194

\bibitem[{Yang {$et~al$.}(2018)Yang, Lowell, Zoglauer, Tomsick, Chiu, Kierans,
  Sleator, Boggs, Chang, Jean, McBride, Mochizuki, Amman, von Ballmoos, Chang,
  Chu, Liang, \& Lin}]{yang18}
Yang, C.-Y., Lowell, A., Zoglauer, A., {$et~al$.} 2018, in Space Telescopes and
  Instrumentation 2018: Ultraviolet to Gamma Ray, ed. J.-W.~A. den Herder,
  S.~Nikzad, \& K.~Nakazawa, Vol. 10699, International Society for Optics and
  Photonics (SPIE), 642 -- 651

\bibitem[{{Yonetoku} {$et~al$.}(2006){Yonetoku}, {Murakami}, {Masui},
  {Kodaira}, {Aoyama}, {Gunji}, {Tokanai}, \& {Mihara}}]{yonetoku06}
{Yonetoku}, D., {Murakami}, T., {Masui}, H., {$et~al$.} 2006, in Society of
  Photo-Optical Instrumentation Engineers (SPIE) Conference Series, Vol. 6266,
  Society of Photo-Optical Instrumentation Engineers (SPIE) Conference Series

\bibitem[{{Yonetoku} {$et~al$.}(2011){Yonetoku}, {Murakami}, {Gunji}, {Mihara},
  {Toma}, {Sakashita}, {Morihara}, {Takahashi}, {Toukairin}, {Fujimoto},
  {Kodama}, \& {Kubo}}]{yonetoku11}
{Yonetoku}, D., {Murakami}, T., {Gunji}, S., {$et~al$.} 2011, Astrophysical
  Journal Letter, 743, L30

\bibitem[{{Yonetoku} {$et~al$.}(2012){Yonetoku}, {Murakami}, {Gunji}, {Mihara},
  {Toma}, {Morihara}, {Takahashi}, {Wakashima}, {Yonemochi}, {Sakashita},
  {Toukairin}, {Fujimoto}, \& {Kodama}}]{yonetoku12}
---. 2012, \apjl, 758, L1

\bibitem[{{Zdziarski} {$et~al$.}(2014){Zdziarski}, {Pjanka}, {Sikora}, \&
  {Stawarz}}]{zdziarski14}
{Zdziarski}, A.~A., {Pjanka}, P., {Sikora}, M., \& {Stawarz}, {\L}. 2014,
  \mnras, 442, 3243

\bibitem[{{Zhang} \& {Yan}(2011)}]{zhang11}
{Zhang}, B., \& {Yan}, H. 2011, \apj, 726, 90

\bibitem[{{Zhang} {$et~al$.}(2012){Zhang}, {Herman}, {He}, {De Geronimo},
  {Vernon}, \& {Fried}}]{zhang12_3DCZT}
{Zhang}, F., {Herman}, C., {He}, Z., {$et~al$.} 2012, IEEE Transactions on
  Nuclear Science, 59, 236

\bibitem[{{Zhang}(2017)}]{zhang17}
{Zhang}, H. 2017, Galaxies, 5, 32

\bibitem[{{Zhang} \& {B{\"o}ttcher}(2013)}]{zhang13}
{Zhang}, H., \& {B{\"o}ttcher}, M. 2013, Astrophysical Journal, 774, 18

\bibitem[{{Zhang} {$et~al$.}(2014){Zhang}, {Chen}, \& {B{\"o}ttcher}}]{zhang14}
{Zhang}, H., {Chen}, X., \& {B{\"o}ttcher}, M. 2014, \apj, 789, 66

\bibitem[{{Zhang} {$et~al$.}(2016){Zhang}, {Feroci}, {Santangelo}, {Dong},
  {Feng}, {Lu}, {Nandra}, {Wang}, {Zhang}, {Bozzo}, {Brandt}, {De Rosa}, {Gou},
  {Hernanz}, {van der Klis}, {Li}, {Liu}, {Orleanski}, {Pareschi}, {Pohl},
  {Poutanen}, {Qu}, {Schanne}, {Stella}, {Uttley}, {Watts}, {Xu}, {Yu}, {in 't
  Zand}, {Zane}, {Alvarez}, {Amati}, {Baldini}, {Bambi}, {Basso},
  {Bhattacharyya S.}, {}, {Belloni}, {Bellutti}, {Bianchi}, {Brez}, {Bursa},
  {Burwitz}, {Budtz-J{\o}rgensen}, {Caiazzo}, {Campana}, {Cao}, {Casella},
  {Chen}, {Chen}, {Chen}, {Chen}, {Chen}, {Chen}, {Civitani}, {Coti Zelati},
  {Cui}, {Cui}, {Dai}, {Del Monte}, {de Martino}, {Di Cosimo}, {Diebold},
  {Dovciak}, {Donnarumma}, {Doroshenko}, {Esposito}, {Evangelista}, {Favre},
  {Friedrich}, {Fuschino}, {Galvez}, {Gao}, {Ge}, {Gevin}, {Goetz}, {Han},
  {Heyl}, {Horak}, {Hu}, {Huang}, {Huang}, {Hudec}, {Huppenkothen}, {Israel},
  {Ingram}, {Karas}, {Karelin}, {Jenke}, {Ji}, {Korpela}, {Kunneriath},
  {Labanti}, {Li}, {Li}, {Li}, {Liang}, {Limousin}, {Lin}, {Ling}, {Liu},
  {Liu}, {Liu}, {Lu}, {Lund}, {Lai}, {Luo}, {Luo}, {Ma}, {Mahmoodifar},
  {Marisaldi}, {Martindale}, {Meidinger}, {Men}, {Michalska}, {Mignani},
  {Minuti}, {Motta}, {Muleri}, {Neilsen}, {Orlandini}, {Pan}, {Patruno},
  {Perinati}, {Picciotto}, {Piemonte}, {Pinchera}, {Rachevski A.}, {Rapisarda},
  {Rea}, {Rossi}, {Rubini}, {Sala}, {Shu}, {Sgro}, {Shen}, {Soffitta}, {Song},
  {Spandre}, {Stratta}, {Strohmayer}, {Sun}, {Svoboda}, {Tagliaferri},
  {Tenzer}, {Hong}, {Taverna}, {Torok}, {Turolla}, {Vacchi}, {Wang}, {Walton},
  {Wang}, {Wang}, {Wang}, {Wang}, {Weng}, {Wilms}, {Winter}, {Wu}, {Wu},
  {Xiong}, {Xu}, {Xue}, {Yan}, {Yang}, {Yang}, {Yang}, {Yuan}, {Yuan}, {Yuan},
  {Zampa}, {Zampa}, {Zdziarski}, {Zhang}, {Zhang}, {Zhang}, {Zhang}, {Zhang},
  {Zhang}, {Zheng}, {Zhou}, \& {Zhou X.~L.}}]{zhang16_extp}
{Zhang}, S.~N., {Feroci}, M., {Santangelo}, A., {$et~al$.} 2016, in Society of
  Photo-Optical Instrumentation Engineers (SPIE) Conference Series, Vol. 9905,
  Space Telescopes and Instrumentation 2016: Ultraviolet to Gamma Ray, 99051Q

\bibitem[{{Zhang} {$et~al$.}(2019){Zhang}, {Kole}, {Bao}, {Batsch},
  {Bernasconi}, {Cadoux}, {Chai}, {Dai}, {Dong}, {Gauvin}, {Hajdas}, {Lan},
  {Li}, {Li}, {Li}, {Liu}, {Liu}, {Marcinkowski}, {Produit}, {Orsi}, {Pohl},
  {Rybka}, {Shi}, {Song}, {Sun}, {Szabelski}, {Tymieniecka}, {Wang}, {Wang},
  {Wen}, {Wu}, {Wu}, {Wu}, {Xiao}, {Xiong}, {Zhang}, {Zhang}, {Zhang}, {Zhang},
  \& {Zwolinska}}]{zhang19}
{Zhang}, S.-N., {Kole}, M., {Bao}, T.-W., {$et~al$.} 2019, Nature Astronomy, 3,
  258

\bibitem[{{Zharkova} {$et~al$.}(2010){Zharkova}, {Kuznetsov}, \&
  {Siversky}}]{zharkova10}
{Zharkova}, V.~V., {Kuznetsov}, A.~A., \& {Siversky}, T.~V. 2010, Astronomy and
  Astrophysics, 512, A8

\bibitem[{{Zhitnik} {$et~al$.}(2006){Zhitnik}, {Logachev}, {Bogomolov},
  {Denisov}, {Kavanosyan}, {Kuznetsov}, {Morozov}, {Myagkova}, {Svertilov},
  {Ignat'ev}, {Oparin}, {Pertsov}, \& {Tindo}}]{zhitnik06}
{Zhitnik}, I.~A., {Logachev}, Y.~I., {Bogomolov}, A.~V., {$et~al$.} 2006, Solar
  System Research, 40, 93

\end{thebibliography}



\end{document}